\documentclass[pdflatex]{sn-jnl}%

\usepackage[
  style=apa,               %
  backend=biber,           %
  sorting=nyt,             %
  doi=true,
  url=true,
]{biblatex}%

\usepackage{array}%
\usepackage{booktabs}%
\usepackage{calc}%
\usepackage{graphicx}%
\usepackage{amsmath,amssymb,amsfonts,amstext}%
\usepackage{calc}%
\usepackage{amsthm}%
\usepackage{csquotes}%
\usepackage{enumitem}%
\usepackage{float}%
\usepackage{mathrsfs}%
\usepackage[nameinlink]{cleveref}
\usepackage[title]{appendix}%
\usepackage[table, xcdraw]{xcolor}%
\usepackage{textcomp}%
\usepackage{manyfoot}%
\usepackage{booktabs}%
\usepackage{algorithm}%
\usepackage{algorithmicx}%
\usepackage{algpseudocode}%
\usepackage{fancyvrb}%
\usepackage{listings}%
\usepackage{longtable}
\usepackage{caption}
\usepackage{etoolbox}
\usepackage{subcaption}
\usepackage{pgfplots}%
\usepackage{multicol}%
\usepackage{multirow}%
\usepackage{pslatex}%
\usepackage{utfsym}%
\usepackage{rotating}%
\usepackage{tikz}%
\usepackage{wrapfig}%
\usepackage{xspace}%
\usepackage{xparse}%
\usetikzlibrary{shapes.geometric, arrows, positioning, backgrounds,patterns}

\usepackage{orcidlink}

\newcommand{\eg}{\textit{e.g.,}~}
\newcommand{\ie}{\textit{i.e.,}~}
\newcommand{\etc}{\textit{etc.}}
\newcommand{\cf}{\textit{cf.}~}
\newcommand{\one}{({\em i})\xspace}
\newcommand{\two}{({\em ii})\xspace}
\newcommand{\three}{({\em iii})\xspace}
\newcommand{\four}{({\em iv})\xspace}
\newcommand{\five}{({\em v})\xspace}
\newcommand{\six}{({\em vi})\xspace}

\newcommand{\ccitetext}[1]{\xspace\textcite{#1}}
\newcommand{\cciteparen}[1]{~\parencite{#1}}
\newcommand{\ccitecf}[1]{~\parencite[\cf][]{#1}}
\newcommand{\cciteeg}[1]{~\parencite[\eg][]{#1}}
\newcommand{\ccitepage}[2]{~\parencite[][~p.~#2]{#1}}

\newlist{question}{enumerate}{2}
\setlist[question,1]{label=RQ\arabic*.,ref=RQ\arabic*,leftmargin=1.3cm,labelwidth=*}
\setlist[question,2]{label=(\alph*),ref=\thequestioni(\alph*)leftmargin=10cm,labelwidth=*}

\newlist{hypothesis}{enumerate}{2}
\setlist[hypothesis,1]{label=H\arabic*.,ref=H\arabic*,leftmargin=1.3cm,labelwidth=*}
\setlist[hypothesis,2]{label=(\alph*),ref=\thehypothesisi(\alph*)leftmargin=10cm,labelwidth=*}

\newlist{inclusioncriteria}{enumerate}{2}
\setlist[inclusioncriteria,1]{label=IC\arabic*.,ref=IC\arabic*,leftmargin=1.3cm,labelwidth=*}
\setlist[inclusioncriteria,2]{label=(\alph*),ref=\theinclusioncriteriai(\alph*)leftmargin=10cm,labelwidth=*}

\pgfplotsset{compat=1.18}

\tikzstyle{slrnode} = [rectangle, rounded corners, 
minimum width=3cm, 
minimum height=1cm,
text centered,
text width=6cm,
draw=black, 
fill=lightgray!30]
\tikzstyle{slrnode1} = [rectangle, rounded corners, 
minimum width=3cm, 
minimum height=1cm,
text centered,
text width=6.1cm,
draw=black, 
fill=lightgray!30]
\tikzstyle{slrnode2} = [rectangle, rounded corners, 
minimum width=4.1cm, 
minimum height=1cm,
text centered,
text width=3.5cm,
draw=black, 
fill=lightgray!30]
\tikzstyle{slrnodewide} = [rectangle, rounded corners, 
minimum width=4cm, 
minimum height=1cm,
text centered,
text width=12.5cm,
draw=black, 
fill=lightgray!30]
\tikzstyle{process} = [rectangle, 
minimum width=3cm, 
minimum height=1cm,
text centered, 
text width=12.5cm,
draw=black, 
fill=darkgray!30]
\tikzstyle{arrow} = [thick,->,>=stealth]

\DeclareQuoteStyle[american]{english}
        {\itshape\textquotedblleft}
        [\textquotedblleft]
        {\textquotedblright}
        [0.05em]
        {\textquoteleft}
        {\textquoteright}

\makeatletter
\disable@package@load{program}{}
\newcommand{\expr}[3]{%
    \textit{\ref*{#1}-E#3}%
    \protected@edef\@currentlabel{\ref*{#1}-E#3}%
    \label{#2}%
}
\makeatother

\makeatletter
\tikzset{inside/.code=\preto\tikz@auto@anchor{\pgf@x-\pgf@x\pgf@y-\pgf@y}}
\makeatother

\newcounter{varcounter}
\makeatletter
\newcommand{\var}[1]{%
    V\the\numexpr\value{varcounter}+1\relax\refstepcounter{varcounter}%
    \protected@edef\@currentlabel{V\thevarcounter}%
    \label{#1}%
}
\makeatother

\ExplSyntaxOn
\NewDocumentCommand{\getNth}{mmm}
  {
    \seq_set_split:Nnx \l_tmpa_seq { #2 } { #1 }
    \seq_item:Nn \l_tmpa_seq { #3 }
  }
\ExplSyntaxOff

\usepackage{pifont}
\newcommand{\x}{\ding{51}}%

\newcommand{\goal}{\ref{v:adaptationGoal} adaptation goal}
\newcommand{\basis}{\ref{v:adaptationBasis} adaptation basis}
\newcommand{\effect}{\ref{v:reportedEffects} reported effects}

\newcommand{\abknowledge}{learners' knowledge}
\newcommand{\abpath}{learning path}
\newcommand{\abaffect}{affect}
\newcommand{\abmetacognition}{metacognition}
\newcommand{\abstyle}{learning style}

\newcommand{\all}{ITS Instruction Loops\cciteparen{vanlehn2006behavior}}
\newcommand{\bloom}{Blooms taxonomy\cciteparen{bloom1956taxonomy}}
\newcommand{\flippedca}{Flipped classroom approach\cciteparen{bergmann2012flip}}
\newcommand{\mastery}{Mastery Learning\cciteparen{bloom1968learning}}

\newcommand{\sap}{students as partners approach\cciteparen{matthews2017five}}
\newcommand{\srl}{self-regulated learning\cciteparen{zimmerman2000attaining}}

\newcommand{\fslsm}{Felder-Silverman learning style model\cciteparen{felder1988learning}}
\newcommand{\halsm}{Honey-Alonso learning style model\cciteparen{alonso1994estilos}}
\newcommand{\hmlsm}{Honey-Mumford learning style model\cciteparen{mumford1992manual}}
\newcommand{\kolblsm}{Kolb learning style model\cciteparen{kolb1984experiential}}
\newcommand{\mbti}{Myers-Briggs Type Indicator\cciteparen{myers1962myers}}
\newcommand{\varklsm}{VARK learning style model\cciteparen{fleming1992not}}

\begin{document}

\title[Adaptive Learning Mechanisms for Learning Management Systems: A Scoping Review and Practical Considerations]{Adaptive Learning Mechanisms for Learning Management Systems: A Scoping Review and Practical Considerations}

\author[1]{\fnm{Sebastian} \sur{Kucharski}\,\orcidlink{0009-0003-4210-5281}}\email{sebastian.kucharski@tu-dresden.de}
\author[1]{\fnm{Iris} \sur{Braun}\,\orcidlink{0009-0000-0900-2158}}\email{iris.braun@tu-dresden.de}
\author[2]{\fnm{Gregor} \sur{Damnik}\,\orcidlink{0000-0001-9829-6994}}\email{gregor.damnik@tu-dresden.de}
\author[1]{\fnm{Matthias} \sur{W\"ahlisch}\,\orcidlink{0000-0002-3825-2807}}\email{m.waehlisch@tu-dresden.de}
\affil[1]{\orgdiv{Chair of Distributed and Networked Systems}, \orgname{TUD Dresden University of Technology}, \orgaddress{\street{Helmholtzstr. 10}, \city{Dresden}, \postcode{01069}, \country{Germany}}}
\affil[2]{\orgdiv{Center for Teacher Education and Educational Research}, \orgname{TUD Dresden University of Technology}, \orgaddress{\street{Zellescher Weg 20}, \city{Dresden}, \postcode{01217}, \country{Germany}}}

\abstract{\textbf{Background:} Traditional Learning Management Systems (LMSs) usually offer one-size-fits-all solutions that cannot be customized to meet specific learner needs. To address this issue, adaptive learning mechanisms (ALMs) are integrated either by LMS-specific approaches into individual LMSs or by system-independent approaches into various existing LMSs.

\textbf{Objective:} We conducted a scoping review of the literature addressing the following research questions. How are ALMs integrated into LMSs system-independently? How are they provided, how are they specified, and what data do they process as the basis for adaptation? What are the educational surrounding conditions and outcomes of their application? A priori, we proposed three hypotheses. First, the focused ALMs rarely consider existing data. Second, they usually support a limited number of data processing mechanisms. Third, the users intended to provide them, are rarely given the ability to adapt how they work. Furthermore, to investigate the differences between system-independent and LMS-specific approaches, we also included the latter.

\textbf{Design:} We used Scopus, Web of Science, and Google Scholar to identify 3370 papers, excluding duplicates, that had been published between 2003 and 2023 for screening, and conducted a snowball search.

\textbf{Results:} We identified 61 relevant approaches and extracted eleven variables for them through in-depth reading. The results support the proposed hypotheses.

\textbf{Conclusion:} Based on the challenges raised by these hypotheses with regard to the relevant target user groups, we defined two future research directions - developing a conceptual model for the system-independent specification of ALMs and a corresponding architecture for the provision, and supporting the authoring of these mechanisms.} 

\maketitle

\section{Introduction}
\label{sec:introduction}

Learning Management Systems (LMSs) are well-established tools in digital education\cciteparen{ashrafi_exploring_2022} and a continuous target of research in the field of technology-enhanced learning\cciteparen{prahani_learning_2022}.
They have two major advantages\cciteparen{prahani_learning_2022}.
First, they provide learning materials that can be accessed anywhere, anytime.
Second, they create a centralized communication channel between instructors and learners in asynchronous setups.
With the functionalities they provide, they can support all aspects of digital learning processes\cciteparen{watson_argument_2007}.
Thus, the COVID-19 pandemic has even increased the need for and the focus on LMSs\cciteparen{camilleri_acceptance_2022}.
However, LMSs are frequently referred to as one-size-fits-all solutions.
As such, their lack of adaptability to different characteristics of the heterogeneous learners, has been identified as a major drawback\cciteparen{giuffra_palomino_intelligent_2014,kaouni2023design}.

The goal of research related to adaptive learning is overcoming the drawbacks of using one-size-fits-all approaches to learning\cciteparen{ochukut_research_2023}.
In this context, various adaptive learning systems (ALSs) were developed that can adapt the learning process to the learner's needs, such as Adaptive Hypermedia Systems\cciteeg{brusilovsky1998methods}, Intelligent Tutoring Systems\cciteeg{almasri2019intelligent}, Educational Recommender Systems\cciteeg{da2023systematic}, Personalized Learning Systems\cciteeg{tang2020personalized}, or Adaptive Instructional Systems\cciteeg{betts2021personalized}.
These systems aim to adapt certain characteristics of the learning process, such as the learning content or the user interface\cciteparen{mikic_personalisation_2022}, to certain characteristics of the learners, such as learning style, prior knowledge, or interests\cciteparen{bernacki2021systematic}.

The mechanisms that implement the adaptation of the learning process are referred to as adaptive learning mechanisms (ALMs).
To ensure flexibility in the ALMs that can be applied, ALSs are often developed as stand-alone systems.
This procedure has two drawbacks.
First, existing learning content cannot be used.
Second, didactic adaptation concepts implemented in one ALS cannot be reused for other systems.
Both drawbacks result in the need for re-authoring.
To address the drawbacks of one-size-fits-all solutions, research has been conducted on integrating ALMs into LMSs. 
To address the drawbacks of re-authoring, this research has moved toward ALMs that are not integrated into a single LMS.
Instead, these ALMs are provided by independent ALSs, which interact with various LMSs.
We refer to this as system-independent ALMs for LMSs. 

We conducted a scoping review according to the guidelines proposed by \ccitetext{tricco_prisma_2018} to provide a structured overview of this research.
In addition, we illustrate differences between system-specific and system-independent approaches by including LMS-specific approaches in our review as well.
This paper presents the procedure and the results.

\begin{figure}[H]
  \centering
  \resizebox{.9\linewidth}{!}{%
\begin{tikzpicture}[->,shorten >=1pt,auto,node distance=2em,every node/.style={inner sep=0.4em, node distance=2em]},thick,main node/.style={rounded corners=3pt,dashed,minimum width=60em,font=\sffamily\Large\bfseries,node distance=1cm},every label/.style={font=\sffamily\Large\bfseries},inner node/.style={rounded corners=3pt,draw,minimum width=18em,font=\sffamily\large\bfseries, minimum height=1.5em,text height=1.5ex,text depth=.25ex},multiline node/.style={rounded corners=3pt,draw,minimum width=18em,font=\sffamily\large\bfseries},subcat node/.style={draw,dashed,rounded corners=3pt,minimum width=55em,font=\sffamily\bfseries,}]
    \node[main node] (1) [fill=gray!25,inner sep=1em, minimum height=3.5em] {\Cref{sec:introduction}: Introduction};
    \node[main node] (2) [fill=gray!5,below of=1, yshift=-6.5em, minimum height=15.5em, label={[inside,inner sep=1em]north:{\Cref{sec:background}: Background}}]{};
    \node[subcat node] (201) [below of=2,yshift=4em, minimum height=5em, label={[inside,inner sep=0.5em,style={font=\sffamily\large\bfseries}]north west:{Theoretical Foundation}}] {};
    \node[inner node] (2011) [below of=201,minimum width=17em, xshift=-18em, yshift=1em] {Terminology};
    \node[inner node] (2012) [right of=2011,minimum width=17em, node distance=18em] {Descriptive Framework};
    \node[inner node] (2013) [right of=2012,minimum width=17em, node distance=18em] {Educational Framework};
    \node[subcat node] (202) [below of=2,yshift=-2em, minimum height=5em, label={[inside,inner sep=0.5em,style={font=\sffamily\large\bfseries}]north west:{Review Objectives}}] {};
    \node[inner node] (2021) [below of=202,minimum width=17em, xshift=-18em, yshift=1em] {Rationale};
    \node[inner node] (2022) [right of=2021,minimum width=17em, node distance=18em] {Research Questions};
    \node[inner node] (2023) [right of=2022,minimum width=17em, node distance=18em] {Hypotheses};
    \node[main node] (3) [fill=gray!25,below of=2,yshift=-11em, minimum height=13.5em, label={[inside,inner sep=1em]north:{\Cref{sec:relatedWork}: Related Work}}]{};
    \node[subcat node] (301) [below of=3,yshift=2em, minimum height=6.5em, label={[inside,inner sep=0.5em,style={font=\sffamily\large\bfseries}]north west:{Related Reviews}}] {};
    \node[multiline node] (3011) [below of=301,minimum width=11em,text width=11em,align=center,xshift=-20em, yshift=1em] {Adaptive Learning Mechanisms};
   	\node[multiline node] (3012) [right of=3011,minimum width=11em,text width=11em,align=center,node distance=13em] {Intelligent Tutoring Systems};
   	\node[multiline node] (3013) [right of=3012,minimum width=11em,text width=11em,align=center,node distance=13em] {Learning Management Systems};
   	\node[multiline node] (3014) [right of=3013,minimum width=11em,text width=11em,align=center,node distance=13em] {Didactic and Pedagogical Properties};
    \node[inner node] (302) [below of=3,minimum width=20em, yshift=-3em] {Delimitation and Contribution};
    \node[main node] (4) [fill=gray!5,below of=3, yshift=-6.5em, minimum height=6em, label={[inside,inner sep=1em]north:{\Cref{sec:method}: Method}}] {};
    \node[inner node] (401) [below of=4,minimum width=13em,minimum height=1.5em,yshift=1em,xshift=-21em]{Inclusion Criteria};
    \node[inner node] (402) [right of=401,minimum width=13em,minimum height=1.5em,node distance=14em] {Literature Databases};
    \node[inner node] (403) [right of=402,minimum width=13em,minimum height=1.5em,node distance=14em] {Search Strategy};
    \node[inner node] (404) [right of=403,minimum width=13em,minimum height=1.5em,node distance=14em] {Data Charting};
    \node[main node] (5) [fill=gray!25,below of=4, yshift=-8.5em, minimum height=16em, label={[inside,inner sep=1em]north:{\Cref{sec:results}: Results}}] {};
    \node[inner node] (501) [below of=5,minimum width=55em,minimum height=1.5em,yshift=6em]{Implementation of Adaptive Learning Mechanisms};
    \node[inner node] (502) [below of=501,minimum width=55em,minimum height=1.5em,node distance=2.5em] {Supported Adaptive Learning Mechanisms};
    \node[inner node] (503) [below of=502,minimum width=55em,minimum height=1.5em,node distance=2.5em] {Authoring and Initiation of Adaptive Learning Mechanisms};
    \node[inner node] (504) [below of=503,minimum width=55em,minimum height=1.5em,node distance=2.5em] {Considered Learning Content and Learning Data};
    \node[inner node] (505) [below of=504,minimum width=55em,minimum height=1.5em,node distance=2.5em] {Educational Surrounding Conditions and Outcomes};
    \node[main node] (6) [fill=gray!5,below of=5, yshift=-9em, minimum height=7em, label={[inside,inner sep=1em]north:{\Cref{sec:discussion}: Discussion}}] {};
    \node[inner node] (601) [below of=6,minimum width=27em,minimum height=1.5em,yshift=1em,xshift=-14em]{Implications};
    \node[inner node] (602) [right of=601,minimum width=27em,minimum height=1.5em,node distance=28em] {Limitations};
    \node[main node] (7) [fill=gray!25,below of=6, yshift=-2em,inner sep=1em, minimum height=3.5em] {\Cref{sec:futureProspects}: Future Prospects};
    \node[main node] (8) [fill=gray!5,below of=7, yshift=-0.5em,inner sep=1em, minimum height=3.5em] {\Cref{sec:conclusion}: Conclusion};
  \end{tikzpicture}   }
  \caption{Content structure of this review.}
  \label{fig:paperStructure}
\end{figure}

The paper is structured as shown in \Cref{fig:paperStructure}.
\Cref{sec:background} describes the terminology used and the underlying theoretical framework, introduces the objectives and the hypotheses of the review, and formulates the research questions addressed.
\Cref{sec:relatedWork} lists related reviews, discusses differences from this review, and highlights new insights.
\Cref{sec:method} describes the methodology used during the review, including the definition of inclusion criteria, the construction of the search string, and the specification of the extracted variables.
\Cref{sec:results} presents the results of the review.
\Cref{sec:discussion} discusses these results in terms of the hypothesis underlying the review and describes their implications as well as the limitations of our review.
\Cref{sec:futureProspects} highlights further research directions and summarizes the next steps of our research.
\Cref{sec:conclusion} concludes this review.
\section{Background and Objectives}
\label{sec:background}

Different terms are used to refer to methods, functionalities, and systems that adapt learning processes to learner characteristics.
In the absence of commonly agreed definitions, these terms are sometimes used synonymously and sometimes with different meanings\cciteparen{shemshack_systematic_2020}.
In \Cref{sec:terminology}, we define the terms that we use.
Afterwards, we describe the theoretical framework of our work. 
It consists of two parts.
First, in \Cref{sec:descriptiveFramework}, we explain the descriptive framework on which our understanding of ALMs is based.
Second, in \Cref{sec:educationalFramework}, we describe the educational framework that we use to contextualize the compiled approaches.
This allows us to connect these approaches to established educational theories and analyze their educational impact.
Finally, in \Cref{sec:rationale}, we describe the objectives of this review, the research questions we address, and the underlying hypotheses.

\subsection{Terminology}
\label{sec:terminology}

\ccitetext{shemshack_systematic_2020} provide a systematic review of the most commonly used terms regarding the consideration of specific learner characteristics for digital learning.
They define personalized learning as \textquote{a complex activity approach that is the product of self-organization (Chatti, 2010; Miliband, 2006) or learning and customized instruction that considers individual needs and goals}\ccitepage{shemshack_systematic_2020}{2}.
Furthermore, they state that \textit{personalized learning} is an umbrella term, and many other terms are used as substitues, even though they do not have the same meaning.
\textit{Adaptive learning} is one of these terms\cciteparen{xie2019trends}.
Furthermore,\ccitetext{shemshack_systematic_2020} characterize ALSs as \textquote{computerized learning systems that adapt learning content, presentation styles, or learning paths based on individual students' profiles, learning status, or human factors}\ccitepage{shemshack_systematic_2020}{5}.
This definition characterizes technology-enhanced learning based on the adapted learning means.
In contrast, our understanding of the differences between personalized and adaptive learning is based on the focus of the corresponding research, similar to\ccitetext{peng2019personalized}.
While research related to personalized learning focuses on the personal aspects considered during the learning process with the key points \textit{individual differences}, \textit{personal development}, and \textit{personal needs}; adaptive learning focuses on the technological side of the learning process and the systems used to adapt it with the key points \textit{adaptive adjustment}, \textit{individual differences}, and \textit{individual performance}\cciteparen{peng2019personalized}.
We will use the term \textit{adaptive learning} since the focus of this review is on the systems that provide personalized learning.
Nevertheless, \textit{personalized learning} is a valid umbrella term frequently used in relevant papers.
Therefore, we include it in our search term.

\textit{Intelligent tutoring} is another keyword that is frequently used in Intelligent Tutoring Systems (ITSs)-related research.
According to\ccitetext{paladines2020systematic}, ITSs are defined as follows:
\begin{displayquote}
``computer programs that provide instruction adapted to the needs of individual students; \ie they perform functions inherent to the tutorial process (presenting information that must be learned, asking questions or assigning tasks, providing feedback, \etc) to cause a cognitive and motivational change in the student.''\ccitepage{paladines2020systematic}{1}
\end{displayquote}
According to this definition, intelligent tutoring is synonymous with individualized instruction, which is \textquote{a product of personalized learning}\cciteparen{shemshack_systematic_2020}.
Subsequently, the term adaptive learning includes intelligent tutoring.
Nevertheless, it is reasonable to use the term separately because research on ITS frequently uses this term without using the keywords personalized or adaptive learning.

\subsection{Descriptive Framework}
\label{sec:descriptiveFramework}

\begin{figure}[H]
  \centering
  \includegraphics[width=.5\textwidth]{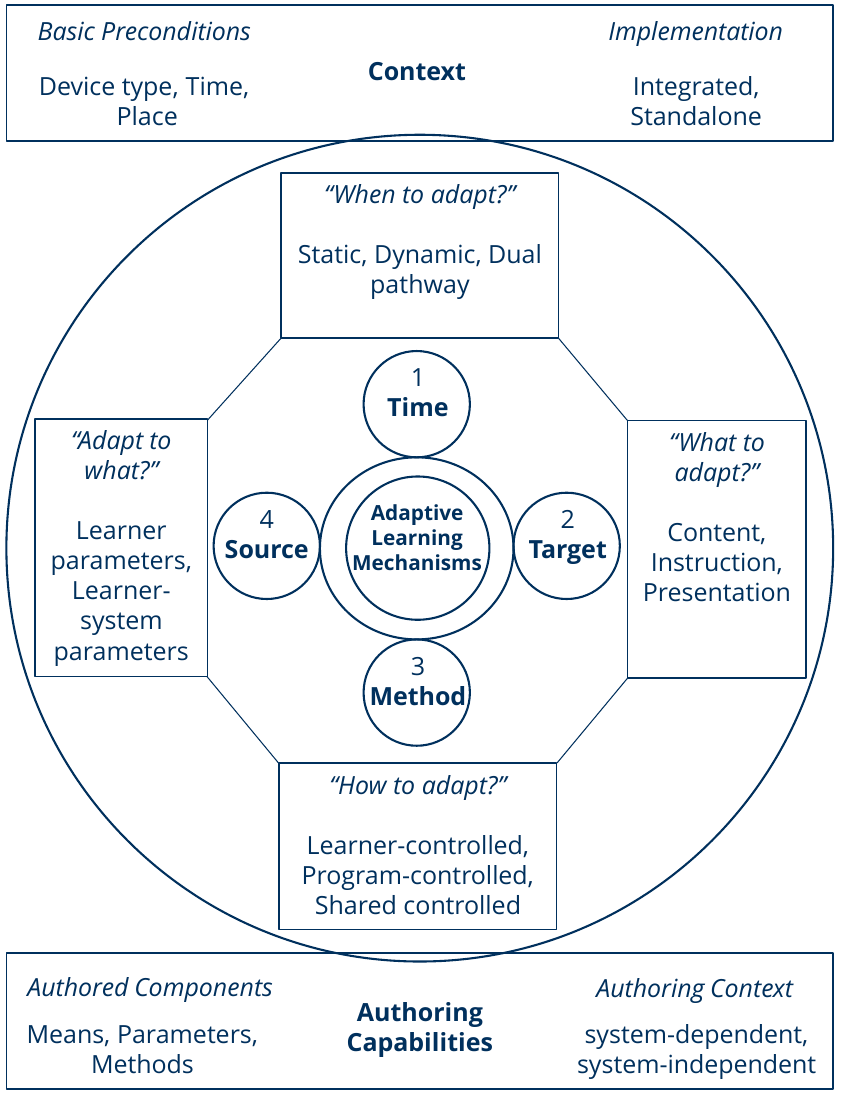}
  \caption{Descriptive framework based on\ccitetext{van2021overview}.}
  \label{fig:descriptiveFramework}
\end{figure}

Our descriptive framework~(see \Cref{fig:descriptiveFramework}) is an abstract schema that specifies the information necessary to describe the analyzed ALMs.
It is based on the four dimensions that\ccitetext{vandewaetere2014advanced} proposed and that\ccitetext{van2021overview} adopted.

\paragraph{Four Dimensions of ALMs}
The first dimension is the time of adaptation.
\ccitetext{vandewaetere2014advanced} distinguish between static adaptation before the learning process, dynamic adaptation during the learning process, or a mixture of both (\ie dual pathway).
The second dimension is the target of adaptation.
The learning content itself, the way it is presented, or the instructions provided to learners during the learning process can be adapted.
The third dimension is the method of adaptation.
The learning process adaptation can be enforced by the ALMs~(\ie program-controlled), suggested to the learners~(\ie learner-controlled), or implemented by letting the learners choose from a restricted set of learning options~(\ie shared controlled).
The fourth dimension is the source of the ALMs.
The framework distinguishes between taking into account parameters of the learners only, such as learning style, and taking into account parameters of the learner's interaction with the learning environment, such as the result of a completed test.

\paragraph{Extension of the Dimensions by a Context Aspect}
These four dimensions provide a comprehensive description of ALMs.
However, two extensions of this framework are needed in terms of our work.
While these dimensions solely focus on the ALMs themselves, they do not consider how these ALMs are integrated in the learning process.
This is crucial to evaluate their practical value.
Therefore,\ccitetext{van2021overview} adds the provision context as an additional aspect with two dimensions.
The first dimension describes the basic preconditions that must be met for learners to take advantage of ALMs.
The framework distinguishes between three different preconditions.
The first is the type of device required to participate in the learning process.
The second is the amount of time the learner has to interact before the ALMs take effect.
The third is the place where the learner has to interact with the ALMs.
The second dimension describes the implementation.
The framework distinguishes between ALMs that are implemented in a standalone ALS, and ALMs that are integrated into an existing learning environment.

\paragraph{Extension of the Dimensions by an Authoring Aspect}
 
Additionally, how ALMs are specified is another crucial aspect to evaluate their practical value.
This information is needed to answer two questions that are part of this review.
First, what effort is required by the stakeholders (\eg instructors) to integrate ALMs into the intended learning environments?
Second, to what extent can the ALMs be adapted by the stakeholders to suit their needs and the educational context?
Therefore, we propose to add the authoring as an additional aspect, with two dimensions to be specified.
The first dimension describes the components to be authored.
These are the components related to the ALMs that can or has to be authored.
We distinguish three components.
The first is the means that are the target of the ALMs, such as the learning content.
The second is the parameters of the ALMs.
Authoring the parameters that control the application of the ALMs means that, although the mechanisms are fixed, the author can influence their application.
The third is the ALMs themselves.
This can be done, for example, by providing an algorithmic specification of when and how to adapt the learning process.
The second dimension is the authoring context, \ie the context in which an authored ALM is applicable.
We distinguish two authoring contexts.
First, system-dependent authoring means that the authored ALMs are tied to a specific learning environment.
In contrast, system-independent authoring allows the authored ALMs to be reused.

\subsection{Educational Framework}
\label{sec:educationalFramework}

An educational contextualization of the compiled ALMs is required to connect them to existing learning theories.
This can be achieved by analyzing the implemented learning models or frameworks.
Learning frameworks describe how learning should occur to achieve specific learning goals, depending on the learning context and target learner group.
The reviewed approaches to adaptive learning in LMSs can implement or partially implement these frameworks, which can specify two aspects of the implemented ALMs\cciteparen{aleven2016instruction}.
First, they can specify which learner parameters should serve as the adaptation basis~(\ie\textit{"what to adapt to"}).
Second, they can specify how the learning process is adapted (\ie\textit{"how to adapt"}).
However, the "how" can be described in terms of the presented descriptive framework.

\paragraph{Self-Regulated Learning as a Lens for Analysis}
Regarding established learning theories and models, we consider the different models related to self-regulated learning to be especially relevant for integrating ALMs into LMSs.
The underlying rationale for this assertion is that self-regulated learning skills are considered essential for learning success in general\cciteparen{zimmerman2002becoming}.
Thus, one of a teacher's main goals should be to empower learners to become self-aware of the strengths and limitations of the learning process, so they can take corrective actions\cciteparen{zimmerman2002becoming}.
In learning scenarios where LMSs are used, such as MOOCs, there is often no teacher available to provide the necessary support to help learners develop self-regulatory skills.
This leaves the provision of such support as an essential task for ALMs.
Therefore, analyzing the extent to which ALMs can support self-regulatory processes is useful for assessing their value in effectively supporting the learning process and improving learning success.

\begin{figure}[htbp]
    \centering
    \begin{subfigure}{0.54\textwidth}
        \centering
        \includegraphics[width=\linewidth]{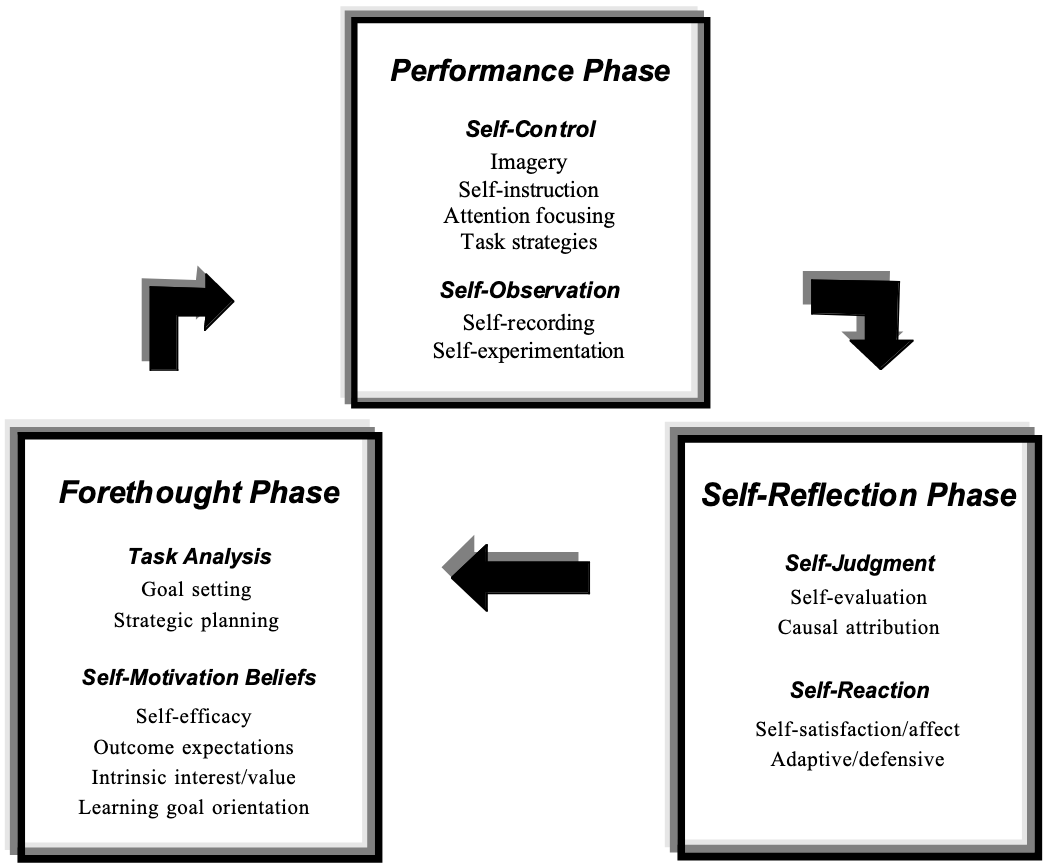}
        \caption{The phases of the self-regulated learning process, according to\ccitetext{zimmerman2000attaining}, taken from\ccitetext{zimmerman2002becoming}, p.~4.}
        \label{fig:zimmermanSrlModels}
        \vspace*{\baselineskip}
    \end{subfigure}
    \hfill
    \begin{subfigure}{0.4\textwidth}
        \centering
        \includegraphics[width=\linewidth]{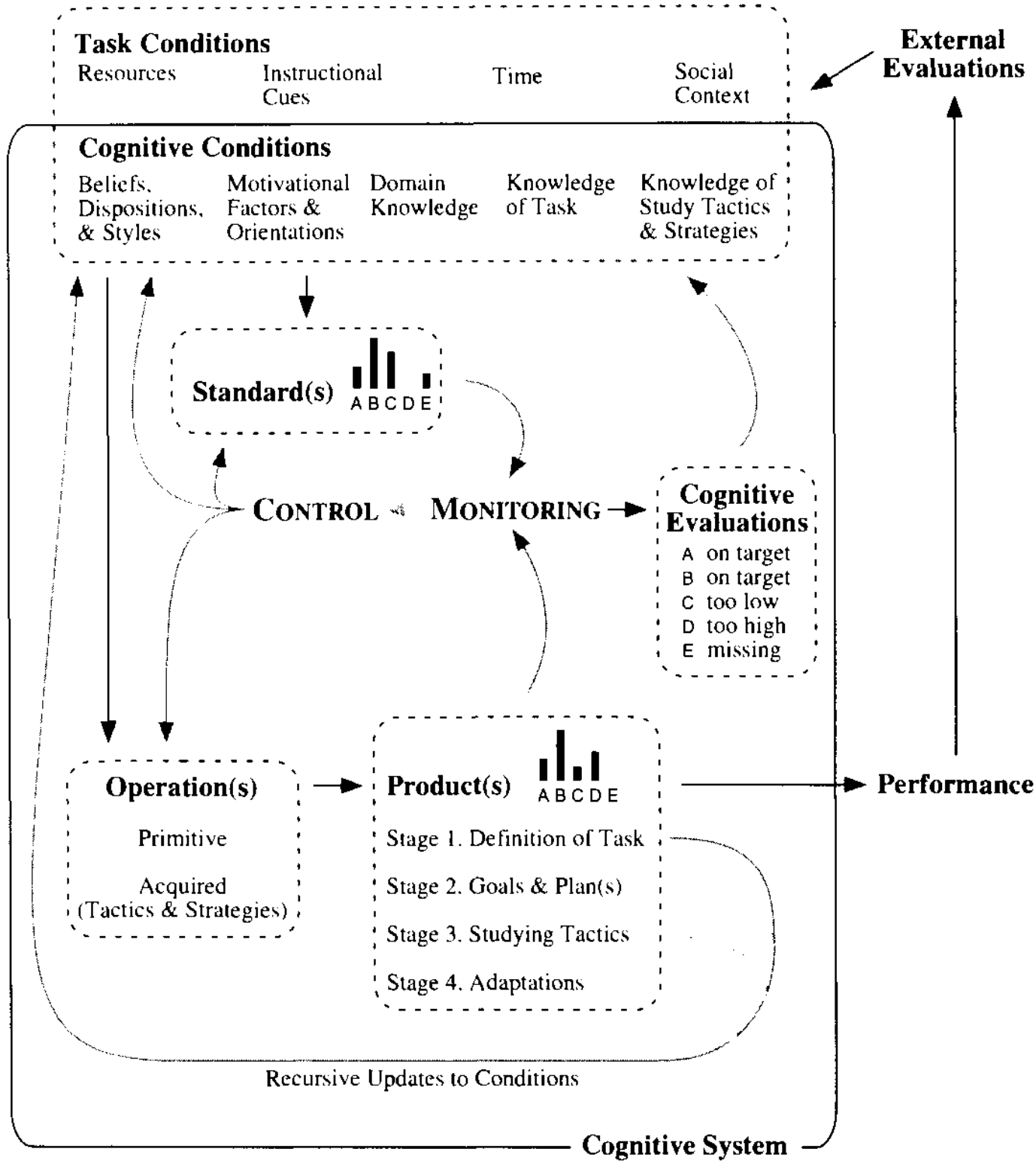}
        \caption{The COPES model defining four stages of learning, according to\ccitetext{winnie1998studying}, taken from\ccitetext{winnie1998studying}, p.~6.}
        \label{fig:copesModel}
    \end{subfigure}

    \caption{The two models of self-regulated learning considered by this review.}
    \label{fig:selfRegulatedLearningModels}
\end{figure}

For this purpose, we consider two models of self-regulated learning in our review.
These are shown in \Cref{fig:selfRegulatedLearningModels}.
The first one is the model proposed by\ccitetext{zimmerman2000attaining}.
It specifies a three-phase, cyclic process that should be followed to successfully implement self-regulation for learning.
The first forethought phase occurs before learning.
Key elements of this phase include setting learning goals, planning the learning process (task analysis), and formulating expected learning outcomes (self-motivation beliefs).
The second phase, the performance phase, occurs during the learning process.
It encompasses self-control, such as focus and time management, as well as self-observation, such as monitoring progress.
The third phase, referred to as the self-reflection phase, happens after learning.
This phase includes self-judgment, for example in the form of self-evaluation, and self-reaction, which is reacting to the outcome of the self-evaluation through adjusting learning behavior or continuing learning as before.

The second one is the COPES model proposed by\ccitetext{winnie1998studying}.
Similar to the model proposed by\ccitetext{zimmerman2000attaining}, the COPES model specifies recursive process steps for learning to be conducted and supported, referred to as stages.
A total of four stages are proposed.
The first stage is called task definition.
During the task definition stage, the learners develop an understanding of the learning task and the associated constraints and resources.
The second stage is called goal and plan and is based on the perception developed during the first stage.
At this stage, learners select or generate idiosyncratic goals and construct a plan to address the learning task.
Both stages can be linked to the forethought phase of Zimmerman's self-regulated learning model\ccitecf{zimmerman2002becoming}.
The third stage is called enactment.
At this stage, the learner carries out the learning plan and tactics developed in stage 2.
This stage aligns with the performance phase of the self-regulated learning model proposed by\ccitetext{zimmerman2002becoming}.
The fourth stage is called adaptation.
At this stage, the cognitive structure of learners, including their knowledge, skills, beliefs, and motivational factors, is modified based on experiences from previous stages.
This stage aligns with the self-reflection phase of the self-regulated learning model proposed by\ccitetext{zimmerman2002becoming}.

The COPES model defines five facets for each stage that can be used to describe the surrounding conditions, the actions taken by the learner, and how the stage can be supported.
While the self-regulated learning model proposed by\ccitetext{zimmerman2002becoming} lists different actions to be conducted or supported in each phase, the COPES model specifies these actions in greater detail, by describing the aspects and conditions to be considered during conduction or when providing support.
This is particularly important when analyzing the extent to which ALMs support the self-regulated learning process, which is why we consider both models.
The first letters of these facets form the name of the model, COPES.
The first facet is conditions.
Conditions determine the environment in which cognitive activities related to learning occur and affect how the learning task is approached.
There are two types of conditions.
First, there are task conditions related to the learning task, such as the available learning resources.
Second, there are conditions related to the learners themselves, such as their prior knowledge and learning style.
These are referred to as cognitive conditions.
The second facet is the operations that learners apply to complete the learning task.
Operations can refer to single learning actions or coordinated actions over time referred to as learning tactics or strategies.
The third facet is the products resulting from learners completing the corresponding stage.
Products can be changes in cognitive profiles or attributes, or they can manifest themselves externally through changes in the learners' behavior or performance.
The fourth facet is about evaluations.
These evaluations specify how products of stages are monitored and compared to standards of the learning process.
Evaluations can be performed internally by the learners themselves, which is essential for self-regulation\ccitecf{zimmerman2002becoming}, or externally, resulting in internal or external feedback.
The fifth facet is the aforementioned standards that are examined during the evaluation process.
Standards represent intended learning states.

\paragraph{Educational Contextualization Requires a Meta-Framework}
There are dozens of implications for ALMs in both models of self-regulated learning.
For instance, in the COPES model, task conditions such as task time, instructional cues, and resources can adaptively change based on learners' cognitive conditions, such as their prior knowledge, learning styles, and motivational factors.
To focus this review on the broader aspects of learning rather than the microanalytical aspects of (self-regulated) learning processes, we use the educational meta-framework proposed by\ccitetext{aleven2016instruction} to contextualize the reviewed approaches in relation to their adaptation goals and basis.
Another reason for this is that, after conducting a general literature search prior to our scoping review, we concluded that using the described self-regulated learning models might be inappropriate for data charting.
This is because not all approaches to integrating ALMs into LMSs might consider self-regulated learning.
However, there is also no other overarching learning framework that encompasses all the approaches and strategies included in the reviewed papers\cciteparen{du2024personalized}. 
This is due to the number of established learning frameworks appropriate for different learning scenarios.
Additionally, a number of approaches have developed their own strategies for adapting the learning process instead of adopting an established framework.
Therefore, rather than adopting a specific learning theory or model for data charting, we adopt the classification of ALMs proposed by\ccitetext{aleven2016instruction}, which we refer to as our educational framework.
In addition, we employ the introduced self-regulated learning models as a cross-cutting analytical lens to connect our review to the theory of self-regulated learning during the specification of the data charting variables and the analysis of the results.

Our educational framework categorizes ALMs according to the characteristics of learners that function as their basis for adaptation\cciteparen{aleven2016instruction}.
It delineates five categories of adaptation basis.
While this categorization is not contingent upon a particular learning framework, established learning frameworks can be associated with this categorization in terms of their adaptation bases.
This facilitates the alignment of approaches with established learning frameworks during analysis, even in cases where the approaches propose their own adaptation strategies.

The first basis of adaptation is learners' prior knowledge, as well as the knowledge they develop during the learning process (\ie knowledge growth).
The mastery learning approach\cciteparen{bloom1968learning,bloom1984the2} is an example of a learning framework that considers this basis.
It assesses whether learners have gained the knowledge associated with small learning units.
If so, it allows learners to proceed to the next unit.
Otherwise, it provides support until they have mastered the material.
The second basis of adaptation is the learner's path through a learning problem, including their solution strategy, errors, and requests for help.
This is considered by the learning framework of the two instruction loops of ITSs\cciteparen{vanlehn2006behavior,koedinger2013new}.
The outer loop selects a learning problem to work on.
The inner loop supports the learning process based on learning successes and errors.
The third basis of adaptation is the learner's motivation or affect.
This is considered by the learning framework proposed by\ccitetext{woolf2009affect}, which introduces the concept of affect-aware tutors.
These apply tutoring strategies based on cognitive and affective information.
The fourth basis of adaptation involves the metacognitive and self-regulatory processes of learners.
Both can be supported by ALMs.
The described self-regulated learning models are learning frameworks that structure this support.
The fifth basis of adaptation is learners' learning styles.
The Felder-Silverman learning style model proposes four categories to describe learning preferences\cciteparen{felder1988learning}.
The learning process can then be adapted according to how learners are categorized.

\subsection{Rationale}
\label{sec:rationale}

The authors of previous research have investigated how to integrate ALMs into LMSs\cciteeg{cobos_learning_2007,ivanova_adaptive_2011,moutafi2013mining,pratas_adapt_2014}.
While some of the concepts focus on specific LMSs\cciteeg{despotovic2012providing}, the development is moving toward service-oriented ALS architectures\cciteparen{sottilare_exploring_2019}.
Decoupling the components that apply ALMs from the LMSs~(\ie provide the ALMs system-independently), has two advantages\cciteparen{kucharski_adaptive_2023}.
First, it facilitates the integration of these components in different systems and thus can enable the reuse of existing learning content.
Second, it allows the reuse of applied didactic concepts.

ITSs, as one type of ALS, are usually already implemented in well-differentiated modules\cciteparen{nwana1990intelligent} that loosely interact with each other.
Thus, they can be more easily coupled with other systems to take advantage of this coupling\cciteparen{aleven_beginning_2015}.
Therefore, the integration of ITSs into established learning infrastructures is mentioned as a likely future of these systems\cciteparen{nye_skope-it_2018}.
However, in order to bring ALMs to LMSs in a system-independent manner, not only ITSs are investigated.
It is also researched how, for example, mechanisms used in adaptive hypermedia systems\cciteparen{de_bra_grapple1_2013} or personalized learning systems in general\cciteparen{apoki_modular_2022} can be integrated into LMSs.

\paragraph{Scoping Literature Review}
Research on integrating ALMs into LMSs addresses the following questions -- what ALMs are supported, how are they defined, how are they integrated into an LMS, what data do they work with, what educational goals do they have, and what positive impacts do they have on the learning process?
However, to the best of our knowledge, there is no structured overview that maps the available literature on these points.
For this purpose, we conduct a scoping review to systematically analyze related existing literature.
We intend to make four main contributions with this.

\begin{enumerate}
  \item The scoping review provides a structured overview, informing the reader about how state-of-the-art approaches integrate ALMs into LMSs, the mechanisms they integrate, the learning content they work with, and the degree to which these mechanisms can be authored or parameterized. This overview can serve as the foundation for further analysis. Furthermore, it can serve as a means of classification for practitioners and researchers to select approaches for continued investigation based on their own criteria.
  \item The review provides insight into the extent to which approaches to integrating ALMs into LMSs support self-regulated learning according to the described self-regulated learning models.
  \item The review identifies patterns with regard to implementing ALMs into LMSs that have been reported to be beneficial for the learning process. These findings can serve as a foundation for further research on this topic and guide practitioners in implementing adaptive learning in LMSs.
  \item The review identifies gaps in the specification and implementation of ALMs that practitioners and researchers must address to effectively provide ALMs for LMSs.
\end{enumerate}

Furthermore, we intend to investigate the differences between system-specific and system-independent ALMs, apart from the implementation context.
The results are expected to provide insights into how both types of approaches can benefit from considering certain aspects of each other.

\paragraph{Target User Group-Based Analysis}
There are two groups of users who are interested in the integration of ALMs into LMSs\ccitecf{kurni_intelligent_2023}.
The first group are instructors who want to take advantage of ALMs.
Users in this group typically have two requirements.
First, they want to integrate ALMs with their existing LMS and use them with their learning content.
This way, they won't need to adapt to a new system or re-author their learning content.
Often, it is not even possible to change LMSs.
This is because they are tied to a specific system by the institution they belong to.
Second, they want to understand what kind of support the integrated ALMs provide so they can build trust in these mechanisms.
Some instructors may also want to do some basic parameterization.
The second group of users are researchers or instructional designers who want to investigate which ALMs are advantageous.
Users in this group are often interested in reusing existing learning content because their focus is on the details of the ALMs used rather than the content.
Therefore, they require the ability to define these mechanisms and make fine-grained adjustments themselves.
To facilitate the dissemination of an approach to integrating ALMs into LMSs, the requirements of all associated target user groups must be met.
Due to the lack of a structured review, there is no analysis of related approaches to determine whether these requirements are met.
Thus, we intend to make the following additional contribution.

\begin{enumerate}
  \setcounter{enumi}{4}
  \item The scoping review provides information on the extent to which approaches to integrating ALMs into LMSs fulfill the target user groups' requirements.
\end{enumerate}

\subsection{Objectives}
\label{sec:objectives}

\paragraph{Research Questions}
Our review addresses the following research questions to make the intended contributions:

\begin{question}
    \item\label{rq:implementation} Which implementation approaches are proposed for the system-independent integration of ALMs into LMSs?
    \item\label{rq:provision} Which ALMs are supported and which components of the learning process are adapted by approaches proposed by studies on the system-independent integration of ALMs into LMSs?
    \item\label{rq:definition} How do the approaches proposed by studies on the system-independent integration of ALMs in LMSs specify: a) in which way the adaptive mechanisms work and b) which mechanisms are appropriate in which situations?
    \item\label{rq:database} What type of data do the adaptive mechanisms proposed by studies on the system-independent integration of ALMs into LMSs process?
    \item\label{rq:educationalContext} What are the educational surrounding conditions and outcomes of approaches proposed by studies on the system-independent integration of ALMs into LMSs, and how do these align with the introduced models of self-regulated learning?
\end{question}

\ref{rq:implementation} focuses on the technical implementation of ALMs for LMSs.
We intend to explore how ALMs are integrated into LMSs, and which data processing mechanisms are implemented to apply these ALMs.
\ref{rq:provision} focuses on the adaptation process.
We intend to explore which ALMs are implemented and which components of the learning process are adapted.
\ref{rq:definition} focuses on the specification of the ALMs.
We intend to explore which authoring capabilities are proposed to support the specification of ALMs and to which level.
Furthermore, we intend to investigate the proposed triggers that determine which ALMs are appropriate.
\ref{rq:database} focuses on the data basis of the ALMs.
We intend to explore what data is stored for and processed by the ALMs.
\ref{rq:educationalContext} focuses on the educational contextualization.
We intend to explore the educational parameters on which ALMs are based and the adopted established learning frameworks.
Furthermore, we intend to investigate the educational goals that ALMs aim to achieve and the reported effects related to achieving these goals.
Finally, we are interested in which degree the reviewed approaches, intentionally or unintentionally, support or can support self-regulated learning according to the introduced models.

A further objective of this review is to illustrate the differences between system-specific and system-independent approaches in terms of the research questions posed.
We do so by analyzing the differences between the systematically extracted system-independent approaches and the LMS-specific approaches that we identified during the review process.

\paragraph{Review Hypotheses}
We formulated three hypotheses based on a general literature review that we conducted prior to the review.
They express our beliefs about characteristics of state-of-the-art approaches to system-independently integrating ALMs into LMSs.
They contradict the requirements of our target user groups and hinder the support of self-regulated learning according to the introduced models.
It is another objective of this review to gather evidence that either supports or refutes these hypotheses.

\begin{hypothesis}
    \item\label{h:insufficientDataBasisSupport} ALMs rarely consider existing data (learning content data, interaction data).
    \item\label{h:insufficientProcessingMechanismsSupport} Most systems' ALM data processing mechanisms are limited to a single type.
    \item\label{h:insufficientAuthoringCapabilitiesSupport} Users are rarely given the ability to model different levels of ALMs.
\end{hypothesis}

\ref{h:insufficientDataBasisSupport} states that existing LMS content is rarely used for adaptive learning.
Instead, learning content must be created specifically for the ALMs integrated into the LMSs.
The same applies to interaction data.
Integrated ALMs often only consider interaction data they generate themselves, rather than considering interaction data generated natively by the LMSs.
From a self-regulated learning perspective, especially the consideration of native LMS interaction data, may be crucial.
The reason for this is that supporting the performance phase of the self-regulated learning process requires continuous monitoring throughout the entire learning process to support learners' self-observation\ccitecf{zimmerman2000attaining}.
Thus, not considering native interaction data restricts monitoring to specialized learning activities or learning content, impeding the possibility to extensively support self-regulated learning.

\ref{h:insufficientProcessingMechanismsSupport} states that ALMs usually apply only a single type of learning process data processing mechanisms that determine appropriate adaptations.
This results from certain assumptions~(\eg a fixed domain) that usually underlie the conceptualization of ALMs.
A limited number of applied processing mechanisms leads to simpler authoring tools due to the reduced number of processing options.
However, this also limits the ALMs that can be implemented.
Furthermore, from a self-regulated learning perspective, different processing mechanisms are required to support all facets of the different learning process stages\ccitecf{winnie1998studying}.
For example, detecting when learners finish a learning task in order to support self-evaluation of learning results requires trigger-based processing of interaction data.
Instead, determining whether learners have understood the underlying domain knowledge of a test set in order to provide external feedback may require a complex evaluation of their entire learning state.
Therefore, supporting only one processing mechanism, or a limited number of them, hinders the extent to which self-regulated learning can be supported.

\ref{h:insufficientAuthoringCapabilitiesSupport} states that most approaches only support authoring to the extent of editing the learning means or parameters of fixed ALMs.
This results in simpler authoring tools.
However, it also limits the applicability and potential use cases of the corresponding ALMs.
Furthermore, failing to support authoring ALMs may impede the support of self-regulated learning.
The reason for this is that, to support the development of self-regulated learning skills, the provision of support is as important as omitting support in specific learning scenarios\ccitecf{zimmerman2002becoming}.
These learning scenarios may depend on the different task and cognitive conditions introduced by\ccitetext{winnie1998studying} and may be known to the instructor applying these ALMs but not to the developers of the ALSs.
Furthermore, since fading support based on these scenarios is an essential concept of self-regulated learning, it may also be one of the main focuses of instructional designers.
Therefore, both user groups require the appropriate authoring capabilities.
\section{Related Work}
\label{sec:relatedWork}

We conducted an informal search prior to this review.
We list relevant reviews that we identified during this search, as well as related work that we identified while conducting the review, below.
Related reviews can be divivded into four categories.
In three of the four categories, there are reviews that are conducted primarily from a technical point of view, such as our review too.
These three categories include reviews related to ALMs (see \Cref{sec:adaptiveLearningMechanisms}), ITSs (see \Cref{sec:intelligentTutoringSystems}), and LMSs (see \Cref{sec:learningManagementSystems}).
The fourth category of reviews focuses primarily on the didactics and the pedagogical properties of ALMs (see \Cref{sec:didacticAndPedagogicalProperties}).
In \Cref{sec:delimitationAndContribution}, we compare related reviews based on their inclusion criteria and analytical focus.
Based on this, we highlight the differences between the related reviews and our own.
Furthermore, we discuss the contributions of our review and the insights that can be derived from it.

\subsection{Adaptive Learning Mechanisms}
\label{sec:adaptiveLearningMechanisms}

Related work that focuses on reviewing research on ALMs and their characteristics can be divided into five groups.
The first group focuses on adaptive learning-related models.
\ccitetext{ochukut_research_2023} and\ccitetext{abyaa2019learner} investigated the learner characteristics being modeled and the modeling techniques.
\ccitetext{normadhi2019identification} analyzed the traits being processed in learner models, as well as the techniques used to identify these traits.

The second group focuses on the application of ALMs.
\ccitetext{ilic2023intelligent} examined the AI techniques used, their purpose, and their impact on learners.
\ccitetext{mikic_personalisation_2022} studied personalization techniques, the impact on attitudes, engagement, and performance, and the addressed key implementation challenges.
\ccitetext{raj2022systematic} analyzed the personalization parameters and models of educational content recommender systems, the recommendation techniques, and the evaluation methods.
\ccitetext{xie2019trends} investigated the ALMs provided, their effect on learning outcomes, the subjects to which they are applied, and the devices used for e-learning.
\ccitetext{bimba2017adaptive} examined the means, targets, goals, and strategies of adaptive feedback provision.

The third group focuses on use case-specific ALMs.
\ccitetext{fariani2023systematic} analyzed ALMs in the context of higher education, focusing on their personalization features, modeling techniques, theoretical underpinnings, and observed learning effects.
\ccitetext{van2021overview} investigated differences among studies on personalized learning support in primary and secondary education in terms of documentation and description, the tools used to implement ALMs, and the evidence of their impact on learning outcomes.

The fourth group focuses on the characteristics of research on ALSs.
\ccitetext{ahtesham2022software} studied applied software patterns and the quality attributes that guide the development.
\ccitetext{koutsantonis2022bibliometric} investigated the authors involved, the keywords used, and the related work that is referenced most often.
\ccitetext{bernacki2021systematic} analyzed which researchers are studying personalized learning, which learner populations they consider, which learner characteristics they study, and which operational definitions and contexts they use.

The fifth group focuses on AI-based ALMs.
\ccitetext{hardaker2025artificial} examined AI methods used to adapt the learning process at micro and macro levels in higher education institutions, as well as their impact on the learning experience.
\ccitetext{guo2024artificial} investigated countries conducting research on AI in education, the grade level of the focused learners, the keywords used, and the focused applications.
\ccitetext{halkiopoulos2024leveraging} analyzed how different cognitive neuropsychological principles can be used to improve AI methods with respect to personalized learning and adaptive assessment.
\ccitetext{li2023artificial} analyzed sources of publications, considered applications, and the emerging research trends.
\ccitetext{nadimpalli2023systematic} discussed the advantages and disadvantages of methods for organizing and recommending learning content based on identified learner characteristics in LMSs.
\ccitetext{hashim2022trends} analyzed focused technologies, the purpose of the application of AI, and the target user groups.
\ccitetext{chen2022two} examined the annual number of publications related to AI in education, the countries and institutions involved, the scientific collaborations, and the realized applications.
\ccitetext{kabudi2020systematic} studied the problems that AI techniques are designed to solve.

\subsection{Intelligent Tutoring Systems}
\label{sec:intelligentTutoringSystems}

Related work that focuses on reviewing research on ITSs as a specific type of ALS can be divided into three groups.
The first group focuses on the application of ITSs and the provided ALMs.
\ccitetext{alrakhawi2023intelligent} studied the domains of use, the main purposes, the creation tools, the developed interfaces, and the impact of using ITSs.
\ccitetext{mousavinasab_intelligent_2021} analyzed the domains of use, the AI techniques used and their main purposes, the types of user interfaces utilized, and the evaluation methods.
\ccitetext{almasri2019intelligent} examined ITSs developed from 2000 to 2018 and provided a brief description of each.

The second group focuses on authoring tools and deployment options.
\ccitetext{rokhman_intelligent_2022} examined the components and the types of ITSs that can be authored, and the authoring technologies developed for non-programmers.
\ccitetext{soofi_systematic_2019} analyzed the referenced domains, the techniques and tools developed to author ITSs, the delivery modes, and the validation techniques.
\ccitetext{dermeval2018authoring} investigated the components and the types of ITSs that can be authored, the features that support the authoring process, the authoring technologies, the benefits of use, and when authoring occurs.

The third group focuses on the characteristics of research on ITSs.
\ccitetext{han2019intelligent} analyzed publication locations, research purposes, domains, and designs of research on ITSs using text mining techniques.

\subsection{Learning Management Systems}
\label{sec:learningManagementSystems}

Related work that focuses on reviewing research on characteristics, ALMs, development and use of LMSs can be divided into two groups.
The first group focuses on the properties and functionalities of LMSs.
\ccitetext{aldahwan2020use} analyzed the problems that AI functions in LMSs attempt to solve and the AI models used for this purpose.
\ccitetext{kasim2016choosing} studied the characteristics of LMSs to be used in the context of higher education, such as flexibility, ease of use, accessibility, and usability.

The second group focuses on the characteristics of research on LMSs.
\ccitetext{prahani_learning_2022} investigated research metrics such as publication language, research country, funding, and most cited authors.
\ccitetext{oliveira2016learning} analyzed the research strategy, research design, and investigated LMS categories.

\subsection{Didactic and Pedagogical Properties}
\label{sec:didacticAndPedagogicalProperties}

The fourth category includes reviews focusing on the didactic and pedagogical properties of ALMs, as well as their influence on learners and learning outcomes.
\ccitetext{liu2025advancing} examined the technological and pedagogical benefits of ITSs and robot tutoring systems from an ethical perspective, analyzing the engagement strategies used.
\ccitetext{du2024personalized} investigated the impact of personalized learning on the academic performance and engagement of learners in higher education.
\ccitetext{lin2024personalized} analyzed the impact of ALMs on learner interest and cognitive load, as well as the retention and transfer of learning content.
\ccitetext{sharma2024self} studied the challenges and opportunities of ALSs that support self-regulated and collaborative learning.
\ccitetext{cevikbas2022promoting} analyzed the strategies and tools used in flipped classroom research to improve personalized learning, as well as the role of personalized flipped classroom approaches in enhancing teaching and learning.
\ccitetext{alqahtani2021review} investigated the types of research and methods used to compare the benefits of non-adaptive versus adaptive learning support tools.
\ccitetext{khamparia2020association} examined the relationship between learning styles and e-learning problems, focusing on ALMs that categorize learners based on their learning style.

\subsection{Delimitation and Contribution}
\label{sec:delimitationAndContribution}

{
\setlength{\aboverulesep}{0pt}
\setlength{\belowrulesep}{0pt}
\newcolumntype{C}{>{\centering\arraybackslash}m{(\textwidth)/15}}
\begin{sidewaystable}[ph!]
  \resizebox{ \textwidth}{!}{%
  \def \arraystretch{2.4}
  \begin{tabular}{|l||C|C|C|C|C|C|C|C|C||C|C|C|C|C|C|C|C|C|C|}
    \toprule
    \multirow{3}{*}{\parbox[t]{7cm}{\raggedright Related Reviews}} & %
    \multicolumn{9}{|c||}{\parbox[t]{3cm}{\raggedright Inclusion criteria}} & %
    \multicolumn{10}{|c|}{\parbox[t]{3cm}{\raggedright Analysis objectives}} \\
    \cmidrule{2-20}
    & %
    \multicolumn{3}{|c|}{\parbox[t]{3cm}{\raggedright Learning context}} & %
    \multicolumn{3}{|c|}{\parbox[t]{3.8cm}{\raggedright Learning support functionalities}} & %
    \multicolumn{3}{|c||}{\parbox[t]{3.5cm}{\raggedright What is learned and by whom}} & %
    \multicolumn{4}{|c|}{\parbox[t]{4cm}{\raggedright Learning context and learning support functionalities\\}} & %
    \multicolumn{2}{|c|}{\parbox[t]{3cm}{\raggedright Authoring of learning support functionalities}} & %
    \multicolumn{3}{|c|}{\parbox[t]{3cm}{\raggedright Didactic and pedagogical properties}} & %
    \multirow{2}{*}{\rotatebox[origin=c]{270}{\hspace*{.2cm}\parbox[t]{3cm}{\raggedright Meta-aspects of conducted research}}} \\
    \cmidrule{2-19}
    & %
    \rotatebox[origin=c]{270}{\hspace*{.2cm}\parbox[t]{2cm}{\raggedright Only LMSs}} & %
    \rotatebox[origin=c]{270}{\hspace*{.2cm}\parbox[t]{2cm}{\raggedright Only ALSs}} & %
    \rotatebox[origin=c]{270}{\hspace*{.2cm}\parbox[t]{2cm}{\raggedright No restrictions on the learning context}} & %
    \rotatebox[origin=c]{270}{\hspace*{.2cm}\parbox[t]{2cm}{\raggedright Only learning support functionalities}} & %
    \rotatebox[origin=c]{270}{\hspace*{.2cm}\parbox[t]{2cm}{\raggedright Only specific learning support functionalities}} & %
    \rotatebox[origin=c]{270}{\hspace*{.2cm}\parbox[t]{2.4cm}{\raggedright No restrictions on the learning support functionalities}} & %
    \rotatebox[origin=c]{270}{\hspace*{.2cm}\parbox[t]{2cm}{\raggedright Only specific learners}} & %
    \rotatebox[origin=c]{270}{\hspace*{.2cm}\parbox[t]{2cm}{\raggedright Only specific learning subjects}} & %
    \rotatebox[origin=c]{270}{\hspace*{.2cm}\parbox[t]{2cm}{\raggedright No restrictions on learners or learning subject}} & %
    \rotatebox[origin=c]{270}{\hspace*{.2cm}\parbox[t]{2cm}{\raggedright Learning support integration}} & %
    \rotatebox[origin=c]{270}{\hspace*{.2cm}\parbox[t]{2cm}{\raggedright Learning context properties}} & %
    \rotatebox[origin=c]{270}{\hspace*{.2cm}\parbox[t]{2cm}{\raggedright Learning support functionalities properties}} & %
    \rotatebox[origin=c]{270}{\hspace*{.2cm}\parbox[t]{2cm}{\raggedright Learning support functionalities data source}} & %
    \rotatebox[origin=c]{270}{\hspace*{.2cm}\parbox[t]{2cm}{\raggedright Provided authoring capabilities}} & %
    \rotatebox[origin=c]{270}{\hspace*{.2cm}\parbox[t]{2cm}{\raggedright Authorable properties}} & %
    \rotatebox[origin=c]{270}{\hspace*{.2cm}\parbox[t]{2cm}{\raggedright Educational goals}} & %
    \rotatebox[origin=c]{270}{\hspace*{.2cm}\parbox[t]{2cm}{\raggedright Underlying didactics}} & %
    \rotatebox[origin=c]{270}{\hspace*{.2cm}\parbox[t]{2cm}{\raggedright Evaluation of learning support functionalities}} & %
    \\
    \hline
    \hline
    \rowcolor{gray!50} %
   \ccitetext{kasim2016choosing}
    & \x %
    &  %
    &  %
    &  %
    &  %
    & \x %
    & \x %
    &  %
    &  %
    &  %
    & \x %
    &  %
    &  %
    &  %
    &  %
    &  %
    &  %
    &  %
    &  %
    \\
    \hline
   \ccitetext{oliveira2016learning}
    & \x %
    &  %
    &  %
    &  %
    &  %
    & \x %
    &  %
    &  %
    & \x %
    &  %
    & \x %
    &  %
    &  %
    &  %
    &  %
    &  %
    &  %
    &  %
    & \x %
    \\
    \hline
    \rowcolor{gray!50} %
   \ccitetext{bimba2017adaptive}
    &  %
    &  %
    & \x %
    &  %
    & \x %
    &  %
    &  %
    &  %
    & \x %
    &  %
    &  %
    & \x %
    & \x %
    &  %
    &  %
    & \x %
    & \x %
    & \x %
    & \x %
    \\
    \hline
   \ccitetext{dermeval2018authoring}
    &  %
    & \x %
    &  %
    &  %
    &  %
    & \x %
    &  %
    &  %
    & \x %
    &  %
    & \x %
    &  %
    &  %
    & \x %
    & \x %
    &  %
    &  %
    &  %
    & \x %
    \\
    \hline
    \rowcolor{gray!50} %
   \ccitetext{abyaa2019learner}
    &  %
    & \x %
    &  %
    &  %
    &  %
    & \x %
    &  %
    &  %
    & \x %
    &  %
    &  %
    &  %
    & \x %
    & \x %
    & \x %
    &  %
    &  %
    &  %
    & \x %
    \\
    \hline
   \ccitetext{normadhi2019identification}
    &  %
    & \x %
    &  %
    &  %
    &  %
    & \x %
    &  %
    &  %
    & \x %
    &  %
    &  %
    &  %
    & \x %
    &  %
    &  %
    &  %
    & \x %
    &  %
    & \x %
    \\
    \hline
    \rowcolor{gray!50} %
   \ccitetext{almasri2019intelligent}
    &  %
    & \x %
    &  %
    &  %
    &  %
    & \x %
    &  %
    &  %
    & \x %
    &  %
    &  %
    & \x %
    &  %
    &  %
    &  %
    &  %
    &  %
    &  %
    &  %
    \\
    \hline
   \ccitetext{han2019intelligent}
    &  %
    & \x %
    &  %
    &  %
    &  %
    & \x %
    &  %
    &  %
    & \x %
    &  %
    &  %
    & \x %
    &  %
    &  %
    &  %
    &  %
    &  %
    & \x %
    & \x %
    \\
    \hline
    \rowcolor{gray!50} %
   \ccitetext{soofi_systematic_2019}
    &  %
    & \x %
    &  %
    &  %
    &  %
    & \x %
    &  %
    &  %
    & \x %
    &  %
    & \x %
    & \x %
    &  %
    &  %
    &  %
    &  %
    &  %
    & \x %
    & \x %
    \\
    \hline
   \ccitetext{xie2019trends}
    &  %
    &  %
    & \x %
    &  %
    & \x %
    &  %
    &  %
    &  %
    & \x %
    &  %
    & \x %
    & \x %
    & \x %
    &  %
    &  %
    &  %
    &  %
    & \x %
    & \x %
    \\
    \hline
    \rowcolor{gray!50} %
   \ccitetext{aldahwan2020use}
    & \x %
    &  %
    &  %
    &  %
    & \x %
    &  %
    &  %
    &  %
    & \x %
    &  %
    &  %
    & \x %
    &  %
    &  %
    &  %
    &  %
    &  %
    &  %
    & \x %
    \\
    \hline
   \ccitetext{kabudi2020systematic}
    &  %
    &  %
    & \x %
    &  %
    & \x %
    &  %
    &  %
    &  %
    & \x %
    & \x %
    &  %
    &  %
    &  %
    &  %
    &  %
    &  %
    &  %
    &  %
    & \x %
    \\
    \hline
    \rowcolor{gray!50} %
   \ccitetext{khamparia2020association}
    &  %
    &  %
    & \x %
    &  %
    & \x %
    &  %
    &  %
    &  %
    & \x %
    &  %
    & \x %
    & \x %
    &  %
    &  %
    &  %
    &  %
    & \x %
    &  %
    & \x %
    \\
    \hline
   \ccitetext{alqahtani2021review}
    &  %
    &  %
    & \x %
    & \x %
    &  %
    &  %
    &  %
    &  %
    & \x %
    &  %
    &  %
    & \x %
    &  %
    &  %
    &  %
    &  %
    &  %
    & \x %
    &  %
    \\
    \hline
    \rowcolor{gray!50} %
   \ccitetext{bernacki2021systematic}
    &  %
    &  %
    & \x %
    & \x %
    &  %
    &  %
    &  %
    &  %
    & \x %
    &  %
    &  %
    & \x %
    & \x %
    &  %
    &  %
    & \x %
    & \x %
    & \x %
    & \x %
    \\
    \hline
   \ccitetext{mousavinasab_intelligent_2021}
    &  %
    & \x %
    &  %
    &  %
    &  %
    & \x %
    &  %
    &  %
    & \x %
    &  %
    & \x %
    & \x %
    & \x %
    &  %
    &  %
    &  %
    &  %
    &  %
    & \x %
    \\
    \hline
    \rowcolor{gray!50} %
   \ccitetext{van2021overview}
    &  %
    &  %
    & \x %
    & \x %
    &  %
    &  %
    & \x %
    &  %
    &  %
    &  %
    &  %
    & \x %
    & \x %
    &  %
    &  %
    &  %
    &  %
    & \x %
    & \x %
    \\
    \hline
   \ccitetext{ahtesham2022software}
    &  %
    & \x %
    &  %
    &  %
    &  %
    & \x %
    &  %
    &  %
    & \x %
    & \x %
    &  %
    &  %
    &  %
    &  %
    &  %
    &  %
    &  %
    &  %
    & \x %
    \\
    \hline
    \rowcolor{gray!50} %
   \ccitetext{cevikbas2022promoting}
    &  %
    &  %
    & \x %
    &  %
    & \x %
    &  %
    &  %
    &  %
    & \x %
    & \x %
    &  %
    & \x %
    & \x %
    &  %
    &  %
    &  %
    &  %
    & \x %
    & \x %
    \\
    \hline
   \ccitetext{hashim2022trends}
    &  %
    &  %
    & \x %
    &  %
    & \x %
    &  %
    &  %
    &  %
    & \x %
    & \x %
    &  %
    & \x %
    &  %
    &  %
    &  %
    &  %
    &  %
    &  %
    & \x %
    \\
    \hline
    \bottomrule
    \end{tabular}
  }
  \caption{Inclusion criteria and analysis focus of related reviews. The table lists the first 20 related reviews. The reviews are ordered by publication year, with the most recent one listed last.}
  \label{tab:related-work-comparison-table-1}
\end{sidewaystable}
} %
{
\setlength{\aboverulesep}{0pt}
\setlength{\belowrulesep}{0pt}
\newcolumntype{C}{>{\centering\arraybackslash}m{(\textwidth)/15}}
\begin{sidewaystable}[ph!]
  \resizebox{ \textwidth}{!}{%
  \def \arraystretch{2.4}
  \begin{tabular}{|l||C|C|C|C|C|C|C|C|C||C|C|C|C|C|C|C|C|C|C|}
    \toprule
    \multirow{3}{*}{\parbox[t]{7cm}{\raggedright Related Reviews}} & %
    \multicolumn{9}{|c||}{\parbox[t]{3cm}{\raggedright Inclusion criteria}} & %
    \multicolumn{10}{|c|}{\parbox[t]{3cm}{\raggedright Analysis objectives}} \\
    \cmidrule{2-20}
    & %
    \multicolumn{3}{|c|}{\parbox[t]{3cm}{\raggedright Learning context}} & %
    \multicolumn{3}{|c|}{\parbox[t]{3.8cm}{\raggedright Learning support functionalities}} & %
    \multicolumn{3}{|c||}{\parbox[t]{3.5cm}{\raggedright What is learned and by whom}} & %
    \multicolumn{4}{|c|}{\parbox[t]{5cm}{\raggedright Learning context and learning support functionalities\\}} & %
    \multicolumn{2}{|c|}{\parbox[t]{3cm}{\raggedright Authoring of learning support functionalities}} & %
    \multicolumn{3}{|c|}{\parbox[t]{3cm}{\raggedright Didactic and pedagogical properties}} & %
    \multirow{2}{*}{\rotatebox[origin=c]{270}{\hspace*{.2cm}\parbox[t]{3cm}{\raggedright Meta-aspects of conducted research}}} \\
    \cmidrule{2-19}
    & %
    \rotatebox[origin=c]{270}{\hspace*{.2cm}\parbox[t]{2cm}{\raggedright Only LMSs}} & %
    \rotatebox[origin=c]{270}{\hspace*{.2cm}\parbox[t]{2cm}{\raggedright Only ALSs}} & %
    \rotatebox[origin=c]{270}{\hspace*{.2cm}\parbox[t]{2cm}{\raggedright No restrictions on the learning context}} & %
    \rotatebox[origin=c]{270}{\hspace*{.2cm}\parbox[t]{2cm}{\raggedright Only learning support functionalities}} & %
    \rotatebox[origin=c]{270}{\hspace*{.2cm}\parbox[t]{2cm}{\raggedright Only specific learning support functionalities}} & %
    \rotatebox[origin=c]{270}{\hspace*{.2cm}\parbox[t]{2.4cm}{\raggedright No restrictions on the learning support functionalities}} & %
    \rotatebox[origin=c]{270}{\hspace*{.2cm}\parbox[t]{2cm}{\raggedright Only specific learners}} & %
    \rotatebox[origin=c]{270}{\hspace*{.2cm}\parbox[t]{2cm}{\raggedright Only specific learning subjects}} & %
    \rotatebox[origin=c]{270}{\hspace*{.2cm}\parbox[t]{2cm}{\raggedright No restrictions on learners or learning subject}} & %
    \rotatebox[origin=c]{270}{\hspace*{.2cm}\parbox[t]{2cm}{\raggedright Learning support integration}} & %
    \rotatebox[origin=c]{270}{\hspace*{.2cm}\parbox[t]{2cm}{\raggedright Learning context properties}} & %
    \rotatebox[origin=c]{270}{\hspace*{.2cm}\parbox[t]{2cm}{\raggedright Learning support functionalities properties}} & %
    \rotatebox[origin=c]{270}{\hspace*{.2cm}\parbox[t]{2cm}{\raggedright Learning support functionalities data source}} & %
    \rotatebox[origin=c]{270}{\hspace*{.2cm}\parbox[t]{2cm}{\raggedright Provided authoring capabilities}} & %
    \rotatebox[origin=c]{270}{\hspace*{.2cm}\parbox[t]{2cm}{\raggedright Authorable properties}} & %
    \rotatebox[origin=c]{270}{\hspace*{.2cm}\parbox[t]{2cm}{\raggedright Educational goals}} & %
    \rotatebox[origin=c]{270}{\hspace*{.2cm}\parbox[t]{2cm}{\raggedright Underlying didactics}} & %
    \rotatebox[origin=c]{270}{\hspace*{.2cm}\parbox[t]{2cm}{\raggedright Evaluation of learning support functionalities}} & %
    \\
    \hline
    \hline
    \rowcolor{gray!50} %
   \ccitetext{koutsantonis2022bibliometric}
    &  %
    &  %
    & \x %
    & \x %
    &  %
    &  %
    &  %
    &  %
    & \x %
    &  %
    &  %
    &  %
    &  %
    &  %
    &  %
    &  %
    &  %
    &  %
    & \x %
    \\
    \hline
   \ccitetext{mikic_personalisation_2022}
    &  %
    &  %
    & \x %
    & \x %
    &  %
    &  %
    &  %
    &  %
    & \x %
    & \x %
    &  %
    & \x %
    &  %
    &  %
    &  %
    &  %
    &  %
    & \x %
    &  %
    \\
    \hline
    \rowcolor{gray!50} %
   \ccitetext{prahani_learning_2022}
    & \x %
    &  %
    &  %
    &  %
    &  %
    & \x %
    &  %
    &  %
    & \x %
    &  %
    & \x %
    &  %
    &  %
    &  %
    &  %
    &  %
    &  %
    &  %
    & \x %
    \\
    \hline
   \ccitetext{raj2022systematic}
    &  %
    &  %
    & \x %
    &  %
    & \x %
    &  %
    &  %
    &  %
    & \x %
    &  %
    &  %
    & \x %
    & \x %
    &  %
    &  %
    &  %
    &  %
    & \x %
    &  %
    \\
    \hline
    \rowcolor{gray!50} %
   \ccitetext{rokhman_intelligent_2022}
    &  %
    & \x %
    &  %
    &  %
    &  %
    & \x %
    &  %
    &  %
    & \x %
    &  %
    & \x %
    &  %
    &  %
    & \x %
    & \x %
    &  %
    &  %
    &  %
    &  %
    \\
    \hline
   \ccitetext{chen2022two}
    &  %
    &  %
    & \x %
    &  %
    & \x %
    &  %
    &  %
    &  %
    & \x %
    & \x %
    &  %
    & \x %
    &  %
    &  %
    &  %
    &  %
    &  %
    &  %
    & \x %
    \\
    \hline
    \rowcolor{gray!50} %
   \ccitetext{alrakhawi2023intelligent}
    &  %
    & \x %
    &  %
    &  %
    &  %
    & \x %
    &  %
    &  %
    & \x %
    &  %
    & \x %
    & \x %
    &  %
    &  %
    &  %
    &  %
    &  %
    & \x %
    & \x %
    \\
    \hline
   \ccitetext{fariani2023systematic}
    &  %
    &  %
    & \x %
    & \x %
    &  %
    &  %
    & \x %
    &  %
    &  %
    & \x %
    &  %
    & \x %
    & \x %
    &  %
    &  %
    &  %
    & \x %
    & \x %
    & \x %
    \\
    \hline
    \rowcolor{gray!50} %
   \ccitetext{ilic2023intelligent}
    &  %
    &  %
    & \x %
    &  %
    & \x %
    &  %
    &  %
    &  %
    & \x %
    &  %
    &  %
    & \x %
    &  %
    &  %
    &  %
    &  %
    &  %
    & \x %
    &  %
    \\
    \hline
   \ccitetext{li2023artificial}
    &  %
    &  %
    & \x %
    &  %
    & \x %
    &  %
    &  %
    &  %
    & \x %
    & \x %
    &  %
    & \x %
    &  %
    &  %
    &  %
    &  %
    &  %
    &  %
    & \x %
    \\
    \hline
    \rowcolor{gray!50} %
   \ccitetext{nadimpalli2023systematic}
    & \x %
    &  %
    &  %
    &  %
    & \x %
    &  %
    &  %
    &  %
    & \x %
    &  %
    &  %
    & \x %
    & \x %
    &  %
    &  %
    &  %
    &  %
    &  %
    &  %
    \\
    \hline
   \ccitetext{ochukut_research_2023}
    &  %
    & \x %
    &  %
    &  %
    &  %
    & \x %
    &  %
    &  %
    & \x %
    &  %
    &  %
    & \x %
    & \x %
    &  %
    &  %
    &  %
    &  %
    & \x %
    &  %
    \\
    \hline
    \rowcolor{gray!50} %
   \ccitetext{du2024personalized}
    &  %
    &  %
    & \x %
    & \x %
    &  %
    &  %
    & \x %
    &  %
    &  %
    &  %
    &  %
    & \x %
    &  %
    &  %
    &  %
    &  %
    &  %
    & \x %
    & \x %
    \\
    \hline
   \ccitetext{guo2024artificial}
    &  %
    &  %
    & \x %
    &  %
    & \x %
    &  %
    &  %
    &  %
    & \x %
    & \x %
    & \x %
    & \x %
    &  %
    &  %
    &  %
    &  %
    &  %
    &  %
    & \x %
    \\
    \hline
    \rowcolor{gray!50} %
   \ccitetext{halkiopoulos2024leveraging}
    &  %
    &  %
    & \x %
    &  %
    & \x %
    &  %
    &  %
    &  %
    & \x %
    &  %
    &  %
    & \x %
    &  %
    &  %
    &  %
    & \x %
    & \x %
    & \x %
    & \x %
    \\
    \hline
   \ccitetext{lin2024personalized}
    &  %
    &  %
    & \x %
    &  %
    & \x %
    &  %
    &  %
    &  %
    & \x %
    &  %
    &  %
    &  %
    &  %
    &  %
    &  %
    &  %
    &  %
    & \x %
    & \x %
    \\
    \hline
    \rowcolor{gray!50} %
   \ccitetext{sharma2024self}
    &  %
    &  %
    & \x %
    &  %
    & \x %
    &  %
    &  %
    &  %
    & \x %
    &  %
    & \x %
    & \x %
    &  %
    &  %
    &  %
    & \x %
    & \x %
    & \x %
    & \x %
    \\
    \hline
   \ccitetext{hardaker2025artificial}
    &  %
    &  %
    & \x %
    &  %
    & \x %
    &  %
    &  %
    &  %
    & \x %
    &  %
    &  %
    & \x %
    &  %
    &  %
    &  %
    &  %
    &  %
    & \x %
    &  %
    \\
    \hline
    \rowcolor{gray!50} %
   \ccitetext{liu2025advancing}
    &  %
    & \x %
    &  %
    &  %
    &  %
    & \x %
    &  %
    &  %
    & \x %
    & \x %
    &  %
    & \x %
    &  %
    &  %
    &  %
    &  %
    &  %
    & \x %
    & \x %
    \\
    \hline
    \hline
    \textbf{this review}
    & \x %
    &  %
    &  %
    & \x %
    &  %
    &  %
    &  %
    &  %
    & \x %
    & \x %
    & \x %
    & \x %
    & \x %
    & \x %
    & \x %
    & \x %
    & \x %
    & \x %
    & %
    \\
    \hline
    \bottomrule
    \end{tabular}
  }
  \caption{Inclusion criteria and analysis focus of related reviews. The table lists the last 19 related reviews. The reviews are ordered by publication year, with the most recent one listed last. Furthermore, this review is included as the last row in the table.}
  \label{tab:related-work-comparison-table-2}
\end{sidewaystable}
} 
\Cref{tab:related-work-comparison-table-1} and \Cref{tab:related-work-comparison-table-2} list related reviews in chronological order and compare them to our review, which is listed last.
We analyzed related reviews based on two aspects.
First, we examined the specified inclusion criteria, which define the papers they include for data charting.
They specify inclusion criteria regarding three aspects of the learning process:

\begin{enumerate}
  \item \textit{Learning context}: The learning context describes the environment that is provided for learning. Some reviews only include papers that focus on a specific LMS or LMSs in general. Other reviews only include papers that focus on ALSs or a specific type of ALS, such as ITSs. Other reviews do not specify inclusion criteria for the learning context. Instead, they focus on a specific learning support functionality provided by any type of learning context.
  \item \textit{Learning support functionalities}: The learning support functionalities, such as ALMs, are elaborated by the reviewed approaches to support learning. Some reviews only include papers that focus on specific learning support functionalities, such as content recommendation. Others include papers that focus on providing learning support mechanisms in general. Other reviews do not specify inclusion criteria for learning support functionalities. Instead, they focus on specific properties of the learning context.
  \item \textit{What is learned and by whom}: Some reviews only include papers that focus on supporting specific groups of learners (\eg those in higher education) or subjects (\eg STEM education).
\end{enumerate}

Second, we examined the analysis focus.
Related reviews analyze the following four aspects of the included papers:

\begin{enumerate}
  \item \textit{Learning context and learning support functionalities}: Some reviews analyze the learning context and the learning support functionalities that the reviewed approaches work on. This can include analyzing how learning support functionalities are integrated into the focused learning context (\eg system-independent or LMS-specific) and the devices used to provide learning support. Furthermore, this can include analyzing the properties of the focused learning context, \eg analyzing the characteristics of an LMS enriched with ALMs. Furthermore, this can include analyzing the properties of elaborated learning support functionalities, such as which functionalities are provided, how they are implemented, and whether they are intended to adapt the learning process to learners' needs (\ie ALMs). Finally, this can include analyzing the data sources that are processed to provide learning support functionalities.
  \item \textit{Authoring of learning support functionalities}: Some reviews analyze the provided authoring capabilities. This can entail examining which user groups are intended to specify learning support functionalities and how the authoring process works. Furthermore, this can include analyzing which properties of the provided learning support functionalities can be authored.
  \item \textit{Didactic and pedagogical properties}: Some reviews analyze the included papers from an educational point of view. This can include analyzing the approaches' goals, such as the intended effect on the learning process. This can also include analyzing the approaches' underlying didactics, such as their learning frameworks. Finally, this can include analyzing how the learning support functionalities are evaluated, including which metrics are used to measure reported effects on the learning process.
  \item \textit{Meta-aspects of reviewed research}: Some reviews analyze meta-aspects of the reviewed research, such as the research methods used or the researchers involved.
\end{enumerate}

Our review differs from related reviews in five main aspects.
First, few of the described reviews focus on LMSs.
Of these reviews, only two focus on learning support functionalities for LMSs.
Unlike our review, these two reviews analyze the properties of the integrated learning support functionalities but do not examine their integration into the corresponding LMS, authoring capabilities, or the educational context of the related research.
Second, most reviews focus on either no learning support functionalities or only specific ones, such as AI-powered tutoring. 
Thus, they do not provide an overview of the various types of functionalities available.
Third, the integration of learning support functionalities, which is one of the main focuses of this review, is rarely analyzed.
Whether learning support functionalities are integrated system-independently or system-specifically is not analyzed at all.
Fourth, the capabilities provided for authoring are rarely analyzed.
With regard to our target user groups, authoring is one of the main focuses of this review.
Fifth, while several related reviews analyze how reviewed approaches evaluate learning support functionalities, the educational goals and underlying learning frameworks are rarely analyzed.
In contrast, our review explicitly focuses on investigating the educational context of the approaches under review.

In summary, our review includes approaches that provide ALMs to support learning within LMSs, regardless of the type of learner or subject matter.
To answer our research questions, we analyze the included approaches in terms of all described content-related aspects, omitting the analysis of meta-aspects of the corresponding research.
Subsequently, our review provides the following novel insights, which, to the best of our knowledge, cannot be extracted from related reviews:

\begin{itemize}
  \item Which ALMs are provided to support learning within LMSs? How do these mechanisms work, and what data do they operate on?
  \item What related authoring capabilities are provided?
  \item To what extent can elaborated ALMs be used across different LMSs?
  \item How are these ALMs connected to learning theory? What are the educational objectives of applying them? What are the reported effects on the learning process?
\end{itemize}
\section{Method}
\label{sec:method}

In this section, we present our method to conduct a comprehensive and structured scoping review.
We describe the challenges encountered during the different processing steps and the methods employed to address them.
Our review is based on the PRISMA\cciteparen{page_prisma_2021} extension for scoping reviews proposed by\ccitetext{tricco_prisma_2018}.
We defined the research questions, eligibility criteria, and literature databases a priori in team discussions.
We have not published a formal review protocol.
To avoid redundancy with prior surveys, we conducted a general literature search beforehand.

\subsection{Inclusion Criteria}
\label{sec:inclusionCriteria}

Papers had to meet the following criteria to be included in this review.

\begin{inclusioncriteria}
    \item\label{ic:writtenInEnglish} The paper is written in English.
    \item\label{ic:containsAbstract} The paper contains an abstract.
    \item\label{ic:isAccessible} The paper is accessible.
    \item\label{ic:isAResearchPaper} The paper is a peer-reviewed research paper (\eg no workshop or expert opinion).
    \item\label{ic:writtenAfter2002} The paper was written in or after 2003 and in or before July 2023.
    \item\label{ic:mechanismsDescribed} The considered ALMs are clearly described.
    \item\label{ic:focusedOnPractice} The research presented focuses on practical aspects.
    \item\label{ic:focusedOnLearner} The research presented addresses adaptive mechanisms that support learners.
    \item\label{ic:focusedOnLms} The research presented addresses ALMs that are integrated into LMSs.
\end{inclusioncriteria}

The first four inclusion criteria were specified to ensure the accessibility and the quality of the included papers.
We only considered papers written in or before July 2023 (\ie the time up to which papers exist during the review), and only papers written in or after 2003, because we did not find any papers that discuss our research questions and have been published prior to 2003.
This may be due to the rise in popularity of LMSs between 2000 and 2003\cciteparen{davis_evolution_2009}, or the shift to web-based learning systems between 1995 and 2005\cciteparen{nicholson_history_2007}.
Furthermore, we included only papers that clearly describe the proposed ALMs so that we can answer our research questions.
For instance, one paper may describe how an ALM adapts the learning process in detail, yet omit to explain how this adaptation works (\ie what learner data is processed to determine the required adaptation, how this data is processed, \etc).
We could not include this paper because it lacks the necessary information to answer our research questions.
For example, this applies to\ccitetext{gross2013towards}.
To address the challenge that we might not include an approach that does not provide the required information in the considered paper, but in another paper, we also considered related papers~(\cf \Cref{sec:search}).

To ensure they are feasible for integration into LMSs, we only included implemented approaches.
For example, we did not consider pure conceptual work, analysis, or discussion of challenges or potential future developments\cciteeg{gorbunovs2017conceptual}.
To avoid excluding approaches that describe an implementation in another related paper, we considered these related papers as well.
We also considered any reference implementation of a conceptual approach.
Furthermore, we only included research focusing on ALMs that support learners.
Papers describing mechanisms to support instructors\cciteeg{graf_supporting_2009} were not considered.
Finally, we only considered approaches focusing on ALMs for LMSs.

\subsection{Literature Databases}

The screening process was based on two bibliographic databases, \textit{Scopus}\cciteparen{scopus} and \textit{Web of Science}\cciteparen{web_of_science}.
\textit{Google Scholar}\cciteparen{google_scholar} was searched as an additional source to identify potentially relevant literature, including publications not indexed in the primary databases.
We checked whether the two bibliographic databases contain records of relevant journals and conferences on a test basis.
For this purpose, we first compiled a small number of journals and conferences that we considered relevant based on the results of our general literature search during team discussions.
Next, we searched the selected databases using the constructed search string to see if the results include records from the compiled journals and conferences.
The journals considered are the following: \textit{Computers \& Education}\cciteparen{computers_and_education} (2-year impact factor 8.9\cciteparen{computers_and_education}), \textit{Educational Technology Research and Development}\cciteparen{educational_technology_research_and_development} (5-year impact factor 4.8\cciteparen{educational_technology_research_and_development}), \textit{Journal of Computers in Education}\cciteparen{computers_in_education} (5-year impact factor 4.8\cciteparen{computers_in_education}), and \textit{Interactive Learning Environments}\cciteparen{interactive_learning_environments} (5-year impact factor 4.5\cciteparen{interactive_learning_environments}).
The conferences considered are the following: \textit{International Conference on Computer Supported Education}\cciteparen{international_conference_on_computer_supported_education} (CORE rank B\cciteparen{icore_conference_portal}), \textit{International Conference on Advanced Learning Technologies}\cciteparen{international_conference_on_advanced_learning_technologies} (CORE rank B\cciteparen{icore_conference_portal}), and \textit{European Conference on Technology-Enhanced Learning}\cciteparen{
european_conference_on_technology_enhanced_learning} (CORE rank B\cciteparen{icore_conference_portal}).
Our test yielded records from all specified journals and conferences.

\subsection{Search Strategy}
\label{sec:search}

We elaborated the search strategy based on\ccitetext{keele_guidelines_2007,petersen_systematic_2008,mangaroska_learning_2018,kubica2021supporting} and further refined it during team discussions.
We conducted the search in two steps.
First, we constructed the search string.
Second, the papers' first author conducted the selection process.

\subsubsection{Search String Construction}
\label{sec:searchStringConstruction}

We examined the titles, abstracts, and keywords of relevant papers that we found during the general literature search.
We analyzed recurring terms related to our research questions within these meta-fields.
Afterwards, we grouped terms belonging to the same topic into one of three categories.
The synonyms of each category were connected by disjunctions to three logical terms, which were connected by conjunctions to the final search string~(see \Cref{fig:search-strings}).

\begin{figure}[H]
    \centering
    \begin{BVerbatim}
(computer-aided education OR computer-assisted education
  OR intelligent assistance for learning OR intelligent learning
  OR intelligent tutoring OR personalized assistance for learning
  OR personalized learning) 
AND (adaptab OR adaptive OR authoring OR cross OR generalizab
  OR generalized OR generic OR independent OR interoperable)
AND (environment OR platform OR system)

    \end{BVerbatim}
    \caption{Search string used in this review to find papers in common literature databases.}
    \label{fig:search-strings}
\end{figure}

\paragraph{Category 1 --- Adaptive Learning}
The first group of our search string contains terms related to the implementation of adaptive learning, listed in alphabetical order.
The terms used to describe ALMs have varied considerably over time\cciteparen{joksimovic_history_2015}.
This is due to the evolution of ALS capabilities resulting from technological advances and scientific findings.
Additionally, they vary depending on the type of ALS considered.
While research on ITS generally uses the keyword \textit{intelligent tutoring} rather than the keyword \textit{personalized learning}, the keyword \textit{personalized learning} is used in other ALM-related research.

\paragraph{Category 2 --- Adaptable Approach}
The second group of our search string contains terms related to the integration of ALMs into LMSs, listed in alphabetical order.
It is challenging to specify terms that describe the portability of ALMs since some of those terms are also used to describe their characteristics.
For example, the term \textit{adaptive} is used to describe the capabilities of ALS to support adaptive learning as well as to integrate ALMs into different learning systems.

\paragraph{Category 3 --- Implementation Approach}
The third group of our search string contains terms related to the practical implementation of conceptualized ALMs, listed in alphabetical order.
We did not add the terms \textit{Learning Management System} or \textit{LMS} because we found that these terms are not always included in the considered meta-fields of relevant papers.
This is especially true when a paper focuses on a specific LMS and includes its name in the title, for example, instead of the type of learning system.
Thus, to prevent missing relevant papers, we decided not to further restrict the search string and accepted a potentially higher number of papers to screen.

\subsubsection{Selection Process}
\label{sec:selectionProcess}

We applied the defined search string to the title, abstract, and keywords in Scopus and extracted 2672 papers.
From Web of Science we extracted 2053 using the same procedure.
For Google Scholar, we used the search string on the title only since Google Scholar does not support searching abstracts or keywords.
This resulted in another 248 papers, making a total of 4973 to screen.
A complete list of the bib entries can be requested from one of the authors.
By comparing the considered meta-fields, we programmatically removed duplicates.
If certain duplicates could not be detected due to spelling inconsistencies we removed them manually during screening.
Since we screened the papers in alphabetical order, duplicates appeared among each other.
During the screening of the meta-fields, we removed a total of 1603 duplicates.
Furthermore, we filtered out 3187 papers because they did not meet our inclusion criteria.
During this step, we could definitively ascertain whether the screened papers meet \ref{ic:containsAbstract}, \ref{ic:isAResearchPaper}, or \ref{ic:writtenAfter2002}.
By contrast, we could not expect to find information that would allow us to determine unequivocally whether criteria \ref{ic:mechanismsDescribed}, \ref{ic:focusedOnPractice}, and \ref{ic:focusedOnLearner}, and \ref{ic:focusedOnLms} are met.
Thus, during this step, we only excluded papers that clearly state that they do not meet one of these inclusion criteria and included the others for full-text screening.

Afterwards, we screened the full texts of the remaining 183 papers in three steps.
The first step was to check if the text could actually be processed.
If the full text of a paper was not accessible (\ie\ref{ic:isAccessible}) or was not in English although the abstract was written in English (\ie\ref{ic:writtenInEnglish}), the paper was not included.
For 38 papers, we could not access the full texts.
Two other papers were filtered out because they were not written in English.
16 papers were related to the research described in another paper included in the full-text screening.
Subsequently, they were considered as related papers.

In the second step, we screened the full texts of the remaining 127 papers.
We proceeded as follows during the reading.
Because our research questions cover the entire process, from the specification to the provision of ALMs, we could not expect to find all the necessary information in one paper.
To address this challenge, the first author assigned a score, $S$, to each paper.
This score indicated the subjective assessment of whether the research described in the paper is relevant with respect to the defined inclusion criteria.
We calculated this score in accordance with the methodology outlined by\ccitetext{da2023systematic}, whereby three questions (\ie $Q_{1}$ to $Q_{3}$) according to the three inclusion criteria \ref{ic:mechanismsDescribed}, \ref{ic:focusedOnPractice}, and \ref{ic:focusedOnLearner} were delineated and assigned a numerical value for each extracted response (\ie $QA_{1}$ to $QA_{3}$ for $Q_{1}$ to $Q_{3}$).
We formulated the following questions -- \textit{"Are the considered ALMs clearly described?"} ($Q_{1}$), \textit{"Does the presented reasearch focuses on practical aspects?"} ($Q_{2}$), and \textit{"Are the presented ALMs focused on supporting learners?"} ($Q_{3}$).
If we answered \textit{yes} to question $Q_{x}$ during full-text screening, we assigned a value of $1$ to $QA_{x}$.
If we answered \textit{no}, we assigned a value of $-2$.
This way we ensured that failing to meet one of the inclusion criteria would be enough to exclude the paper.
If we could not find the information needed to answer $Q_{x}$, we assigned a value of $0$ to $QA_{x}$.
It was hypothesized that each paper would explicitly delineate the application context of the investigated ALMs.
Therefore, an additional question for \ref{ic:focusedOnLms} was not included.
Instead, we excluded a paper right away if we could determine that it does not focus on LMSs, or if we could not determine whether it does.

\begin{equation}
S = \sum_{i=1}^{3} QA_{i}
\label{eq:score}
\end{equation}

The rating was primarily conducted by the first author of this paper.
Any ambiguity was discussed within the team until consensus was reached.
After answering all questions for a specific paper, we calculated the paper's score according to \Cref{eq:score} adopted from\ccitetext{da2023systematic}.
This procedure resulted in a score ranging from $-6$ to $3$.
We used this to group the papers into three categories.
We assigned all papers with scores between $-6$ and $0$ to category 1.
Category 1 indicates that the research under review does not meet the inclusion criteria.
This may be because the paper clearly indicates that at least one inclusion criterion is not met or does not indicate that at least one is met.
We excluded the papers in category 1 right away.
We assigned all papers with a score of $3$ to category 3.
Category 3 indicates that the paper meets all inclusion criteria.
We included the papers in category 3 right away.
We assigned all papers with scores of $1$ or $2$ to category 2.
Category 2 indicates that the paper meets at least one inclusion criterion or that no information was found indicating that it does not meet one of the criteria.
Consequently, we could expect to find the requisite information in a related paper that focuses on the same research as the considered paper.
Thus, in the third step, we searched for these related papers.
We did this by checking which papers reference the papers of category 2 and which papers are referenced by the considered papers.
Then, we screened the related papers as well.
In this step, we excluded 78 papers because their research did not meet our inclusion criteria.
This resulted in a total of 134 papers excluded during full-text screening and a total of 49 papers for data charting.

Finally, we reviewed the references of the selected papers to identify additional relevant papers and checked by which papers these papers are referenced (\ie snowball technique\cciteparen{wohlin2014guidelines}).
We did this by first checking the title.
If the paper seemed relevant based on the title, we screened the abstract.
If the the paper still appeared to be relevant, we screened the full text, including searching for related papers as described above.
Through this process, we identified 12 additional relevant papers.
In total, we extracted 61 papers presenting distinct approaches to integrating ALMs into LMSs for data charting.
The complete selection process is visualized as a flow chart in \Cref{fig:selectionProcess}.

\begin{figure}[h]
  \centering
  \resizebox{.65\linewidth}{!}{%
\begin{tikzpicture}[node distance=1.4cm, every node/.style={align=left,text=black}]
  \pgfdeclarelayer{background}
  \pgfdeclarelayer{foreground}
  \pgfsetlayers{background,main,foreground}
  \begin{pgfonlayer}{foreground}
  \node (scopus) [slrnode2] {Scopus papers from 2003 to 2023 (n~=~2672)};
  \node (wos) [slrnode2, right=of scopus.west, anchor=west, xshift=2.9cm] {Web of Science papers from 2003 to 2023 (n~=~2053)};
  \node (googleScholar) [slrnode2, right=of wos.west, anchor=west, xshift=2.9cm] {Google Scholar papers from 2003 to 2023 (n~=~248)};
  \node (papers) [slrnodewide, below=of scopus.west, anchor=west, yshift=-0.1cm] {Total papers from 2003 to 2023 (n~=~4973)};
  \node (eliminateDuplicates) [process, below=of scopus.west, anchor=west, yshift=-1.4cm] {Elimination of duplicates};
  \node (papersWithoutDuplicates) [slrnode, below=of eliminateDuplicates.west, anchor=west] {Total papers without duplicates (n~=~3370)};
  \node (duplicates) [slrnode1, right=of papersWithoutDuplicates.west, anchor=west, xshift=5cm] {Duplicates (n~=~1603)};
  \node (abstractScreening) [process, below=of papersWithoutDuplicates.west, anchor=west] {Screening of titles, abstract and keywords};
  \node (papersSelectedForFulltexScreening) [slrnode, below=of abstractScreening.west, anchor=west] {Papers selected for full-text screening (n~=~183)};
  \node (papersExcludedByAbstract) [slrnode1, right=of papersSelectedForFulltexScreening.west, anchor=west, xshift=5cm] {Excluded papers (n~=~3187)};
  \node (fulltextScreening) [process, below=of papersSelectedForFulltexScreening.west, anchor=west] {Full-text screening};
  \node (papersSelectedForDataCharting) [slrnode, below=of fulltextScreening.west, anchor=west] {Papers selected for data charting (n~=~49)};
  \node (papersExcludedByFulltext) [slrnode1, right=of papersSelectedForDataCharting.west, anchor=west, xshift=5cm] {Excluded papers (n~=~134)};
  \node (snowballSearch) [process, below=of papersSelectedForDataCharting.west, anchor=west] {Snowball search};
  \node (papersIdentifiedDuringSnowballSearch) [slrnode, below=of snowballSearch.west, anchor=west, xshift=3.3cm] {Papers found during snowball search (n~=~12)};
  \node (totalPapersSelectedForDataCharting) [slrnode, below=of snowballSearch.west, anchor=west, yshift=-1.5cm] {Papers selected for data charting (n~=~61)};
  \end{pgfonlayer}
  \begin{pgfonlayer}{background}
  \draw [arrow] (scopus) -- (papers);
  \draw [arrow] (wos) -- (papers);
  \draw [arrow] (googleScholar) -- (papers);
  \draw [arrow] (papers) -- (eliminateDuplicates);
  \draw [arrow] (eliminateDuplicates) -- (duplicates);
  \draw [arrow] (eliminateDuplicates) -- (papersWithoutDuplicates);
  \draw [arrow] (papersWithoutDuplicates) -- (abstractScreening);
  \draw [arrow] (abstractScreening) -- (papersSelectedForFulltexScreening);
  \draw [arrow] (abstractScreening) -- (papersExcludedByAbstract);
  \draw [arrow] (papersSelectedForFulltexScreening) -- (fulltextScreening);
  \draw [arrow] (fulltextScreening) -- (papersSelectedForDataCharting);
  \draw [arrow] (fulltextScreening) -- (papersExcludedByFulltext);
  \draw [arrow] (papersSelectedForDataCharting) -- (snowballSearch);
  \draw [arrow] (papersSelectedForDataCharting) -- (totalPapersSelectedForDataCharting);
  \draw [arrow] (snowballSearch) -- (papersIdentifiedDuringSnowballSearch);
  \draw [arrow] (papersIdentifiedDuringSnowballSearch) -- (totalPapersSelectedForDataCharting);
  \end{pgfonlayer}
  \end{tikzpicture} %
  }
  \caption{Selection process shows the number of papers processed in each step.} \label{fig:selectionProcess}%
\end{figure}

\subsection{Data Charting}
\label{sec:dataCharting}

We developed a data charting scheme through several team discussions.
We specified the variables of this scheme based on our research questions and theoretical framework.
We further refined the possible variable expressions during the charting process.
Variable expressions are numbered in alphabetical order.
The papers' first author extracted the variable values during in-depth reading of the extracted papers and the identified related papers.

\paragraph{Implementation Approach}
\textit{\var{v:implementationApproach} implementation approach} and \textit{\var{v:processingMechanism} processing mechanism} were extracted to answer \ref{rq:implementation}.
\ref{v:implementationApproach} refers to the implementation dimension of our descriptive framework.
The first two expressions represent LMS-specific implementation options.
\expr{v:implementationApproach}{v:implementationApproach:e:lmsSpecificImplementation}{3} \textit{direct implementation} means extending a specific LMS or implementing a new adaptive LMS.
\expr{v:implementationApproach}{v:implementationApproach:e:plugin}{5} \textit{plugin implementation} means implementing a plugin for a particular LMS.
Three expressions represent system-independent implementation options.
\expr{v:implementationApproach}{v:implementationApproach:e:adapterImplementation}{2} \textit{direct adapter implementation} refers to implementing an adaptation layer in a specific LMS to communicate with components that implement ALMs system-independently.
For example, this could refer to the implementation of certain API calls.
Another option is to implement this adaptation layer as a plugin using the plugin architectures that many LMSs offer (\expr{v:implementationApproach}{v:implementationApproach:e:adapterPlugin}{1} \textit{adapter plugin}).
The third option is to use the Learning Tools Interoperability (LTI) specification to implement ALMs as an LTI provider (\expr{v:implementationApproach}{v:implementationApproach:e:lti}{4} \textit{LTI}) that can interact with any LTI consumer.

From a self-regulated learning perspective, the context of support is a condition of the learning process~(\cf COPES model \Cref{sec:educationalFramework}).
Modifying this condition during the learning process causes learners to start the learning stages again, which involves analyzing existing conditions, setting personal learning goals, and developing learning strategies to achieve them.
Therefore, changes in the support context may weaken support for self-regulated learning and disrupt the self-regulated learning process.
The implementation approach described by \ref{v:implementationApproach} is an indicator of continuity in support.
While ALMs integrated using LTI, support the learning process in all LMSs the same way, ALMs that are directly implemented in different LMSs may vary from LMS to LMS.

\paragraph{Processing Mechanism}
\ref{v:processingMechanism} refers to the processing of learning data to apply ALMs.
It is related to the method dimension of our descriptive framework and provides evidence that either supports or refutes \ref{h:insufficientProcessingMechanismsSupport}.
The variable expressions differ according to two aspects.
First, they differ based on what users need to specify when parameterizing ALMs.
Second, they differ based on the ALS's responsibilities during data processing.

Processing mechanisms corresponding to \expr{v:processingMechanism}{v:processingMechanism:e:executionBased}{2} \textit{execution based} require the user to define triggers for certain learner interactions.
These can be active interactions, such as answering a question, or passive interactions, such as spending a certain amount of time on a learning content page.
The ALS is responsible for determining which ALMs, if any, should be initiated by a particular interaction and for initiating the appropriate ones.
From a self-regulated learning perspective, execution based processing mechanisms can be used to recognize transitions between learning stages or phases.
Processing mechanisms corresponding to \expr{v:processingMechanism}{v:processingMechanism:e:reasoningBased}{4} \textit{reasoning based} require users to specify learning states along with appropriate ALMs.
These states are typically represented by a set of rules or logical terms that are dissolved to true or false.
The ALS is responsible for determining which states match the current learning state to deduce the ALMs that need to be initiated.
From a self-regulated learning perspective, reasoning based processing mechanisms can be used to evaluate learners' performance externally\ccitecf{winnie1998studying} or to support self-evaluation during the self-reflection phase\ccitecf{zimmerman2002becoming}.
Processing mechanisms corresponding to \expr{v:processingMechanism}{v:processingMechanism:e:calculationBased}{1} \textit{calculation based} require users to specify formulas that describe the learning process. These formulas are typically based on variables that change over time or on learning success.
Based on the calculation results of these formulas, predefined ALMs are initiated.
The ALS is responsible for calculating the formulas and initiating the associated ALMs.
These mechanisms can also be used to evaluate learning processes from the perspective of self-regulated learning.
Processing mechanisms corresponding to \expr{v:processingMechanism}{v:processingMechanism:e:optimizationBased}{3} \textit{optimization based} require users to specify parameters of the ALMs and associated optimization goals.
The ALS is responsible for optimizing parameter values and initiating ALMs with those values.
These processing mechanisms also include AI methods that optimize adaptation processes based on a set of underlying training data.
From a self-regulated learning perspective, processing mechanisms based on optimization are important to support the adaptation stage.
For example, these mechanisms can optimize the mapping between learners' products and standards by updating applicable learning operations.

In summary, analyzing the reviewed approaches with regard to their implemented processing mechanisms provides insight into the extent to which they support self-regulated learning.
Implementing single processing mechanisms results in supporting certain learning stages or single learning operations.
However, extensive support of self-regulated learning requires supporting most or all of the described processing mechanisms.

\paragraph{Adapted Learning Process Components}
\textit{\var{v:adaptedComponent} adapted component} and \textit{\var{v:adaptiveLearningMechanism} adaptive learning mechanism} were extracted to answer \ref{rq:provision}.
\ref{v:adaptedComponent} refers to the target dimension of our descriptive framework.
\expr{v:adaptedComponent}{v:adaptedComponent:e:learningContent}{3} \textit{learning content} means adapting the learning content, including what is presented and how it is presented.
\expr{v:adaptedComponent}{v:adaptedComponent:e:supportiveOrInstructionalOutput}{4} \textit{supportive or instructional output} refers to adapting output that is provided by the ALMs to support or guide the learning process.
For example, this can include adapting prompts that are provided based on the learner's performance.
\expr{v:adaptedComponent}{v:adaptedComponent:e:interaction}{1} \textit{interaction} means adapting the interaction between learners and the ALS.
For example, this can include adapting a dialogue between learners and the ALS, which is common with ITSs.
\expr{v:adaptedComponent}{v:adaptedComponent:e:interface}{2} \textit{interface} means adapting the entire user interface of the learning environment.

The adapted learning components can indicate which self-regulated learning phases\ccitecf{zimmerman2002becoming} are supported.
ALMs that adapt the learning content or the learning interface can personalize the learning environment to learners' needs based on their interests or learning styles during the forethought phase.
ALMs that adapt instructions provided to learners during interaction or as prompts can support the performance phase by providing guidance in developing self-regulated learning skills.

\paragraph{Supported ALMs}
\ref{v:adaptiveLearningMechanism} is related to the target dimension of our descriptive framework as well.
It captures which ALMs adapt the previously described targets.
From a self-regulated learning perspective, this variable offers detailed information about the supported learning process operations\ccitecf{winnie1998studying}.

Three types of ALMs can adapt instructions.
First, ALSs can provide feedback based on learners' interactions (\expr{v:adaptiveLearningMechanism}{v:adaptiveLearningMechanism:e:feedbackProvision}{4} \textit{feedback provision}).
Second, ALSs can provide general prompts, which may include recommendations for action, general advice, or motivational support (\expr{v:adaptiveLearningMechanism}{v:adaptiveLearningMechanism:e:promptProvision}{5} \textit{prompt provision}).
Third, ALSs can support scaffolding to guide learners to a specific learning goal (\expr{v:adaptiveLearningMechanism}{v:adaptiveLearningMechanism:e:scaffoldingSupport}{6} \textit{scaffolding support}).
Three types of ALMs can adapt learning content or its presentation.
Since \ref{v:adaptiveLearningMechanism} focuses on the functional provision of ALMs rather than on the pedagogical intent, all of these can be used to adapt both the content and its presentation.
First, ALSs can adapt learning content by modifying it (\expr{v:adaptiveLearningMechanism}{v:adaptiveLearningMechanism:e:contentAdaptation}{1} \textit{content adaptation}).
Second, ALSs can attempt to adapt learning content by recommending it (\expr{v:adaptiveLearningMechanism}{v:adaptiveLearningMechanism:e:contentRecommendation}{2} \textit{content recommendation}).
Learners may or may not follow these recommendations.
Third, ALSs can select learning content to force the learner to work with it (\expr{v:adaptiveLearningMechanism}{v:adaptiveLearningMechanism:e:contentSelection}{3} \textit{content selection}).

\paragraph{Authoring Capabilities}
\textit{\var{v:authoringCapability} authoring capability} and \textit{\var{v:adaptiveLearningMechanismTrigger} adaptive learning mechanism trigger} were extracted to answer \ref{rq:definition}.
\ref{v:authoringCapability} refers to the authored components dimension of our descriptive framework.
It captures the level at which our target user groups are intended to specify the function of ALMs.
Since our target user groups cannot be assumed to have a high level of technical expertise, it does not focus on the ability of a programmer to customize freely available open source code.
Analyzing the extent to which our target users are supported in authoring ALMs determines how effectively self-regulated learning can be supported, particularly with regard to fading support.
Furthermore, it provides evidence that either supports or refutes \ref{h:insufficientAuthoringCapabilitiesSupport}.

\expr{v:authoringCapability}{v:authoringCapability:e:noAuthoring}{4} \textit{no authoring} means that no authoring capabilities are provided.
Such approaches implement specific learning strategies tailored to particular content.
\expr{v:authoringCapability}{v:authoringCapability:e:authoringOfAdaptiveLearningMeans}{1} \textit{authoring of adaptive learning means} means that the authoring of the target or means of ALMs is supported, while their function is fixed.
For example, this could be the ability to create the learning content that the ALMs are working with.
\expr{v:authoringCapability}{v:authoringCapability:e:authoringOfAdaptiveLearningMechanismParameters}{3} \textit{authoring of ALM parameters} means supporting the authoring of parameters for fixed ALMs.
Thus, although the ALMs are predetermined, the users can influence how they work.
\expr{v:authoringCapability}{v:authoringCapability:e:authoringOfAdaptiveLearningMechanisms}{2} \textit{authoring of ALMs} means supporting the authoring of ALMs themselves.
In this case, neither their function nor their means are predetermined.

\paragraph{Adaptation Triggers}
\ref{v:adaptiveLearningMechanismTrigger} refers to when ALMs are initiated.
It is related to the time and the method dimension of our descriptive framework.
\expr{v:adaptiveLearningMechanismTrigger}{v:adaptiveLearningMechanismTrigger:e:assessmentAnswering}{1} \textit{assessment answering} means that ALSs initiate ALMs based on assessment results.
\expr{v:adaptiveLearningMechanismTrigger}{v:adaptiveLearningMechanismTrigger:e:learningSessionBegin}{2} \textit{learning session begin} means that ALSs initiate ALMs when a learner enters the LMS or starts the learning process.
\expr{v:adaptiveLearningMechanismTrigger}{v:adaptiveLearningMechanismTrigger:e:navigation}{3} \textit{navigation} means that ALSs initiate ALMs when learners navigate the learning system.
\expr{v:adaptiveLearningMechanismTrigger}{v:adaptiveLearningMechanismTrigger:e:proactiveInitiation}{4} \textit{proactive initiation} means that the learners proactively request the initiation of ALMs.
One example of this is proactively asking for help.

From a self-regulated learning perspective, trigger expressions are necessary to analyze whether adaptations align with the boundaries of self-regulated learning phases or if solely event-driven triggers disrupt sustained support of all phases.
The forethought phase requires pre-learning activation. 
The performance phase requires ongoing monitoring triggers. 
The self-reflection phase requires post-learning evaluation prompts.
Proactive initiation of adaptation by the learners is essential for self-regulated learning as well.
This is because reflecting on their learning methods, analyzing what kind of help they need, and asking for it specifically supports the development of self-regulated learning skills.

\paragraph{Data Basis}
\textit{\var{v:learningContent} learning content} and \textit{\var{v:adaptationSourceData} adaptation source data} were extracted to answer \ref{rq:database} and to gather evidence that either supports or refutes \ref{h:insufficientDataBasisSupport}.
\ref{v:learningContent} refers to the learning content with which the learner works while the ALMs are provided.
This is related to the target dimension of our descriptive framework.
\expr{v:learningContent}{v:learningContent:e:nonLmsContent}{2} \textit{non-LMS content} refers to the use of external content brought to the LMSs with the ALMs.
For example, this can be the case when an ITS with predefined content for a specific domain is connected to an LMS.
\expr{v:learningContent}{v:learningContent:e:particularLmsContent}{3} \textit{particular LMS content} refers to the use of a specific content format.
\expr{v:learningContent}{v:learningContent:e:nativeLmsContent}{1} \textit{native LMS content} refers to the use of various types of native LMS content.

\paragraph{Adaptation Source Data}
\ref{v:adaptationSourceData} refers to the data that is processed as the basis of applying ALMs.
This variable does not take the learning content data already considered by \ref{v:learningContent} into account, but rather, only the data from the source dimension of our descriptive framework \ie learner data and interaction data.
\expr{v:adaptationSourceData}{v:adaptationSourceData:e:customInteractionData}{1} \textit{custom interaction data} refers to interaction data that is collected by mechanisms brought to the LMSs with the ALMs.
\expr{v:adaptationSourceData}{v:adaptationSourceData:e:existingInteractionData}{3} \textit{existing interaction data} refers to interaction data collected by the LMSs itself or by similar mechanisms that are reused by the proposed ALMs.
\expr{v:adaptationSourceData}{v:adaptationSourceData:e:requestedLearnerData}{4} \textit{requested learner data} refers to learner data that is collected by directly asking the learner for information \eg in the form of questionnaires.
The last expression \expr{v:adaptationSourceData}{v:adaptationSourceData:e:derivedLearnerData}{2} \textit{derived learner data} refers to how the described data is used.
It captures whether ALMs consider data derived from recorded raw data (\ie by appropriate processing mechanisms) for adaptation, or if they consider the collected raw data directly.

Analyzing the data basis of ALMs provides information about the extent to which the complete learning process can be supported.
This is crucial with regard to the cyclic self-regulated learning process.
Preventing working with existing learning content, or considering the native interaction data generated during learning within the LMS, reduces the effectiveness of supporting self-regulated learning by restricting support to only specific learning activities.
Furthermore, analyzing how information is processed during the learning process is also important with regard to self-regulated learning.
Providing extensive and effective support for self-regulated learning requires a broad range of mechanisms.
According to the COPES model\cciteparen{winnie1998studying}, evaluations of the learning process can be performed internally by the learners themselves or externally by an instructor or an ALS.
To evaluate the learning process externally, it is necessary to monitor the learning products and process the monitored information by comparing it to learning process standards.
To support internal evaluations, ALMs must first be aware of learners' current state of internal evaluation.
This can only be achieved by asking the learner to self-assess this process.

\paragraph{Educational Surrounding Conditions and Outcomes}
\textit{\var{v:adaptationGoal} adaptation goal}, \textit{\var{v:adaptationBasis} adaptation basis}, and \textit{\var{v:reportedEffects} reported effects} were extracted to answer \ref{rq:educationalContext}.
They are required to extract the information necessary for an educational contextualization of the reviewed approaches according to our educational framework.

\ref{v:adaptationGoal} refers to the educational goal pursued by the implemented ALMs.
This can be analyzed to determine which learning phases or stages can be supported from a self-regulated learning perspective.
\ref{v:reportedEffects} refers to the educational effects reported by the reviewed approaches during their evaluation or usage.
Because of the variety of goals and reported effects, we did not provide explicit expressions for these two variables.
nstead, we collect the corresponding information as free text.
\ref{v:adaptationBasis} refers to the learner characteristics that function as the basis of ALMs.
According to our educational framework, there are five different bases -- \textit{affect} (\expr{v:adaptationBasis}{v:adaptationBasis:e:affect}{1}), \textit{learners' knowledge} (\expr{v:adaptationBasis}{v:adaptationBasis:e:knowledge}{2}), \textit{learning path} (\expr{v:adaptationBasis}{v:adaptationBasis:e:path}{3}), \textit{learning style} (\expr{v:adaptationBasis}{v:adaptationBasis:e:style}{4}), and \textit{metacognition} (\expr{v:adaptationBasis}{v:adaptationBasis:e:metacognition}{5}).
Furthermore, we investigated whether the reviewed approaches reference established learning frameworks for the considered adaptation bases.
If this is the case, then this information is also captured by \ref{v:adaptationBasis}.
This allows us to analyze the extent to which the reviewed approaches support certain aspects of self-regulated learning and the extent to which these approaches intentionally support these aspects.
\section{Results}
\label{sec:results}

In this section, we discuss the 61 approaches that we selected for data charting.
For 20 of these approaches, we found an implementation that is system-independent.
For each research question, we first present the extracted information.
Afterwards, we discuss the results in terms of the patterns, gaps and contradictions we identified across the reviewed approaches and with regard to self-regulated learning according to the introduced models.
The results are summarized in \Cref{tab:scopingReviewResultsPart1}, \Cref{tab:scopingReviewResultsPart2}, \Cref{tab:scopingReviewResultsPart3}, and \Cref{tab:scopingReviewResultsEducationalContext}.

{
\newcolumntype{C}{>{\centering \arraybackslash}m{(\textwidth-(6\arrayrulewidth)-(10\tabcolsep))/34}}
\begin{sidewaystable}[ph!]
  \resizebox{\textwidth}{!}{%
    \def\arraystretch{2.4}
    \begin{tabular}{|l||C|C|C|C|C||C|C|C|C||C|C|C|C||C|C|C|C|C|C||C|C|C|C||C|C|C|C||C|C|C||C|C|C|C|}
      \hline
      \multicolumn{1}{|c||}{\parbox[t]{7cm}{\raggedright Variables}} & %
      \multicolumn{5}{c||}{\parbox[t]{4.2cm}{\raggedright \textbf{\ref{v:implementationApproach}}~~~\parbox{2cm}{\raggedright Implementation approach}}} & %
      \multicolumn{4}{c||}{\parbox[t]{3.3cm}{\raggedright \textbf{\ref{v:processingMechanism}}~~~\parbox{3cm}{\raggedright Processing mechanism}\\~\\~\\}} & %
      \multicolumn{4}{c||}{\parbox[t]{3.3cm}{\raggedright \textbf{\ref{v:adaptedComponent}}~~~\parbox{3cm}{\raggedright Adapted component}}} & %
      \multicolumn{6}{c||}{\parbox[t]{5.2cm}{\raggedright \textbf{\ref{v:adaptiveLearningMechanism}}~~~\parbox{5.2cm}{\raggedright Adaptive learning mechanism}}} & %
      \multicolumn{4}{c||}{\parbox[t]{3.3cm}{\raggedright \textbf{\ref{v:authoringCapability}}~~~\parbox{3.2cm}{\raggedright Authoring capability}}} & %
      \multicolumn{4}{c||}{\parbox[t]{3.3cm}{\raggedright \textbf{\ref{v:adaptiveLearningMechanismTrigger}}~~~\parbox{3cm}{\raggedright Adaptive learning mechanism trigger}}} & %
      \multicolumn{3}{c||}{\parbox[t]{2.35cm}{\raggedright \textbf{\ref{v:learningContent}}~~~\parbox{2cm}{\raggedright Learning content}}} & %
      \multicolumn{4}{c|}{\parbox[t]{3.4cm}{\raggedright \textbf{\ref{v:adaptationSourceData}}~~~\parbox{3cm}{\raggedright Adaptation\\source data}}} \\
      \hline
      \multicolumn{1}{|c||}{\parbox[t]{7cm}{\raggedright Explanations}} & %
      \multicolumn{5}{c||}{\parbox[t]{4.2cm}{\raggedright How are the ALMs technically integrated into LMS?\\~\\}} & %
      \multicolumn{4}{c||}{\parbox[t]{3.3cm}{\raggedright What data processing mechanisms are applied to implement the ALMs?}} & %
      \multicolumn{4}{c||}{\parbox[t]{3.3cm}{\raggedright Which components of the learning process are being adapted?}} & %
      \multicolumn{6}{c||}{\parbox[t]{5.2cm}{\raggedright Which ALMs are supported?}} & %
      \multicolumn{4}{c||}{\parbox[t]{3.3cm}{\raggedright At what level are users supported to author parts of the ALMs on their own?}} & %
      \multicolumn{4}{c||}{\parbox[t]{3.3cm}{\raggedright How are the ALMs triggered?}} & %
      \multicolumn{3}{c||}{\parbox[t]{2.35cm}{\raggedright What content do the ALMs operate on?}} & %
      \multicolumn{4}{c|}{\parbox[t]{3.4cm}{\raggedright What data is processed in order to provide the ALMs?}} \\
      \cmidrule{1-35} %
      \multicolumn{1}{|c||}{\parbox[t]{7cm}{\raggedright Expressions}} & %
      \rotatebox[origin=l]{270}{\textbf{\ref{v:implementationApproach:e:adapterPlugin}}~~~Adapter Plugin} & %
      \rotatebox[origin=l]{270}{\textbf{\ref{v:implementationApproach:e:adapterImplementation}}~~~Direct Adapter Implementation} & %
      \rotatebox[origin=l]{270}{\textbf{\ref{v:implementationApproach:e:lmsSpecificImplementation}}~~~Direct Implementation} & %
      \rotatebox[origin=l]{270}{\textbf{\ref{v:implementationApproach:e:lti}}~~~LTI} & %
      \rotatebox[origin=l]{270}{\textbf{\ref{v:implementationApproach:e:plugin}}~~~Plugin Implementation} & %
      \rotatebox[origin=l]{270}{\textbf{\ref{v:processingMechanism:e:calculationBased}}~~~Calculation based} & %
      \rotatebox[origin=l]{270}{\textbf{\ref{v:processingMechanism:e:executionBased}}~~~Execution based} & %
      \rotatebox[origin=l]{270}{\textbf{\ref{v:processingMechanism:e:optimizationBased}}~~~Optimization based} & %
      \rotatebox[origin=l]{270}{\textbf{\ref{v:processingMechanism:e:reasoningBased}}~~~Reasoning based} & %
      \rotatebox[origin=l]{270}{\textbf{\ref{v:adaptedComponent:e:interaction}}~~~Interaction} & %
      \rotatebox[origin=l]{270}{\textbf{\ref{v:adaptedComponent:e:interface}}~~~Interface} & %
      \rotatebox[origin=l]{270}{\textbf{\ref{v:adaptedComponent:e:learningContent}}~~~Learning content} & %
      \rotatebox[origin=l]{270}{\textbf{\ref{v:adaptedComponent:e:supportiveOrInstructionalOutput}}~~~Supportive or instructional output} & %
      \rotatebox[origin=l]{270}{\textbf{\ref{v:adaptiveLearningMechanism:e:contentAdaptation}}~~~Content adaptation} & %
      \rotatebox[origin=l]{270}{\textbf{\ref{v:adaptiveLearningMechanism:e:contentRecommendation}}~~~Content recommendation} & %
      \rotatebox[origin=l]{270}{\textbf{\ref{v:adaptiveLearningMechanism:e:contentSelection}}~~~Content selection} & %
      \rotatebox[origin=l]{270}{\textbf{\ref{v:adaptiveLearningMechanism:e:feedbackProvision}}~~~Feedback provision} & %
      \rotatebox[origin=l]{270}{\textbf{\ref{v:adaptiveLearningMechanism:e:promptProvision}}~~~Prompt provision} & %
      \rotatebox[origin=l]{270}{\textbf{\ref{v:adaptiveLearningMechanism:e:scaffoldingSupport}}~~~Scaffolding support} & %
      \rotatebox[origin=l]{270}{\textbf{\ref{v:authoringCapability:e:authoringOfAdaptiveLearningMeans}}~~~\parbox{4.5cm}{Authoring of adaptive learning means}} & %
      \rotatebox[origin=l]{270}{\textbf{\ref{v:authoringCapability:e:authoringOfAdaptiveLearningMechanisms}}~~~\parbox{4.5cm}{Authoring of ALMs}} & %
      \rotatebox[origin=l]{270}{\textbf{\ref{v:authoringCapability:e:authoringOfAdaptiveLearningMechanismParameters}}~~~\parbox{5cm}{Authoring of adaptive learning\\mechanism parameters~~}} & %
      \rotatebox[origin=l]{270}{\textbf{\ref{v:authoringCapability:e:noAuthoring}}~~~No authoring} & %
      \rotatebox[origin=l]{270}{\textbf{\ref{v:adaptiveLearningMechanismTrigger:e:assessmentAnswering}}~~~Assessment answering} & %
      \rotatebox[origin=l]{270}{\textbf{\ref{v:adaptiveLearningMechanismTrigger:e:learningSessionBegin}}~~~Learning session begin} & %
      \rotatebox[origin=l]{270}{\textbf{\ref{v:adaptiveLearningMechanismTrigger:e:navigation}}~~~Navigation} & %
      \rotatebox[origin=l]{270}{\textbf{\ref{v:adaptiveLearningMechanismTrigger:e:proactiveInitiation}}~~~Proactive initiation} & %
      \rotatebox[origin=l]{270}{\textbf{\ref{v:learningContent:e:nativeLmsContent}}~~~Native LMS content} & %
      \rotatebox[origin=l]{270}{\textbf{\ref{v:learningContent:e:nonLmsContent}}~~~Non-LMS content} & %
      \rotatebox[origin=l]{270}{\textbf{\ref{v:learningContent:e:particularLmsContent}}~~~Particular LMS content} & %
      \rotatebox[origin=l]{270}{\textbf{\ref{v:adaptationSourceData:e:customInteractionData}}~~~Custom interaction data} & %
      \rotatebox[origin=l]{270}{\textbf{\ref{v:adaptationSourceData:e:derivedLearnerData}}~~~Derived learner data} & %
      \rotatebox[origin=l]{270}{\textbf{\ref{v:adaptationSourceData:e:existingInteractionData}}~~~Existing interaction data} & %
      \rotatebox[origin=l]{270}{\textbf{\ref{v:adaptationSourceData:e:requestedLearnerData}}~~~Requested learner data} \\
      \hline
      \hline
      \rowcolor{gray!50} %
     \ccitetext{santos2004overview} %
      & %
      & %
      & \x %
      & %
      & %
      & %
      & %
      & \x %
      & \x %
      & %
      & %
      & \x %
      & %
      & %
      & \x %
      & %
      & %
      & %
      & %
      & \x %
      & %
      & %
      & %
      & \x %
      & \x %
      & \x %
      & %
      & \x %
      & %
      & %
      & \x %
      & \x %
      & %
      & \x %
      \\
      \hline
     \ccitetext{watson2005steps} %
      & %
      & %
      & \x %
      & %
      & %
      & %
      & %
      & %
      & \x %
      & %
      & %
      & \x %
      & %
      & %
      & \x %
      & %
      & %
      & %
      & %
      & \x %
      & %
      & \x %
      & %
      & \x %
      & %
      & \x %
      & %
      & \x %
      & %
      & %
      & \x %
      & \x %
      & %
      & %
      \\
      \hline
      \rowcolor{gray!50} %
     \ccitetext{arapi2007supporting} %
      & %
      & \x %
      & %
      & %
      & %
      & \x %
      & %
      & %
      & \x %
      & %
      & %
      & \x %
      & %
      & %
      & \x %
      & \x %
      & %
      & %
      & %
      & \x %
      & %
      & \x %
      & %
      & %
      & \x %
      & \x %
      & \x %
      & %
      & \x %
      & %
      & %
      & \x %
      & \x %
      & \x %
      \\
      \hline
     \ccitetext{cobos2007learning} %
      & %
      & %
      & \x %
      & %
      & %
      & %
      & %
      & %
      & \x %
      & %
      & %
      & \x %
      & %
      & %
      & \x %
      & \x %
      & %
      & %
      & %
      & \x %
      & %
      & \x %
      & %
      & \x %
      & %
      & \x %
      & \x %
      & \x %
      & %
      & %
      & \x %
      & \x %
      & %
      & \x %
      \\
      \hline
      \rowcolor{gray!50} %
     \ccitetext{roselli2008integration} %
      & %
      & %
      & \x %
      & %
      & %
      & \x %
      & \x %
      & \x %
      & \x %
      & \x %
      & \x %
      & \x %
      & \x %
      & \x %
      & %
      & \x %
      & \x %
      & \x %
      & \x %
      & \x %
      & \x %
      & \x %
      & %
      & \x %
      & %
      & \x %
      & %
      & \x %
      & %
      & %
      & \x %
      & \x %
      & %
      & \x %
      \\
      \hline
     \ccitetext{serce_moda_2008} %
      & %
      & \x %
      & %
      & %
      & %
      & \x %
      & %
      & %
      & %
      & %
      & %
      & \x %
      & %
      & %
      & %
      & \x %
      & %
      & %
      & %
      & \x %
      & %
      & %
      & %
      & %
      & %
      & \x %
      & \x %
      & \x %
      & %
      & %
      & \x %
      & \x %
      & %
      & \x %
      \\
      \hline
      \rowcolor{gray!50} %
     \ccitetext{pedrazzoli2009artificial} %
      & %
      & %
      & \x %
      & %
      & %
      & %
      & %
      & \x %
      & \x %
      & %
      & %
      & \x %
      & \x %
      & %
      & \x %
      & \x %
      & \x %
      & %
      & %
      & \x %
      & %
      & %
      & %
      & \x %
      & %
      & \x %
      & \x %
      & \x %
      & \x %
      & %
      & %
      & \x %
      & \x %
      & %
      \\
      \hline
     \ccitetext{adorni2010caddie} %
      & %
      & %
      & \x %
      & %
      & %
      & %
      & %
      & %
      & \x %
      & %
      & %
      & \x %
      & %
      & %
      & %
      & \x %
      & %
      & %
      & %
      & \x %
      & %
      & \x %
      & %
      & %
      & \x %
      & %
      & %
      & \x %
      & %
      & %
      & \x %
      & \x %
      & %
      & \x %
      \\
      \hline
      \rowcolor{gray!50} %
     \ccitetext{graf2010flexible} %
      & %
      & %
      & \x %
      & %
      & %
      & \x %
      & %
      & %
      & \x %
      & %
      & %
      & \x %
      & %
      & %
      & \x %
      & %
      & %
      & %
      & %
      & \x %
      & %
      & %
      & %
      & \x %
      & %
      & %
      & %
      & \x %
      & %
      & %
      & %
      & \x %
      & %
      & \x %
      \\
      \hline
     \ccitetext{limongelli2010module} %
      & %
      & %
      & \x %
      & %
      & %
      & %
      & %
      & %
      & \x %
      & %
      & %
      & \x %
      & %
      & %
      & %
      & \x %
      & %
      & %
      & %
      & \x %
      & %
      & \x %
      & %
      & \x %
      & %
      & %
      & %
      & %
      & \x %
      & %
      & \x %
      & \x %
      & %
      & \x %
      \\
      \hline
      \rowcolor{gray!50} %
     \ccitetext{peter2010adaptable} %
      & %
      & %
      & \x %
      & %
      & %
      & %
      & %
      & %
      & \x %
      & %
      & %
      & \x %
      & %
      & %
      & \x %
      & \x %
      & %
      & %
      & %
      & \x %
      & %
      & \x %
      & %
      & %
      & %
      & %
      & \x %
      & \x %
      & %
      & %
      & %
      & \x %
      & %
      & \x %
      \\
      \hline
     \ccitetext{santos_web-based_2010} %
      & %
      & \x %
      & %
      & %
      & %
      & %
      & \x %
      & %
      & %
      & %
      & %
      & \x %
      & \x %
      & %
      & %
      & \x %
      & \x %
      & \x %
      & %
      & %
      & %
      & %
      & \x %
      & \x %
      & %
      & \x %
      & \x %
      & %
      & %
      & \x %
      & \x %
      & \x %
      & %
      & %
      \\
      \hline
      \rowcolor{gray!50} %
     \ccitetext{hammadadaptive} %
      & %
      & %
      & \x %
      & %
      & %
      & %
      & \x %
      & %
      & \x %
      & %
      & %
      & \x %
      & %
      & %
      & %
      & \x %
      & %
      & %
      & %
      & \x %
      & %
      & %
      & %
      & \x %
      & \x %
      & %
      & %
      & \x %
      & %
      & %
      & %
      & \x %
      & \x %
      & \x %
      \\
      \hline
     \ccitetext{rossi2011mapit} %
      & %
      & \x %
      & %
      & %
      & %
      & %
      & %
      & \x %
      & %
      & \x %
      & %
      & %
      & \x %
      & %
      & %
      & %
      & \x %
      & %
      & %
      & %
      & %
      & %
      & \x %
      & %
      & %
      & %
      & \x %
      & \x %
      & %
      & %
      & \x %
      & \x %
      & %
      & %
      \\
      \hline
      \rowcolor{gray!50} %
     \ccitetext{yaghmaie_context-aware_2011} %
      & %
      & \x %
      & %
      & %
      & %
      & %
      & %
      & %
      & \x %
      & %
      & %
      & \x %
      & %
      & %
      & %
      & \x %
      & %
      & %
      & %
      & \x %
      & %
      & %
      & %
      & %
      & \x %
      & %
      & %
      & %
      & %
      & \x %
      & \x %
      & \x %
      & %
      & \x %
      \\
      \hline
     \ccitetext{al2012intelligent} %
      & %
      & %
      & \x %
      & %
      & %
      & %
      & %
      & \x %
      & \x %
      & %
      & %
      & \x %
      & %
      & \x %
      & \x %
      & %
      & %
      & %
      & %
      & \x %
      & %
      & %
      & %
      & %
      & \x %
      & %
      & \x %
      & \x %
      & %
      & %
      & \x %
      & \x %
      & %
      & \x %
      \\
      \hline
      \rowcolor{gray!50} %
     \ccitetext{de_bra_grapple1_2013} %
      & %
      & \x %
      & %
      & %
      & %
      & %
      & \x %
      & %
      & %
      & %
      & \x %
      & \x %
      & %
      & \x %
      & %
      & \x %
      & %
      & %
      & %
      & \x %
      & \x %
      & \x %
      & %
      & \x %
      & %
      & \x %
      & %
      & %
      & \x %
      & %
      & \x %
      & \x %
      & %
      & %
      \\
      \hline
     \ccitetext{moutafi2013mining} %
      & %
      & %
      & \x %
      & %
      & %
      & \x %
      & %
      & %
      & \x %
      & %
      & %
      & \x %
      & %
      & %
      & \x %
      & \x %
      & %
      & %
      & %
      & \x %
      & %
      & %
      & %
      & \x %
      & %
      & %
      & %
      & \x %
      & %
      & %
      & \x %
      & \x %
      & %
      & \x %
      \\
      \hline
      \rowcolor{gray!50} %
     \ccitetext{santos_phd_2013} %
      & %
      & \x %
      & %
      & %
      & %
      & %
      & %
      & \x %
      & \x %
      & %
      & %
      & \x %
      & \x %
      & %
      & \x %
      & %
      & %
      & \x %
      & %
      & \x %
      & %
      & \x %
      & %
      & %
      & \x %
      & \x %
      & \x %
      & \x %
      & %
      & %
      & %
      & \x %
      & \x %
      & \x %
      \\
      \hline
     \ccitetext{batanero2014considering} %
      & %
      & %
      & %
      & %
      & \x %
      & %
      & %
      & %
      & \x %
      & %
      & %
      & \x %
      & %
      & %
      & %
      & \x %
      & %
      & %
      & %
      & \x %
      & %
      & %
      & %
      & %
      & %
      & \x %
      & %
      & \x %
      & %
      & %
      & %
      & \x %
      & %
      & \x %
      \\
      \hline
      \rowcolor{gray!50} %
     \ccitetext{bhaskaran2014research} %
      & %
      & \x %
      & %
      & %
      & %
      & \x %
      & %
      & \x %
      & %
      & %
      & %
      & \x %
      & %
      & %
      & %
      & \x %
      & %
      & %
      & %
      & \x %
      & %
      & %
      & %
      & \x %
      & %
      & \x %
      & %
      & \x %
      & %
      & %
      & %
      & \x %
      & \x %
      & %
      \\
      \hline
    \end{tabular}
  }
  \caption{Summary of the results of the structured scoping review corresponding to the extracted data charting variables \ref{v:implementationApproach} to \ref{v:adaptationSourceData} and their expressions. The table lists approach 1 of the identified approaches to approach 21. The approaches are ordered by the year of the extracted paper, with the most recent approach coming last. The variable expressions are in alphabetical order. \x indicates that the value of a variable matches the corresponding expression.}
  \label{tab:scopingReviewResultsPart1}
\end{sidewaystable}
} %
{
\newcolumntype{C}{>{\centering \arraybackslash}m{(\textwidth-(6\arrayrulewidth)-(10\tabcolsep))/34}}
\begin{sidewaystable}[ph!]
  \resizebox{\textwidth}{!}{%
  \def\arraystretch{2.4}
  \begin{tabular}{|l||C|C|C|C|C||C|C|C|C||C|C|C|C||C|C|C|C|C|C||C|C|C|C||C|C|C|C||C|C|C||C|C|C|C|}
    \hline
    \multicolumn{1}{|c||}{\parbox[t]{7cm}{\raggedright Variables}} & %
    \multicolumn{5}{c||}{\parbox[t]{4.2cm}{\raggedright \textbf{\ref{v:implementationApproach}}~~~\parbox{2cm}{\raggedright Implementation approach}}} & %
    \multicolumn{4}{c||}{\parbox[t]{3.3cm}{\raggedright \textbf{\ref{v:processingMechanism}}~~~\parbox{3cm}{\raggedright Processing mechanism}\\~\\~\\}} & %
    \multicolumn{4}{c||}{\parbox[t]{3.3cm}{\raggedright \textbf{\ref{v:adaptedComponent}}~~~\parbox{3cm}{\raggedright Adapted component}}} & %
    \multicolumn{6}{c||}{\parbox[t]{5.2cm}{\raggedright \textbf{\ref{v:adaptiveLearningMechanism}}~~~\parbox{5.2cm}{\raggedright Adaptive learning mechanism}}} & %
    \multicolumn{4}{c||}{\parbox[t]{3.3cm}{\raggedright \textbf{\ref{v:authoringCapability}}~~~\parbox{3.2cm}{\raggedright Authoring capability}}} & %
    \multicolumn{4}{c||}{\parbox[t]{3.3cm}{\raggedright \textbf{\ref{v:adaptiveLearningMechanismTrigger}}~~~\parbox{3cm}{\raggedright Adaptive learning mechanism trigger}}} & %
    \multicolumn{3}{c||}{\parbox[t]{2.35cm}{\raggedright \textbf{\ref{v:learningContent}}~~~\parbox{2cm}{\raggedright Learning content}}} & %
    \multicolumn{4}{c|}{\parbox[t]{3.4cm}{\raggedright \textbf{\ref{v:adaptationSourceData}}~~~\parbox{3cm}{\raggedright Adaptation\\source data}}} \\
    \hline
    \multicolumn{1}{|c||}{\parbox[t]{7cm}{\raggedright Explanations}} & %
    \multicolumn{5}{c||}{\parbox[t]{4.2cm}{\raggedright How are the ALMs technically integrated into LMS?\\~\\}} & %
    \multicolumn{4}{c||}{\parbox[t]{3.3cm}{\raggedright What data processing mechanisms are applied to implement the ALMs?}} & %
    \multicolumn{4}{c||}{\parbox[t]{3.3cm}{\raggedright Which components of the learning process are being adapted?}} & %
    \multicolumn{6}{c||}{\parbox[t]{5.2cm}{\raggedright Which ALMs are supported?}} & %
    \multicolumn{4}{c||}{\parbox[t]{3.3cm}{\raggedright At what level are users supported to author parts of the ALMs on their own?}} & %
    \multicolumn{4}{c||}{\parbox[t]{3.3cm}{\raggedright How are the ALMs triggered?}} & %
    \multicolumn{3}{c||}{\parbox[t]{2.35cm}{\raggedright What content do the ALMs operate on?}} & %
    \multicolumn{4}{c|}{\parbox[t]{3.4cm}{\raggedright What data is processed in order to provide the ALMs?}} \\
    \cmidrule{1-35} %
    \multicolumn{1}{|c||}{\parbox[t]{7cm}{\raggedright Expressions}} & %
    \rotatebox[origin=l]{270}{\textbf{\ref{v:implementationApproach:e:adapterPlugin}}~~~Adapter Plugin} & %
    \rotatebox[origin=l]{270}{\textbf{\ref{v:implementationApproach:e:adapterImplementation}}~~~Direct Adapter Implementation} & %
    \rotatebox[origin=l]{270}{\textbf{\ref{v:implementationApproach:e:lmsSpecificImplementation}}~~~Direct Implementation} & %
    \rotatebox[origin=l]{270}{\textbf{\ref{v:implementationApproach:e:lti}}~~~LTI} & %
    \rotatebox[origin=l]{270}{\textbf{\ref{v:implementationApproach:e:plugin}}~~~Plugin Implementation} & %
    \rotatebox[origin=l]{270}{\textbf{\ref{v:processingMechanism:e:calculationBased}}~~~Calculation based} & %
    \rotatebox[origin=l]{270}{\textbf{\ref{v:processingMechanism:e:executionBased}}~~~Execution based} & %
    \rotatebox[origin=l]{270}{\textbf{\ref{v:processingMechanism:e:optimizationBased}}~~~Optimization based} & %
    \rotatebox[origin=l]{270}{\textbf{\ref{v:processingMechanism:e:reasoningBased}}~~~Reasoning based} & %
    \rotatebox[origin=l]{270}{\textbf{\ref{v:adaptedComponent:e:interaction}}~~~Interaction} & %
    \rotatebox[origin=l]{270}{\textbf{\ref{v:adaptedComponent:e:interface}}~~~Interface} & %
    \rotatebox[origin=l]{270}{\textbf{\ref{v:adaptedComponent:e:learningContent}}~~~Learning content} & %
    \rotatebox[origin=l]{270}{\textbf{\ref{v:adaptedComponent:e:supportiveOrInstructionalOutput}}~~~Supportive or instructional output} & %
    \rotatebox[origin=l]{270}{\textbf{\ref{v:adaptiveLearningMechanism:e:contentAdaptation}}~~~Content adaptation} & %
    \rotatebox[origin=l]{270}{\textbf{\ref{v:adaptiveLearningMechanism:e:contentRecommendation}}~~~Content recommendation} & %
    \rotatebox[origin=l]{270}{\textbf{\ref{v:adaptiveLearningMechanism:e:contentSelection}}~~~Content selection} & %
    \rotatebox[origin=l]{270}{\textbf{\ref{v:adaptiveLearningMechanism:e:feedbackProvision}}~~~Feedback provision} & %
    \rotatebox[origin=l]{270}{\textbf{\ref{v:adaptiveLearningMechanism:e:promptProvision}}~~~Prompt provision} & %
    \rotatebox[origin=l]{270}{\textbf{\ref{v:adaptiveLearningMechanism:e:scaffoldingSupport}}~~~Scaffolding support} & %
    \rotatebox[origin=l]{270}{\textbf{\ref{v:authoringCapability:e:authoringOfAdaptiveLearningMeans}}~~~\parbox{4.5cm}{Authoring of adaptive learning means}} & %
    \rotatebox[origin=l]{270}{\textbf{\ref{v:authoringCapability:e:authoringOfAdaptiveLearningMechanisms}}~~~\parbox{4.5cm}{Authoring of ALMs}} & %
    \rotatebox[origin=l]{270}{\textbf{\ref{v:authoringCapability:e:authoringOfAdaptiveLearningMechanismParameters}}~~~\parbox{5cm}{Authoring of adaptive learning\\mechanism parameters~~}} & %
    \rotatebox[origin=l]{270}{\textbf{\ref{v:authoringCapability:e:noAuthoring}}~~~No authoring} & %
    \rotatebox[origin=l]{270}{\textbf{\ref{v:adaptiveLearningMechanismTrigger:e:assessmentAnswering}}~~~Assessment answering} & %
    \rotatebox[origin=l]{270}{\textbf{\ref{v:adaptiveLearningMechanismTrigger:e:learningSessionBegin}}~~~Learning session begin} & %
    \rotatebox[origin=l]{270}{\textbf{\ref{v:adaptiveLearningMechanismTrigger:e:navigation}}~~~Navigation} & %
    \rotatebox[origin=l]{270}{\textbf{\ref{v:adaptiveLearningMechanismTrigger:e:proactiveInitiation}}~~~Proactive initiation} & %
    \rotatebox[origin=l]{270}{\textbf{\ref{v:learningContent:e:nativeLmsContent}}~~~Native LMS content} & %
    \rotatebox[origin=l]{270}{\textbf{\ref{v:learningContent:e:nonLmsContent}}~~~Non-LMS content} & %
    \rotatebox[origin=l]{270}{\textbf{\ref{v:learningContent:e:particularLmsContent}}~~~Particular LMS content} & %
    \rotatebox[origin=l]{270}{\textbf{\ref{v:adaptationSourceData:e:customInteractionData}}~~~Custom interaction data} & %
    \rotatebox[origin=l]{270}{\textbf{\ref{v:adaptationSourceData:e:derivedLearnerData}}~~~Derived learner data} & %
    \rotatebox[origin=l]{270}{\textbf{\ref{v:adaptationSourceData:e:existingInteractionData}}~~~Existing interaction data} & %
    \rotatebox[origin=l]{270}{\textbf{\ref{v:adaptationSourceData:e:requestedLearnerData}}~~~Requested learner data} \\
    \hline
    \hline
    \rowcolor{gray!50} %
     \ccitetext{giuffra_palomino_intelligent_2014} %
      & %
      & \x %
      & %
      & %
      & %
      & \x %
      & \x %
      & %
      & %
      & %
      & %
      & \x %
      & \x %
      & %
      & %
      & \x %
      & \x %
      & \x %
      & %
      & \x %
      & %
      & \x %
      & %
      & \x %
      & %
      & \x %
      & %
      & \x %
      & %
      & %
      & \x %
      & \x %
      & %
      & %
      \\
      \hline
     \ccitetext{pratas2014adapt} %
      & %
      & %
      & \x %
      & %
      & %
      & \x %
      & %
      & \x %
      & \x %
      & %
      & %
      & \x %
      & %
      & %
      & \x %
      & \x %
      & %
      & %
      & %
      & \x %
      & %
      & %
      & %
      & \x %
      & %
      & %
      & \x %
      & \x %
      & %
      & %
      & \x %
      & \x %
      & %
      & \x %
      \\
      \hline
      \rowcolor{gray!50} %
     \ccitetext{radwan2014adaptive} %
      & %
      & %
      & \x %
      & %
      & %
      & \x %
      & %
      & %
      & %
      & %
      & %
      & \x %
      & %
      & %
      & %
      & \x %
      & %
      & %
      & %
      & \x %
      & %
      & %
      & %
      & \x %
      & %
      & %
      & %
      & \x %
      & %
      & %
      & %
      & \x %
      & %
      & \x %
      \\
      \hline
     \ccitetext{surjono2014evaluation} %
      & %
      & %
      & \x %
      & %
      & %
      & %
      & %
      & %
      & \x %
      & %
      & %
      & \x %
      & %
      & %
      & %
      & \x %
      & %
      & %
      & %
      & \x %
      & %
      & \x %
      & %
      & \x %
      & %
      & %
      & %
      & \x %
      & %
      & %
      & %
      & \x %
      & %
      & \x %
      \\
      \hline
      \rowcolor{gray!50} %
     \ccitetext{munoz2015ontosakai} %
      & %
      & %
      & \x %
      & %
      & %
      & %
      & %
      & %
      & \x %
      & %
      & %
      & \x %
      & \x %
      & %
      & \x %
      & %
      & %
      & \x %
      & %
      & \x %
      & %
      & %
      & %
      & %
      & %
      & \x %
      & %
      & \x %
      & %
      & %
      & %
      & \x %
      & \x %
      & %
      \\
      \hline
     \ccitetext{nafea2015novel} %
      & %
      & %
      & \x %
      & %
      & %
      & %
      & %
      & %
      & \x %
      & %
      & %
      & \x %
      & %
      & %
      & %
      & \x %
      & %
      & %
      & %
      & \x %
      & %
      & \x %
      & %
      & %
      & \x %
      & \x %
      & %
      & \x %
      & %
      & %
      & \x %
      & \x %
      & %
      & \x %
      \\
      \hline
      \rowcolor{gray!50} %
     \ccitetext{ovalle2015proposal} %
      & %
      & %
      & \x %
      & %
      & %
      & %
      & \x %
      & %
      & \x %
      & \x %
      & %
      & \x %
      & \x %
      & %
      & \x %
      & %
      & \x %
      & \x %
      & \x %
      & \x %
      & %
      & %
      & %
      & %
      & %
      & \x %
      & %
      & \x %
      & %
      & %
      & %
      & \x %
      & \x %
      & %
      \\
      \hline
     \ccitetext{sein2015design} %
      & %
      & %
      & \x %
      & %
      & %
      & %
      & %
      & %
      & \x %
      & %
      & %
      & \x %
      & \x %
      & %
      & %
      & \x %
      & \x %
      & %
      & %
      & \x %
      & %
      & \x %
      & %
      & \x %
      & %
      & \x %
      & %
      & \x %
      & %
      & %
      & %
      & \x %
      & \x %
      & \x %
      \\
      \hline
      \rowcolor{gray!50} %
     \ccitetext{sweta2015adaptive} %
      & %
      & %
      & \x %
      & %
      & %
      & \x %
      & %
      & %
      & %
      & %
      & %
      & \x %
      & %
      & %
      & \x %
      & %
      & %
      & %
      & %
      & %
      & %
      & %
      & \x %
      & \x %
      & %
      & \x %
      & %
      & \x %
      & %
      & %
      & \x %
      & \x %
      & %
      & \x %
      \\
      \hline
     \ccitetext{aleven_embedding_2016} %
      & %
      & \x %
      & %
      & \x %
      & %
      & %
      & \x %
      & %
      & %
      & \x %
      & \x %
      & %
      & \x %
      & %
      & %
      & %
      & \x %
      & \x %
      & \x %
      & \x %
      & %
      & \x %
      & %
      & \x %
      & %
      & \x %
      & \x %
      & %
      & \x %
      & %
      & \x %
      & \x %
      & %
      & %
      \\
      \hline
      \rowcolor{gray!50} %
     \ccitetext{de_amorim_towards_2016} %
      & \x %
      & %
      & %
      & %
      & %
      & %
      & %
      & %
      & \x %
      & %
      & %
      & \x %
      & %
      & %
      & \x %
      & %
      & %
      & %
      & %
      & \x %
      & %
      & %
      & %
      & \x %
      & %
      & \x %
      & %
      & %
      & %
      & \x %
      & %
      & \x %
      & \x %
      & %
      \\
      \hline
     \ccitetext{luccioni_sti-dico_2016} %
      & %
      & %
      & %
      & \x %
      & %
      & %
      & \x %
      & %
      & \x %
      & \x %
      & \x %
      & \x %
      & \x %
      & %
      & %
      & \x %
      & \x %
      & \x %
      & \x %
      & %
      & %
      & %
      & \x %
      & \x %
      & %
      & \x %
      & %
      & %
      & \x %
      & %
      & \x %
      & \x %
      & %
      & %
      \\
      \hline
      \rowcolor{gray!50} %
     \ccitetext{rani2016ontological} %
      & %
      & %
      & \x %
      & %
      & %
      & %
      & %
      & %
      & \x %
      & %
      & %
      & \x %
      & \x %
      & %
      & \x %
      & \x %
      & \x %
      & %
      & %
      & \x %
      & %
      & %
      & %
      & \x %
      & \x %
      & %
      & %
      & \x %
      & %
      & %
      & \x %
      & \x %
      & %
      & \x %
      \\
      \hline
     \ccitetext{aleven2017integrating} %
      & %
      & %
      & %
      & \x %
      & %
      & %
      & \x %
      & %
      & \x %
      & \x %
      & \x %
      & \x %
      & \x %
      & %
      & %
      & \x %
      & \x %
      & \x %
      & \x %
      & \x %
      & \x %
      & \x %
      & %
      & \x %
      & \x %
      & \x %
      & \x %
      & %
      & \x %
      & %
      & \x %
      & \x %
      & %
      & \x %
      \\
      \hline
      \rowcolor{gray!50} %
     \ccitetext{lagman2017extracting} %
      & %
      & %
      & \x %
      & %
      & %
      & %
      & %
      & %
      & \x %
      & %
      & %
      & \x %
      & %
      & %
      & %
      & \x %
      & %
      & %
      & %
      & \x %
      & %
      & %
      & %
      & \x %
      & %
      & %
      & %
      & \x %
      & %
      & %
      & %
      & \x %
      & \x %
      & %
      \\
      \hline
     \ccitetext{oskouei2017proposing} %
      & %
      & %
      & \x %
      & %
      & %
      & \x %
      & %
      & \x %
      & %
      & %
      & %
      & \x %
      & %
      & %
      & \x %
      & \x %
      & %
      & %
      & %
      & %
      & %
      & %
      & \x %
      & %
      & \x %
      & %
      & %
      & \x %
      & %
      & %
      & %
      & \x %
      & \x %
      & \x %
      \\
      \hline
      \rowcolor{gray!50} %
     \ccitetext{verdu_integration_2017} %
      & %
      & \x %
      & %
      & %
      & %
      & %
      & %
      & %
      & \x %
      & %
      & %
      & \x %
      & \x %
      & %
      & \x %
      & %
      & \x %
      & \x %
      & %
      & \x %
      & %
      & \x %
      & %
      & %
      & %
      & \x %
      & %
      & \x %
      & %
      & %
      & \x %
      & \x %
      & %
      & \x %
      \\
      \hline
     \ccitetext{cai2018adaptive} %
      & %
      & %
      & \x %
      & %
      & %
      & %
      & %
      & %
      & \x %
      & %
      & %
      & \x %
      & %
      & %
      & %
      & \x %
      & %
      & %
      & %
      & \x %
      & %
      & \x %
      & %
      & \x %
      & %
      & %
      & %
      & \x %
      & %
      & %
      & %
      & \x %
      & \x %
      & %
      \\
      \hline
      \rowcolor{gray!50} %
     \ccitetext{karagiannis2018adaptive} %
      & %
      & %
      & \x %
      & %
      & %
      & \x %
      & %
      & \x %
      & \x %
      & %
      & %
      & \x %
      & %
      & %
      & \x %
      & %
      & %
      & %
      & %
      & \x %
      & %
      & \x %
      & %
      & %
      & \x %
      & %
      & %
      & \x %
      & %
      & %
      & %
      & \x %
      & \x %
      & \x %
      \\
      \hline
     \ccitetext{perivsic2018semantic} %
      & %
      & %
      & \x %
      & %
      & %
      & %
      & %
      & %
      & \x %
      & %
      & %
      & \x %
      & %
      & %
      & \x %
      & \x %
      & %
      & %
      & %
      & \x %
      & %
      & %
      & %
      & \x %
      & \x %
      & %
      & %
      & \x %
      & \x %
      & %
      & %
      & \x %
      & \x %
      & \x %
      \\
      \hline
    \end{tabular}
  }
  \caption{Summary of the results of the structured scoping review corresponding to the extracted data charting variables \ref{v:implementationApproach} to \ref{v:adaptationSourceData} and their expressions. The table lists approach 22 of the identified approaches to approach 41. The approaches are ordered by the year of the extracted paper, with the most recent approach coming last. The variable expressions are in alphabetical order. \x indicates that the value of a variable matches the corresponding expression.}
  \label{tab:scopingReviewResultsPart2}
\end{sidewaystable}
} %
{
\newcolumntype{C}{>{\centering \arraybackslash}m{(\textwidth-(6\arrayrulewidth)-(10\tabcolsep))/34}}
\begin{sidewaystable}[ph!]
  \resizebox{\textwidth}{!}{%
  \def\arraystretch{2.4}
  \begin{tabular}{|l||C|C|C|C|C||C|C|C|C||C|C|C|C||C|C|C|C|C|C||C|C|C|C||C|C|C|C||C|C|C||C|C|C|C|}
    \hline
    \multicolumn{1}{|c||}{\parbox[t]{7cm}{\raggedright Variables}} & %
    \multicolumn{5}{c||}{\parbox[t]{4.2cm}{\raggedright \textbf{\ref{v:implementationApproach}}~~~\parbox{2cm}{\raggedright Implementation approach}}} & %
    \multicolumn{4}{c||}{\parbox[t]{3.3cm}{\raggedright \textbf{\ref{v:processingMechanism}}~~~\parbox{3cm}{\raggedright Processing mechanism}\\~\\~\\}} & %
    \multicolumn{4}{c||}{\parbox[t]{3.3cm}{\raggedright \textbf{\ref{v:adaptedComponent}}~~~\parbox{3cm}{\raggedright Adapted component}}} & %
    \multicolumn{6}{c||}{\parbox[t]{5.2cm}{\raggedright \textbf{\ref{v:adaptiveLearningMechanism}}~~~\parbox{5.2cm}{\raggedright Adaptive learning mechanism}}} & %
    \multicolumn{4}{c||}{\parbox[t]{3.3cm}{\raggedright \textbf{\ref{v:authoringCapability}}~~~\parbox{3.2cm}{\raggedright Authoring capability}}} & %
    \multicolumn{4}{c||}{\parbox[t]{3.3cm}{\raggedright \textbf{\ref{v:adaptiveLearningMechanismTrigger}}~~~\parbox{3cm}{\raggedright Adaptive learning mechanism trigger}}} & %
    \multicolumn{3}{c||}{\parbox[t]{2.35cm}{\raggedright \textbf{\ref{v:learningContent}}~~~\parbox{2cm}{\raggedright Learning content}}} & %
    \multicolumn{4}{c|}{\parbox[t]{3.4cm}{\raggedright \textbf{\ref{v:adaptationSourceData}}~~~\parbox{3cm}{\raggedright Adaptation\\source data}}} \\
    \hline
    \multicolumn{1}{|c||}{\parbox[t]{7cm}{\raggedright Explanations}} & %
    \multicolumn{5}{c||}{\parbox[t]{4.2cm}{\raggedright How are the ALMs technically integrated into LMS?\\~\\}} & %
    \multicolumn{4}{c||}{\parbox[t]{3.3cm}{\raggedright What data processing mechanisms are applied to implement the ALMs?}} & %
    \multicolumn{4}{c||}{\parbox[t]{3.3cm}{\raggedright Which components of the learning process are being adapted?}} & %
    \multicolumn{6}{c||}{\parbox[t]{5.2cm}{\raggedright Which ALMs are supported?}} & %
    \multicolumn{4}{c||}{\parbox[t]{3.3cm}{\raggedright At what level are users supported to author parts of the ALMs on their own?}} & %
    \multicolumn{4}{c||}{\parbox[t]{3.3cm}{\raggedright How are the ALMs triggered?}} & %
    \multicolumn{3}{c||}{\parbox[t]{2.35cm}{\raggedright What content do the ALMs operate on?}} & %
    \multicolumn{4}{c|}{\parbox[t]{3.4cm}{\raggedright What data is processed in order to provide the ALMs?}} \\
    \cmidrule{1-35} %
    \multicolumn{1}{|c||}{\parbox[t]{7cm}{\raggedright Expressions}} & %
    \rotatebox[origin=l]{270}{\textbf{\ref{v:implementationApproach:e:adapterPlugin}}~~~Adapter Plugin} & %
    \rotatebox[origin=l]{270}{\textbf{\ref{v:implementationApproach:e:adapterImplementation}}~~~Direct Adapter Implementation} & %
    \rotatebox[origin=l]{270}{\textbf{\ref{v:implementationApproach:e:lmsSpecificImplementation}}~~~Direct Implementation} & %
    \rotatebox[origin=l]{270}{\textbf{\ref{v:implementationApproach:e:lti}}~~~LTI} & %
    \rotatebox[origin=l]{270}{\textbf{\ref{v:implementationApproach:e:plugin}}~~~Plugin Implementation} & %
    \rotatebox[origin=l]{270}{\textbf{\ref{v:processingMechanism:e:calculationBased}}~~~Calculation based} & %
    \rotatebox[origin=l]{270}{\textbf{\ref{v:processingMechanism:e:executionBased}}~~~Execution based} & %
    \rotatebox[origin=l]{270}{\textbf{\ref{v:processingMechanism:e:optimizationBased}}~~~Optimization based} & %
    \rotatebox[origin=l]{270}{\textbf{\ref{v:processingMechanism:e:reasoningBased}}~~~Reasoning based} & %
    \rotatebox[origin=l]{270}{\textbf{\ref{v:adaptedComponent:e:interaction}}~~~Interaction} & %
    \rotatebox[origin=l]{270}{\textbf{\ref{v:adaptedComponent:e:interface}}~~~Interface} & %
    \rotatebox[origin=l]{270}{\textbf{\ref{v:adaptedComponent:e:learningContent}}~~~Learning content} & %
    \rotatebox[origin=l]{270}{\textbf{\ref{v:adaptedComponent:e:supportiveOrInstructionalOutput}}~~~Supportive or instructional output} & %
    \rotatebox[origin=l]{270}{\textbf{\ref{v:adaptiveLearningMechanism:e:contentAdaptation}}~~~Content adaptation} & %
    \rotatebox[origin=l]{270}{\textbf{\ref{v:adaptiveLearningMechanism:e:contentRecommendation}}~~~Content recommendation} & %
    \rotatebox[origin=l]{270}{\textbf{\ref{v:adaptiveLearningMechanism:e:contentSelection}}~~~Content selection} & %
    \rotatebox[origin=l]{270}{\textbf{\ref{v:adaptiveLearningMechanism:e:feedbackProvision}}~~~Feedback provision} & %
    \rotatebox[origin=l]{270}{\textbf{\ref{v:adaptiveLearningMechanism:e:promptProvision}}~~~Prompt provision} & %
    \rotatebox[origin=l]{270}{\textbf{\ref{v:adaptiveLearningMechanism:e:scaffoldingSupport}}~~~Scaffolding support} & %
    \rotatebox[origin=l]{270}{\textbf{\ref{v:authoringCapability:e:authoringOfAdaptiveLearningMeans}}~~~\parbox{4.5cm}{Authoring of adaptive learning means}} & %
    \rotatebox[origin=l]{270}{\textbf{\ref{v:authoringCapability:e:authoringOfAdaptiveLearningMechanisms}}~~~\parbox{4.5cm}{Authoring of ALMs}} & %
    \rotatebox[origin=l]{270}{\textbf{\ref{v:authoringCapability:e:authoringOfAdaptiveLearningMechanismParameters}}~~~\parbox{5cm}{Authoring of adaptive learning\\mechanism parameters~~}} & %
    \rotatebox[origin=l]{270}{\textbf{\ref{v:authoringCapability:e:noAuthoring}}~~~No authoring} & %
    \rotatebox[origin=l]{270}{\textbf{\ref{v:adaptiveLearningMechanismTrigger:e:assessmentAnswering}}~~~Assessment answering} & %
    \rotatebox[origin=l]{270}{\textbf{\ref{v:adaptiveLearningMechanismTrigger:e:learningSessionBegin}}~~~Learning session begin} & %
    \rotatebox[origin=l]{270}{\textbf{\ref{v:adaptiveLearningMechanismTrigger:e:navigation}}~~~Navigation} & %
    \rotatebox[origin=l]{270}{\textbf{\ref{v:adaptiveLearningMechanismTrigger:e:proactiveInitiation}}~~~Proactive initiation} & %
    \rotatebox[origin=l]{270}{\textbf{\ref{v:learningContent:e:nativeLmsContent}}~~~Native LMS content} & %
    \rotatebox[origin=l]{270}{\textbf{\ref{v:learningContent:e:nonLmsContent}}~~~Non-LMS content} & %
    \rotatebox[origin=l]{270}{\textbf{\ref{v:learningContent:e:particularLmsContent}}~~~Particular LMS content} & %
    \rotatebox[origin=l]{270}{\textbf{\ref{v:adaptationSourceData:e:customInteractionData}}~~~Custom interaction data} & %
    \rotatebox[origin=l]{270}{\textbf{\ref{v:adaptationSourceData:e:derivedLearnerData}}~~~Derived learner data} & %
    \rotatebox[origin=l]{270}{\textbf{\ref{v:adaptationSourceData:e:existingInteractionData}}~~~Existing interaction data} & %
    \rotatebox[origin=l]{270}{\textbf{\ref{v:adaptationSourceData:e:requestedLearnerData}}~~~Requested learner data} \\
    \hline
    \hline
    \rowcolor{gray!50} %
     \ccitetext{alsobhi2019adaptation} %
      & %
      & %
      & \x %
      & %
      & %
      & %
      & %
      & %
      & \x %
      & %
      & %
      & \x %
      & %
      & %
      & \x %
      & \x %
      & %
      & %
      & %
      & \x %
      & %
      & %
      & %
      & \x %
      & %
      & %
      & %
      & \x %
      & %
      & %
      & %
      & \x %
      & %
      & \x %
      \\
      \hline
     \ccitetext{joseph2019adaptive} %
      & %
      & %
      & \x %
      & %
      & %
      & %
      & \x %
      & %
      & %
      & %
      & %
      & \x %
      & %
      & %
      & %
      & \x %
      & %
      & %
      & %
      & \x %
      & %
      & \x %
      & %
      & \x %
      & %
      & %
      & %
      & \x %
      & %
      & %
      & %
      & \x %
      & \x %
      & %
      \\
      \hline
      \rowcolor{gray!50} %
     \ccitetext{joy2019ontology} %
      & %
      & %
      & \x %
      & %
      & %
      & %
      & %
      & %
      & \x %
      & %
      & %
      & \x %
      & %
      & %
      & \x %
      & %
      & %
      & %
      & %
      & \x %
      & %
      & %
      & %
      & \x %
      & %
      & %
      & %
      & \x %
      & %
      & %
      & %
      & \x %
      & \x %
      & \x %
      \\
      \hline
     \ccitetext{khosravi2019ripple} %
      & %
      & %
      & %
      & \x %
      & %
      & \x %
      & %
      & %
      & %
      & %
      & %
      & \x %
      & \x %
      & %
      & \x %
      & %
      & %
      & \x %
      & %
      & \x %
      & %
      & %
      & %
      & %
      & %
      & %
      & \x %
      & %
      & \x %
      & %
      & \x %
      & \x %
      & %
      & %
      \\
      \hline
      \rowcolor{gray!50} %
     \ccitetext{troussas2019injecting} %
      & %
      & %
      & \x %
      & %
      & %
      & %
      & %
      & %
      & \x %
      & %
      & %
      & \x %
      & \x %
      & \x %
      & %
      & \x %
      & \x %
      & %
      & %
      & %
      & %
      & %
      & \x %
      & \x %
      & %
      & %
      & %
      & \x %
      & %
      & %
      & \x %
      & \x %
      & %
      & %
      \\
      \hline
     \ccitetext{arsovic_e-learning_2020} %
      & \x %
      & %
      & %
      & %
      & %
      & %
      & \x %
      & %
      & %
      & %
      & %
      & \x %
      & %
      & %
      & %
      & \x %
      & %
      & %
      & %
      & \x %
      & %
      & %
      & %
      & %
      & \x %
      & %
      & %
      & %
      & %
      & \x %
      & \x %
      & \x %
      & %
      & \x %
      \\
      \hline
      \rowcolor{gray!50} %
     \ccitetext{louhab2020novel} %
      & %
      & %
      & %
      & %
      & \x %
      & %
      & \x %
      & %
      & %
      & %
      & %
      & \x %
      & %
      & %
      & %
      & \x %
      & %
      & %
      & %
      & \x %
      & %
      & \x %
      & %
      & \x %
      & %
      & %
      & %
      & \x %
      & %
      & %
      & %
      & \x %
      & \x %
      & %
      \\
      \hline
     \ccitetext{sayed2020towards} %
      & %
      & %
      & \x %
      & %
      & %
      & \x %
      & %
      & \x %
      & %
      & %
      & %
      & \x %
      & %
      & %
      & \x %
      & %
      & %
      & %
      & %
      & %
      & %
      & %
      & \x %
      & %
      & %
      & \x %
      & %
      & \x %
      & %
      & %
      & \x %
      & \x %
      & %
      & %
      \\
      \hline
      \rowcolor{gray!50} %
     \ccitetext{zagorskis_eci_2020} %
      & %
      & \x %
      & %
      & %
      & %
      & %
      & \x %
      & \x %
      & %
      & %
      & %
      & \x %
      & \x %
      & \x %
      & %
      & %
      & %
      & \x %
      & %
      & %
      & %
      & %
      & \x %
      & \x %
      & \x %
      & \x %
      & %
      & %
      & \x %
      & %
      & %
      & \x %
      & \x %
      & \x %
      \\
      \hline
     \ccitetext{fakoya2021automatic} %
      & %
      & %
      & \x %
      & %
      & %
      & \x %
      & %
      & %
      & %
      & %
      & %
      & \x %
      & %
      & \x %
      & %
      & \x %
      & %
      & %
      & %
      & %
      & %
      & %
      & \x %
      & %
      & \x %
      & %
      & %
      & \x %
      & %
      & %
      & %
      & \x %
      & \x %
      & %
      \\
      \hline
      \rowcolor{gray!50} %
     \ccitetext{pagano2021training} %
      & %
      & %
      & %
      & %
      & \x %
      & %
      & %
      & %
      & \x %
      & %
      & %
      & \x %
      & %
      & %
      & %
      & \x %
      & %
      & %
      & %
      & \x %
      & %
      & \x %
      & %
      & \x %
      & %
      & %
      & %
      & \x %
      & %
      & %
      & %
      & \x %
      & \x %
      & %
      \\
      \hline
     \ccitetext{sychev_compprehension_2021} %
      & %
      & %
      & %
      & \x %
      & %
      & %
      & %
      & %
      & \x %
      & \x %
      & %
      & \x %
      & \x %
      & %
      & %
      & \x %
      & \x %
      & %
      & \x %
      & %
      & %
      & %
      & \x %
      & \x %
      & %
      & %
      & %
      & %
      & \x %
      & %
      & \x %
      & \x %
      & %
      & %
      \\
      \hline
      \rowcolor{gray!50} %
     \ccitetext{apoki_modular_2022} %
      & \x %
      & %
      & %
      & %
      & %
      & %
      & %
      & %
      & \x %
      & %
      & %
      & \x %
      & %
      & %
      & %
      & \x %
      & %
      & %
      & %
      & \x %
      & %
      & \x %
      & %
      & %
      & \x %
      & \x %
      & %
      & \x %
      & %
      & %
      & \x %
      & \x %
      & %
      & \x %
      \\
      \hline
     \ccitetext{bradavc2022design} %
      & %
      & %
      & \x %
      & %
      & %
      & %
      & \x %
      & %
      & \x %
      & %
      & %
      & \x %
      & %
      & %
      & %
      & \x %
      & %
      & %
      & %
      & \x %
      & %
      & \x %
      & %
      & \x %
      & \x %
      & %
      & %
      & \x %
      & %
      & %
      & %
      & \x %
      & \x %
      & \x %
      \\
      \hline
      \rowcolor{gray!50} %
     \ccitetext{cardenas2022personalised} %
      & %
      & %
      & \x %
      & %
      & %
      & %
      & %
      & %
      & \x %
      & %
      & %
      & \x %
      & \x %
      & %
      & %
      & \x %
      & \x %
      & %
      & %
      & \x %
      & %
      & %
      & %
      & \x %
      & %
      & %
      & %
      & \x %
      & \x %
      & %
      & %
      & \x %
      & %
      & \x %
      \\
      \hline
     \ccitetext{yilmaz2022smart} %
      & %
      & %
      & \x %
      & %
      & %
      & \x %
      & %
      & \x %
      & %
      & %
      & %
      & \x %
      & \x %
      & %
      & %
      & \x %
      & \x %
      & %
      & \x %
      & %
      & %
      & %
      & \x %
      & \x %
      & \x %
      & \x %
      & %
      & \x %
      & %
      & %
      & \x %
      & \x %
      & %
      & %
      \\
      \hline
      \rowcolor{gray!50} %
     \ccitetext{kaouni2023design} %
      & %
      & %
      &  \x %
      & %
      & %
      & %
      & %
      & \x %
      & %
      & %
      & \x %
      & \x %
      & %
      & \x %
      & \x %
      & %
      & %
      & %
      & %
      & \x %
      & %
      & %
      & %
      & \x %
      & %
      & \x %
      & %
      & \x %
      & %
      & %
      & \x %
      & \x %
      & %
      & \x %
      \\
      \hline
     \ccitetext{kouvara2023discovering} %
      & %
      & %
      & \x %
      & %
      & %
      & \x %
      & %
      & %
      & %
      & %
      & %
      & \x %
      & %
      & %
      & \x %
      & %
      & %
      & %
      & %
      & \x %
      & %
      & %
      & %
      & \x %
      & %
      & %
      & %
      & \x %
      & %
      & %
      & %
      & \x %
      & %
      & \x %
      \\
      \hline
      \rowcolor{gray!50} %
     \ccitetext{pardos2023conducting} %
      & %
      & %
      & %
      & \x %
      & %
      & \x %
      & %
      & %
      & %
      & %
      & %
      & \x %
      & \x %
      & %
      & %
      & \x %
      & \x %
      & \x %
      & \x %
      & \x %
      & %
      & \x %
      & %
      & \x %
      & \x %
      & %
      & \x %
      & %
      & \x %
      & %
      & \x %
      & \x %
      & %
      & %
      \\
      \hline
     \ccitetext{senthil2023towards} %
      & %
      & %
      & \x %
      & %
      & %
      & \x %
      & %
      & \x %
      & %
      & %
      & %
      & \x %
      & %
      & %
      & \x %
      & %
      & %
      & %
      & %
      & %
      & %
      & %
      & \x %
      & %
      & \x %
      & \x %
      & %
      & \x %
      & %
      & %
      & %
      & \x %
      & \x %
      & \x %
      \\
      \hline
    \end{tabular}
  }
  \caption{Summary of the results of the structured scoping review corresponding to the extracted data charting variables \ref{v:implementationApproach} to \ref{v:adaptationSourceData} and their expressions. The table lists approach 42 of the identified approaches to approach 61. The approaches are ordered by the year of the extracted paper, with the most recent approach coming last. The variable expressions are in alphabetical order. \x indicates that the value of a variable matches the corresponding expression.}
  \label{tab:scopingReviewResultsPart3}
\end{sidewaystable}
} 
\subsection{Implementation of Adaptive Learning Mechanisms}
\label{sec:adaptiveLearningMechanismImplementationApproaches}

We extracted \ref{v:implementationApproach} and \ref{v:processingMechanism} to answer \ref{rq:implementation}.
\Cref{tab:scopingReviewResultsPart1}, \Cref{tab:scopingReviewResultsPart2}, and \Cref{tab:scopingReviewResultsPart3} show the results.

\subsubsection{Assessment}

\paragraph{Implementation approach}
With regard to the LMS-specific approaches, 92.68\% are directly implemented as an extension of an existing LMS or in the range of the development of a novel LMS (\ie\ref{v:implementationApproach:e:lmsSpecificImplementation}).
With regard to the system-independent approaches, as well most approaches (60\%) implement functionality to communicate with system-independent services that provide the ALMs, directly in corresponding LMSs (\ie\ref{v:implementationApproach:e:adapterImplementation}).
Examples are the INTUITEL backend\cciteparen{verdu_integration_2017} or TELECI\cciteparen{zagorskis_eci_2020}.
A direct implementation offers the widest range of intervention possibilities, as all data processed by the LMSs can be accessed.
This may not be possible if the functionality is implemented as a plugin (\ie\ref{v:implementationApproach:e:plugin} or \ie\ref{v:implementationApproach:e:adapterPlugin}), for example.
However, even with system-independent approaches, the functionality must be implemented in the LMS in a system-dependent manner and delivered with the system itself.
This makes integration with existing LMSs more difficult.
When implemented as a plugin that can be deployed in different LMSs, ALMs can be easily deployed in existing architectures.
However, the LMS data can still be accessed if the plugin architecture of the LMS allows it.
Another advantage of a plugin implementation is that although it is certainly LMS-specific, the code that implements these functionalities is well separated and can be easily identified and partially reused for plugin implementations for other LMSs.
This facilitates the integration of system-independent approaches in additional LMSs and increases the chance that LMS-specific approaches can later be transferred to system-independent approaches.
A plugin implementation is applied only by 15\% of the system-independent approaches evaluted (\ie\ref{v:implementationApproach:e:adapterPlugin}) and 7.32\% of the LMS-specific approaches evaluated (\ie\ref{v:implementationApproach:e:plugin}).

The last option is to use the LTI standard (\ie\ref{v:implementationApproach:e:lti}).
This option provides the greatest degree of reusability.
It is system-independent by design.
The disadvantage of this option is that it limits the LMS data that can be accessed by the ALMs.
This option is implemented by 30\% of the system-independent approaches evaluated.

\paragraph{Applied Data Processing Mechanisms}
Our review supports the hypothesis that most system-independent approaches apply only a limited number of data processing mechanisms (\ie\ref{h:insufficientProcessingMechanismsSupport}).
Furthermore, this hypothesis can be seen to hold for the LMS-specific approaches as well.
There is only one LMS-specific approach that provides all types of processing mechanisms using the Python programming language for specification\cciteparen{roselli2008integration}.
63.41\% of the LMS-specific approaches support only a single processing mechanism. 
With respect to system-independent approaches, there is no approach that implements all types of processing mechanisms.
At most two types are implemented, for example through\ccitetext{giuffra_palomino_intelligent_2014}.
The most implemented kind of processing mechanisms are reasoning-based (\ie\ref{v:processingMechanism:e:reasoningBased}).
This is the case for both system-independent approaches (45\%) and LMS-specific approaches (73.17\%).
While execution-based processing mechanisms are the second most common for the system-independent approaches (40\%), they are the least common for the LMS-specific approaches (14\%).
The second most common processing mechanisms of LMS-specific approaches are computation-based (31.71\%), followed by optimization-based (26.83\%) processing mechanisms.
A wide range of didactically proven ALMs require no more types of processing mechanisms than reasoning-based and execution-based.
Examples are the selection of learning content based on the learner's learning style\cciteeg{apoki_modular_2022} or the provision of feedback for specific learner errors\cciteeg{aleven2016example}.
However, certain ALMs cannot be implemented using processing mechanisms of this type, such as the estimation of mastery based on mathematical models, as used by\ccitetext{pardos2023conducting}.

\subsubsection{Discussion}

\paragraph{Trade-offs Between Portability and Didactic Depth of ALMs}
While direct implementation of most LMS-specific approaches is reasonable since they do not require system independence, we also found that most system-independent approaches (60\%) directly implement adapters.
The reason for this may be the absence of standards for describing two aspects related to adaptation system-independently.
First, there is a need to describe learning process data that can be processed by ALMs.
Second, it is necessary to describe how ALMs can be applied.
To overcome this obstacle, most reviewed approaches directly implement adapter code to abstract the underlying learning environment.
As a result, they specify the function of ALMs system-independently, while the complete implementation of these approaches remains system-dependent in terms of data access and adaptation.
However, coupling approaches with LMSs through data dependencies rather than code or function dependencies contradicts the original goal of providing system-independent ALMs.

One approach to implementing fully system-independent ALMs is to use the LTI specification.
Furthermore, integrating an LTI tool to adapt the learning process rather than letting the adapter implementation decide how to apply the adaptation, has the advantage of delivering learning process support equally in all associated learning environments.
However, analyzing the learning content reveals that none of the LTI-based approaches work with native LMS content.
Instead, these approaches integrate external learning content together with ALMs into the LMSs.
Furthermore, most of these approaches implement only single execution or trigger-based processing mechanisms.
This restricts the types of ALMs that can be implemented to support the learning process.
One reason for this may be that reducing the functional range is necessary when interacting with the learning environments according to a system-independent specification.
This results in a trade-off between portability and pedagogical depth.
This is also why there is no system-independent approach using LTI that implements optimization-based processing mechanisms.
However, in particular the lack of support for this impedes extensive support for self-regulated learning.
This is because the cyclic self-regulated learning process requires optimizing learning products based on learning standards.
In summary, the LTI-based approaches enable the reuse of didactic concepts across LMSs, preventing the need for re-authoring while providing outstanding, self-contained learning units.
These learning units restrict the capabilities of the associated ALMs to support comprehensive didactic scenarios focusing on the entire learning process, such as self-regulated learning.
Therefore, integrating system-independent, fully featured ALMs into LMSs remains a challenge~(\cf\Cref{sec:implications}).
In the absence of the described necessary standards, two points are advisable for LMS developers.
First, they should adopt established standards, such as LTI, to allow for an extension of their functionality, if necessary.
Second, the LMS-specific functionality relevant to ALMs should be well-documented in a way that ALM developers can use.
For example, they can use OpenAPI specifications.
This is especially important for enabling system-independent support mechanisms that focus on the entire learning process or connect different LMSs to use such systems.

\paragraph{Trade-offs Between Applicability and Didactic Depth of ALMs}
Most of the reviewed approaches apply reasoning-based processing mechanisms.
This may be because they are based on clear, evaluable rules that can easily be specified\ccitecf{verdu_integration_2017}.
Furthermore, corresponding mechanisms offer perceptual directness, meaning users can easily understand what triggers adaptation\ccitecf{de_bra_grapple1_2013}.
Conversely, when using calculation-based (\ie\ref{v:processingMechanism:e:calculationBased}) or optimization-based (\ie\ref{v:processingMechanism:e:optimizationBased}) processing mechanisms, the direct impact on the learners does not have to be obvious.
This may cause users to distrust the resulting opacity.
In contrast, transparent, rule-based systems enable users to predict and comprehend adaptation triggers, thereby fostering trust despite theoretical limitations.
This creates a contradiction between the practical applicability of ALMs and their educational feasibility.
This is because solely supporting even calculation- or optimization-based processing mechanisms is insufficient for complex didactic scenarios.
Rather, cooperation between multiple processing mechanisms is required.
To achieve this, researchers must investigate how to support target user groups in building trust in these mechanisms.
Thus far, this has only been investigated for ALMs with a limited number of execution- or reasoning-based processing mechanisms\cciteeg{aleven_embedding_2016}, leaving this issue to be addressed by future work.

Another reason these calculation- and optimization-based processing mechanisms are used less frequently may be that they are usually characterized by high implementation complexity and error-prone parameter derivations.
Thus, reliably implementing such mechanisms requires either extensive authoring capabilities that can be used to adjust parameter derivations, or comes with increased system dependence.
Regarding the former, many instructors may find tuning semi-technical ALM parameters too technically challenging or impossible due to time constraints.
Regarding the latter, this may be the reason why calculation-based and reasoning-based mechanisms are usually applied by two types of approaches.
First, they are applied by LMS-specific approaches that allow fine-tuning according to the known underlying data\cciteeg{kaouni2023design}.
Second, they are applied by system-independent approaches that work with predefined learning content or learning content in a specialized form that is integrated into the LMSs with the ALMs\cciteeg{pardos2023conducting}.
In order to make the extensive LMS-specific ALMs available system-independently, an appropriate architecture is required to unify the underlying LMS data from the ALMs' perspective.
While research has been done in the direction of unified architectures\cciteeg{verdu_integration_2017,zagorskis_eci_2020}, existing approaches typically have at least one of two drawbacks.
First, they apply unification only to the integration of ALMs\cciteeg{zagorskis_eci_2020}, not to the extraction and representation of existing data.
Or, second, they focus only on unifying specific data properties required for well-defined processing mechanisms\cciteeg{verdu_integration_2017}.
Thus, this point is left open as a research target for future work.

\paragraph{Restrictive Consideration of Multiple Processing Mechanisms Hinders the Extensive Support of Self-Regulated Learning}
The lack of a well-defined specification for accessing learning process data may also be a reason why most approaches support only a limited number of processing mechanisms.
These mechanisms are tightly coupled to the learning process data they can access and the degree to which this data influences adaptation.
Subsequently, well-specified data access capabilities are required.
This is especially true when multiple processing mechanisms focused on the entire learning process have to be coordinated, which requires the complete breadth of learning process data.
Due to a lack of these specifications, most approaches focus on only a limited number of processing mechanisms.
For approaches that implement ALMs with limited functionality and didactic breadth~(\eg learning content recommendations based on requested interests), specific processing mechanisms may be sufficient.
These approaches may also be easier to implement system-independently because of their limited dependence on environmental conditions.
However, this is insufficient to implement complex didactic scenarios, as required to support self-regulated learning.
In summary, the results reflect architectural specialization rather than pedagogical optimization, which limits didactic flexibility.

\subsection{Supported Adaptive Learning Mechanisms}
\label{sec:supportedAdaptiveLearningMechanisms}

\begin{figure}[h]
  \centering
  \includegraphics[width=.9\textwidth]{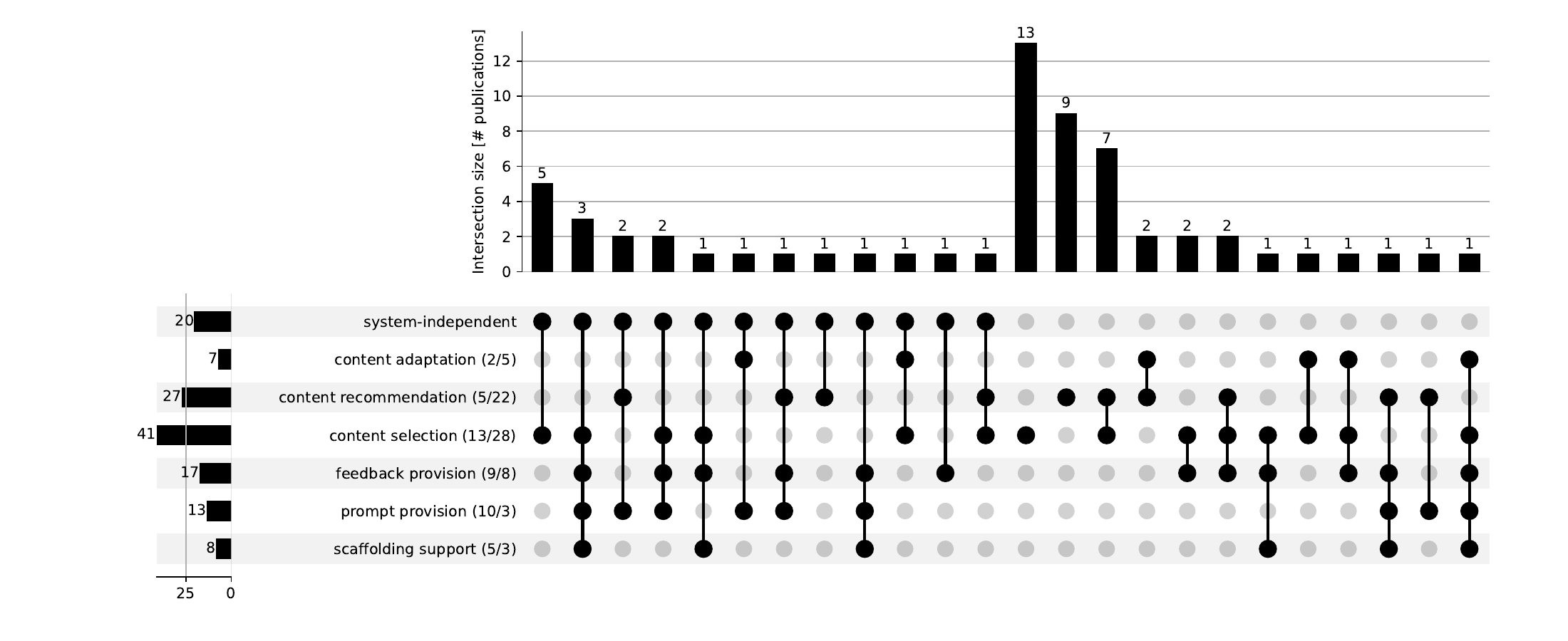}
  \caption{Supported combinations of ALMs in relation to whether they are implemented system-independently or LMS-specifically.}
  \label{fig:upsetPlotAlmSystemdependence}
  \vspace*{-0.5cm}
\end{figure}

We extracted \ref{v:adaptiveLearningMechanism} and \ref{v:adaptedComponent} to answer \ref{rq:provision}.
\Cref{tab:scopingReviewResultsPart1}, \Cref{tab:scopingReviewResultsPart2}, and \Cref{tab:scopingReviewResultsPart3} show the results.
\Cref{fig:upsetPlotAlmSystemdependence} visualizes the supported combinations of ALMs.

\subsubsection{Assessment}

\paragraph{Learning Content Adaptation}
90\% of the system-independent approaches focus on the learning content.
27.78\% of these approaches recommend learning content~(\ie\ref{v:adaptiveLearningMechanism:e:contentRecommendation}).
11.11\% modify learning content~(\ie\ref{v:adaptiveLearningMechanism:e:contentAdaptation}).
72.22\% select learning content~(\ie\ref{v:adaptiveLearningMechanism:e:contentSelection}).
Content selection differs from modification in terms of what is required from our target user groups.
Content modification requires base content that is adapted according to certain rules.
Conversely, content selection requires manually creating all the content to be selected.
Thus, especially when selecting different presentation formats for the same learning content\cciteeg{bradavc2022design}, much more content has to be created than is actually required.
However, the percentage of approaches that modify learning content is lower than the percentage of approaches that select content because the former requires a machine-processable representation of the content and its metadata, such as the one provided by\ccitetext{de_bra_grapple1_2013}.
Such a format usually does not exist for existing systems, especially with regard to native content.

Similar results were found with regard to the LMS-specific approaches.
All of these approaches focus on adapting the learning content.
53.66\% of these approaches recommend learning content, 12.2\% adjust it, and 68.29\% select learning content.
The learning content recommendations allow learners to choose whether to follow the suggestions.
In contrast, the selection forces adaptation of the learning process.
Thus, selection belongs to the program-controlled methods dimension of our descriptive framework.
In total, 73.17\% of the LMS-specific approaches implement mechanisms that apply adaptation program-controlled.
This appears to be similar for the system-independent approaches, where 77.78\% apply program-controlled ALMs.
However, a difference becomes apparent when analyzing the considered learning content.
70.73\% of LMS-specific approaches that work with native LMS learning content apply program-controlled ALMs.
However, only 22.22\% of system-independent approaches that work with native LMS learning content do so.

\paragraph{Guidance Adaptation}
55\% of system-independent approaches adapt the guidance provided to learners in three different forms.
90.91\% of these approaches implement simple prompts~(\ie\ref{v:adaptiveLearningMechanism:e:feedbackProvision}), for example to provide motivational support when boredom is detected\cciteparen{zagorskis_eci_2020}.
81.82\% provide adaptive feedback~(\ie\ref{v:adaptiveLearningMechanism:e:feedbackProvision}).
45.45\% support scaffolding~(\ie\ref{v:adaptiveLearningMechanism:e:scaffoldingSupport}).
Scaffolding support is more complex to specify because, while providing prompts or feedback is usually a response to a single learner action, scaffolding involves multiple cycles of learner interaction with the system.
Therefore, the ALS must also respond to how learners react to ALMs.
This is implemented by typical ITSs\cciteeg{luccioni_sti-dico_2016}, because this is how a human instructor would behave in a one-to-one tutoring situation.

In contrast, only 21.95\% of LMS-specific approaches adapt the guidance that is provided to learners.
This adaptation of guidance is similar to the system-independent approaches distributed to the provision of feedback (88.89\%), the provision of prompts (33.34\%), and the support of scaffolding (33.34\%).

\paragraph{Interface Adaptation}
Going one step further means adapting the entire interface that the learner is working with (\ie\ref{v:adaptedComponent:e:interface}).
20\% of system-independent approaches and 4.88\% of LMS-specific approaches implement this.
Although this type of adaptation provides the greatest flexibility for delivering adaptive content and instruction, it is also the most challenging to apply.
The reason is that, in addition to the content, this requires authoring of the learning environments' user interface as well.
This is feasible when designing an interface for learning content in a specialized domain, when the interface and the content can be integrated into different LMSs\cciteeg{aleven2016example}.
However, this is not feasible when the interface is only usable with a specific LMS.
The reason is as follows.
Since LMSs usually don't support customizing the entire learning user interface, this results in the need to recreate an additional user interface that runs LMS-specifically in parallel.

\subsubsection{Discussion}

\paragraph{Restricting Learner Autonomy by Relying Heavily on Program-Controlled Adaptation Hinders the Development of Self-Regulated Learning Skills}
The appropriateness of selecting or modifying~(\ie program-controlled adaptation), or recommending~(\ie learner-controlled adaptation) learning content depends strongly on the use case.
The former is usually applied when there is a fixed schedule\cciteeg{louhab2020novel} or when adapted learning content is determined to be the only content with which the learner can work, for example due to sensory restrictions\cciteeg{batanero2014considering}.
The latter is typically used for independent learning without a fixed schedule, but with access to a large pool of learning content to explore\cciteeg{senthil2023towards}.
In summary, both methods have reasonable use cases.
However, to support self-regulated learning, it is essential to preserve a certain degree of learner autonomy to enable the development of self-regulated learning skills through self-control and self-observation during the performance phase.
The dominance of program-controlled ALMs over learner-controlled ALMs, reflects a disproportionately high prioritization of behavioral compliance over learner autonomy, which is inconsistent with self-regulated learning.
We found that LMS-specific approaches usually prioritize these ALMs based on their development motivation.
Many LMS-specific approaches focus on problems that occur within a particular LMS.
They intend to address these problems using ALMs instead of focusing on improving learning in general and utilizing the advantages of LMSs for this purpose.
A common problem that is addressed is information overload.
This negatively affects learning outcomes because learners are unsure which content to select to begin or continue learning.
To overcome this, these approaches reduce the amount of available learning content through selection.
However, selecting learning content to overcome insecurity contradicts the idea of self-regulated learning.
Self-regulated learning aims to help learners develop the skills needed to overcome insecurity on their own.

One possible reason for this prioritization by system-independent approaches is the basis for adaptation they consider.
As discussed further in \Cref{sec:adaptiveLearningMechanismEducationalImpact}, few system-independent approaches are aware of considering metacognition.
Most approaches consider learners' learning paths or knowledge in order to support mastery learning, for example.
Sticking with this example, presenting learning content that the learners should consider next, according to their zone of proximal development, while hiding content that is considered too easy or too difficult, may be more appropriate than exploration-based recommendations.

\paragraph{System-Independent Approaches Lack Invasive Learning Adaptation Functionalities}
Although self-regulated learning focuses on making learners self-reliant in optimizing their learning process, improving learning conditions to simplify this process is also considered.
The COPES model describes this as task conditions.
One area for improvement is to adapt learning content to match learners' learning styles prior to learning.
Later on, learners can freely explore a pool of appropriate learning content based on corresponding recommendations to develop self-regulated learning skills.
This is referred to as shared adaptation control.
To implement this, the ALMs must be able to select or adapt learning content first.
As discussed above, both are supported by most system-independent and LMS-specific approaches.
However, only a minority of system-independent approaches that work with native LMS learning content support content selection.
This is because forcing the selection of learning content requires performing certain native LMS functions.
Examples of such functions include restricting access to certain learning content, hiding it, or reordering it\cciteeg{surjono2014evaluation}.
While these functions are easily accessible by extending a specific LMS, this remains a challenge for system-independent approaches.
Thus, system-independent, program-controlled ALMs tend to focus on external content that is under their control\cciteeg{luccioni_sti-dico_2016}.
This prevents the use of existing learning content, which subsequently leads to the need for re-authoring.
To overcome this challenge, a specification format could be used that can describe such adaptations in a unified way, and then allow the integrated adapters to implement them according to the capabilities of the LMSs.
Furthermore, this could also be used to help to make use of LMS-specific ALMs in a system-independent way, by using the LMS-specific implementations of this schema to reverse-specify implemented LMS-specific mechanisms in a system-independent way.
However, existing specifications focus only on a limited set of ALMs.
An example is the format provided by\ccitetext{verdu_integration_2017} for specifying tutorial guidance messages and learning object recommendation messages.
Thus, this point is left open as a research target for future work.

\paragraph{Asymmetric Support for the Adaptation of Guidance}
The adaptation of guidance includes supporting learners to develop self-regulated learning skills by offering external evaluations or by stimulating internal evaluations\ccitecf{winnie1998studying}.
We observe an asymmetry in the degree to which LMS-specific and system-independent approaches support the adaptation of guidance.
System-independent approaches adapt guidance two and a half times more often than LMS-specific ones~(55\% vs. 21.95\%).
We identified the following as the likely reason for this pattern.
Research on system-independent ALMs often comes from ITSs, while research on LMS-specific approaches comes from the LMSs themselves.
From an ITS perspective, the main goal is typically to support the learning of a specific domain or skill with appropriate tutoring processes\cciteeg{sychev_compprehension_2021}.
Subsequently, bringing LMSs and ITSs together means integrating these tutoring techniques into LMSs.
From an LMS perspective, the focus is on the existing system and the goal is to make learning with it more effective by integrating a certain level of adaptivity.
Because many of these systems cause learners to suffer from information overload due to the large amount of content\cciteparen{senthil2023towards}, much of the research in this area focuses on the content itself rather than on how to learn it.

The reported effects may also depend on the distribution of guidance support described.
While more system-independent approaches support the adaptation of guidance than LMS-specific approaches, fewer system-independent approaches report positive effects on the learning process~(\cf\Cref{sec:adaptiveLearningMechanismEducationalImpact}).
This may be because the adaptation of guidance is more difficult to validate.
This is especially true for self-regulated learning, since this involves supporting multiple cyclic learning process steps throughout the entire learning process.
Therefore, a positive effect cannot be expected to be detected immediately.
However, many evaluations of the reviewed approaches focus on short time ranges, sometimes only one learning session.
Thus, while the limited evaluation of the reviewed approaches is an issue in general~(\cf\Cref{sec:adaptiveLearningMechanismEducationalImpact}), future work should focus especially on the long-term evaluations of complex didactic scenarios.
This is necessary to determine whether and to what extent existing approaches based on models intended to support the entire learning process, such as self-regulated learning, can positively influence learning in LMSs.

\paragraph{Limited Scaffolding Support Hinders Support for Self-Regulated Help-Seeking}
An important part of adaptation based on self-evaluation is asking for help when learners encounter learning difficulties they cannot overcome alone.
Through scaffolding, ALMs can support the development of self-regulated learning skills, such as knowing when to request what type of help.
Scaffolding can be used not only to improve learning performance and self-regulated learning skills, but also to develop these skills from the beginning by providing step-by-step guidance and an environment where learners can ask questions, similar to one-to-one tutoring.
However, despite its expected value, scaffolding is rarely supported.
One possible reason for this is that implementing scaffolding requires significant authoring work to create meaningful tutoring workflows.
To determine all possible action responses, either an extensive set of rules must be defined\cciteeg{aleven2016example} or the knowledge about the learning content must be exhaustively specified to ensure that mechanisms such as AI can automatically react appropriately\cciteeg{roselli2008integration}.
This is feasible for a fixed domain or a limited number of domains\cciteeg{pardos2023conducting}, but not for the integration into an existing system with a large amount of existing content and different domains.
To overcome this challenge, tutoring strategies need to be specified more abstractly.
If extensive support for scaffolding is not possible, instructors may consider organizing learning activities according to the level of support students may require\cciteeg{louhab2020novel}.
Thus, while comprehension questions and processing learning difficulties may be addressed together with the instructor, preparation and repetition can be done by learners on their own with effective support from ALMs.

\subsection{Authoring and Initiation of Adaptive Learning Mechanisms}
\label{sec:adaptiveMechanismTriggersAndDefinitionMethods}

We extracted \ref{v:authoringCapability} and \ref{v:adaptiveLearningMechanismTrigger} to answer \ref{rq:definition}.
\Cref{tab:scopingReviewResultsPart1}, \Cref{tab:scopingReviewResultsPart2}, and \Cref{tab:scopingReviewResultsPart3} show the results.

\subsubsection{Assessment}

\paragraph{Authoring Capabilities}

Our review supports the hypothesis that most approaches only allow our target user groups to modify learning means or parameters for fixed ALMs~(\ie\ref{h:insufficientAuthoringCapabilitiesSupport}).
25\% of system-independent approaches do not provide any authoring capabilities~(\ie\ref{v:authoringCapability:e:noAuthoring}).
This is due to one of two reasons.
First, the approaches provide fixed learning content for a fixed domain that can be used but is not intended to be adjusted\cciteeg{luccioni_sti-dico_2016}.
Second, conceptual systems are provided that need to be adapted to the actual use case by appropriate programmers\cciteeg{zagorskis_eci_2020}.
75\% of these approaches provide authoring capabilites that only correspond to the learning means (\ie\ref{v:authoringCapability:e:authoringOfAdaptiveLearningMeans}) or the parameters (\ie\ref{v:authoringCapability:e:authoringOfAdaptiveLearningMechanismParameters}) of fixed ALMs.
In most cases, these approaches offer the ability to customize the learning content for fixed ALMs\cciteeg{serce_moda_2008}.
Only two approaches~(10\%) provide capabilities for authoring the ALMs themselves (\ie\ref{v:authoringCapability:e:authoringOfAdaptiveLearningMechanisms}).

We found similar results for the LMS-specific approaches.
17.03\% of these do not support any authoring.
This is usually due to the provision of fixed learning content supported by a predetermined didactic concept\cciteeg{oskouei2017proposing}.
83.93\% of these approaches support the authoring of learning means.
This percentage is even higher than that of system-independent approaches because LMS-specific adaptation approaches tend to focus on the native learning content that can be authored in LMSs by design.
36.59\% of these approaches focus on the adaptation of ALM parameters. 
Only 2.44\%~(\ie one approach) support authoring the actual ALMs.

\paragraph{Adaptation Triggers}
40\% of system-independent approaches apply \emph{start-of-session triggers} (\ref{v:adaptiveLearningMechanismTrigger:e:learningSessionBegin}).
During the learning session, these approaches support two categories of triggers.
45\% apply \emph{proactive triggers} (\ref{v:adaptiveLearningMechanismTrigger:e:proactiveInitiation}).
Although authors can easily define these, they depend on learners' self-reflection.
Learners must recognize when they need assistance, which makes success more dependent on them.
\emph{System-controlled triggers} are based on observations of learner behavior.
Within this category, 70\% of system-independent approaches apply \emph{navigation-based triggers} (\ref{v:adaptiveLearningMechanismTrigger:e:navigation}) and 55\% apply \emph{assessment-response triggers} (\ref{v:adaptiveLearningMechanismTrigger:e:assessmentAnswering}).
Overall, most system-independent methods rely on automatic, observation-driven triggers.
Fewer than half incorporate learner-initiated, proactive triggers.

In addition, we analyzed the applied triggers with regard to the level of integration into the LMSs that is required to apply them.
Implementing proactive and assessment-based triggers does not necessarily require customizing the LMSs.
These triggers require the ability to request adaptations and a setting in which to conduct assessments, the latter of which can run in parallel.
Though no more than detecting the start of the session, implementing triggers based on the start of a learning session requires interference with the LMS's functionality.
Implementing triggers based on learners' navigation behavior requires the deepest level of interference because it necessitates detecting learners' interactions with the system.
Both triggers require interference with the LMS's functionality and are easier to implement when working with a specific LMS.
Nevertheless, only 36.59\% of LMS-specific approaches apply triggers based on navigation, and only 31.71\% implement triggers based on the beginning of the learning session.
Instead, more LMS-specific approaches implement triggers based on assessment (70.73\%) and triggers based on proactive initiation (12.2\%).

\subsubsection{Discussion}

\paragraph{Authoring of ALMs Through Our Target User Groups Remains a Challenge}
The low number of LMS-specific approaches that support the authoring of ALMs can be explained by the following aspect. 
The more unknowns there are about a learning process during the conceptualization of ALMs, the more our target user groups need to author to apply them later on.
When using LMS-specific approaches, at least the context, the structure of the learning content, and the learner's capabilities to interact with the system are known.
Based on this knowledge, many LMS-specific approaches have developed a didactic concept of how to learn effectively in this context.
This reduces the need for authoring to only the creation of learning content.
Another reason that applies to LMS-specific and system-independent approaches is their development focus.
Many approaches are developed by technicians\cciteparen{rose2019explanatory} and are focused on concepts to model learner behavior and later, with the help of didactic experts, use the gained knowledge to positively influence learning.
Therefore, this research does not focus on enabling authoring for our target user groups and often results in approaches that focus on optimizing learning process data with regard to elaborated data models, missing grounding in didactic theories and models\cciteparen{rose2019explanatory}.

The proposed approaches, which focus on our target user groups, have at least one of two limitations.
First, the possibilities to author ALMs are limited to make the authoring process more user-friendly\cciteeg{aleven2016example,pardos2023conducting}.
Second, although authoring is claimed to be supported, it still requires technical expertise, at least for certain levels of intervention\cciteeg{roselli2008integration,de_bra_grapple_2010}.
Thus, the investigation of authoring capabilities for the entire process, from the conceptualization of ALMs to the provision for our target user groups, remains an open research target for future work.

Furthermore, enabling users who are intended to apply ALMs without providing them the possibility to adjust these mechanisms leads to a contradiction.
Instructors are given control over the parameters of ALMs that they did not design and likely do not fully understand.
However, they are denied the opportunity to author their function in which their expertise would be most valuable pedagogically.
This inverts the expertise hierarchy, with technicians controlling adaptation logic and instructional designers adjusting peripheral parameters.
This issue can be addressed in two ways.
First, interdisciplinary teams of technicians and instructional designers should develop ALMs in cooperation\ccitecf{rose2019explanatory}.
Second, instructional designers should be given user-friendly authoring capabilities to create full-featured ALMs.
Regarding the former option, LMS developers should ensure that technicians can extend existing systems.
There are two ways to ensure this.
First, the LMS is an open-source project that technicians can freely adjust.
Second, the LMS has built-in functionalities that can be used to extend the system.
For example, it has an appropriately documented plugin system.

\paragraph{Event-Based Triggers and Its Connection to Self-Regulated Learning Phase Transitions}
The reviewed approaches primarily apply event-based triggers.
Triggers based on evaluating learning states or transitions between states are not implemented.
To support the entire self-regulated learning process through its multiple phases, it is necessary to detect when learners transition between the described phases\ccitecf{zimmerman2002becoming} or stages\ccitecf{winnie1998studying}.
One reason state-based triggers are rarely supported is that the reviewed approaches determine the state of the learning process on which adaptation is based during the processing of learning process data, rather than within the range of triggers.
This can be achieved by applying reasoning-based processing mechanisms.
This processing is typically initiated by navigation-based triggers to be aware of every interaction of learners with the learning environment that may cause a change in the relevant learning process states.
Although this is technically feasible, the target user groups may have difficulty understanding when a particular adaptation is triggered.
This may, in turn, undermine trust in these mechanisms and impede their widespread adoption.

Another aspect regarding the detected patterns of applied triggers that may hinder effective self-regulated learning support is the less common use of proactive initiation-based triggers.
An important aspect of self-regulated learning is helping learners recognize when they encounter learning difficulties they cannot overcome alone.
Providing them with the opportunity to request help through proactive initiation triggers fosters the development of self-regulated learning skills.

\paragraph{Lack of Navigation-Based Triggers for Native Learning Content}
Although a higher level of learning environment intervention capabilities is required to work with navigation-based triggers, many system-independent approaches use these triggers, whereas fewer LMS-specific approaches do.
Two aspects must be considered to explain this.
First, the provided triggers correlate with the provided ALMs.
As previously described, system-independent approaches tend to support tutoring processes, whereas LMS-specific approaches focus on preventing information overload regarding learning content.
Supporting tutoring processes requires observing learner interactions, which is why 70\% of the evaluated system-independent approaches apply this type of trigger.
Many approaches to selecting content are based on the learner's learning style\cciteeg{pagano2021training}.
These approaches typically work with standardized questionnaires to determine learning styles (\eg the Index of Learning Styles questionnaire proposed by \ccitetext{soloman2005index}).
As a result, 70.73\% of the LMS-specific approaches apply assessment-based triggers.

Another aspect to consider is what type of interaction data and learning content the reviewed approaches work with.
The majority~(85.71\%) of system-independent approaches that apply navigation-based triggers do neither work with the interaction data generated by the LMS nor with the native LMS content.
Instead, these approaches integrate learning content and functionality for observing the learner's interaction with that content into the LMSs together with the ALMs\cciteeg{aleven2017integrating}.
This reduces the level of interference required with the LMS's functionality.
Consequently, navigation-based triggers become feasible for system-independent approaches as well.
Thus, to enable system-independent approaches that implement ALMs with navigation-based triggers and native learning content, LMS developers should implement established standards for representing interaction data, such as the xAPI specification.
Using such standardized data, ALMs could operate without deeply interfering with LMSs to observe user interaction.

\subsection{Considered Learning Content and Learning Data}
\label{sec:adaptiveLearningMechanismDatabase}

We extracted \ref{v:learningContent} and \ref{v:adaptationSourceData} to answer \ref{rq:database}.
\Cref{tab:scopingReviewResultsPart1}, \Cref{tab:scopingReviewResultsPart2}, and \Cref{tab:scopingReviewResultsPart3} show the results.

\subsubsection{Assessment}

\paragraph{Learning Content}
45\% of system-independent approaches work with learning content that is integrated into the LMSs along with the ALMs~(\ie\ref{v:learningContent:e:nonLmsContent}).
35\% work with native learning content (\ie\ref{v:learningContent:e:nativeLmsContent}).
These mechanisms must be generic enough to work with different types of learning content.
Therefore, they are more difficult to design.
However, they reduce development effort and costs by preventing re-authoring.
Furthermore, they increase the likelihood of adoption, because these approaches can be more easily integrated into existing systems.
To facilitate ALM development while at the same time avoiding the need for re-authoring, 20\% of system-independent approaches work with a fixed learning content format~(\ie\ref{v:learningContent:e:particularLmsContent}).
In most cases the Sharable Content Object Reference Model Format (SCORM) is used\cciteeg{santos_web-based_2010}.
This allows existing content to be reused if it is in the specified format.
However, re-authoring is still necessary if it is presented in a different format.

No LMS-specific approach works exclusively with a specific content format.
The reason for this is that most LMS-specific approaches focus on preventing information overload within the LMS.
Therefore, these approaches must work with native LMS learning content, as 97.56\% do.
Some approaches (9.76\%) also consider non-LMS content.
These approaches typically focus on suggesting related content or connecting multiple sources of information in large learning infrastructures\cciteeg{perivsic2018semantic}.

\paragraph{Learning Process Data}

Our review supports the hypothesis that system-independent ALMs rarely consider existing learning data~(\ie\ref{h:insufficientDataBasisSupport}).
We've already seen this with regard to native learning content that is rarely considered.
Regarding the handled interaction data, only 25\% consider existing data (\ie\ref{v:adaptationSourceData:e:existingInteractionData}).
Most of the evaluated approaches (75\%) implement custom methods for tracking learner interactions that generate custom interaction data~(\ie\ref{v:adaptationSourceData:e:customInteractionData}).

All of the system-independent approaches handle either custom interaction data or existing interaction data, while 19.51\% of the LMS-specific approaches handle neither the first nor the second.
This can also be explained by the correlation with the provided ALMs.
Many LMS-specific approaches implement the selection or recommendation of learning content based on the learner's learning style.
Although there are methods for determining learning styles based on learners' interactions\cciteeg{fakoya2021automatic}, many approaches use standardized learning style questionnaires for their simplicity and reliability\cciteparen{peter2010adaptable}.
Consequently, these approaches only require processing of requested learner data, not interaction data.
This is the case for 68.3\% of the LMS-specific approaches.
LMS-specific approaches that process interaction data process 39.02\% custom interaction data and 41.46\% existing interaction data.
While these LMS-specific approaches typically process interaction data generated by the corresponding LMS, a number of approaches propose new LMSs\cciteeg{cobos2007learning}.
This prevents the reuse of existing interaction data as well.

In addition to the question of when and by which components the learner-specific data is collected, \ref{v:adaptationSourceData} also focuses on whether the data is processed prior to being considered~(\ie\ref{v:adaptationSourceData:e:requestedLearnerData}).
In this regard, all approaches work with processed data.
This may be because there are no didactically feasible ALMs that work with raw interaction data or with unprocessed learner answers.
An example of such an approach would be the selection of learning content as defined and requested by the learner.
This would lack the possibility of a pedagogically feasible intervention.

\subsubsection{Discussion}

\paragraph{ALMs Using Non-Native LMS Content Create Learning Environment Islands That Prevent the General Distribution of Didactically Feasible Mechanisms}

A lack of knowledge regarding how learning content is structured in different LMSs hinders the development of system-independent ALMs that can work with native learning content.
Corresponding approaches use two strategies to address this.
First, they simplify ALMs so that they only require generic learning content data, which can be assumed to be accessible in each LMS or is generated by the ALMs themselves (\eg by asking learners).
Second, they prevent the use of native learning content and instead consider external content with a known structure and metadata.
In particular, the first may reflect a prioritization of technical feasibility over didactic usefulness that may result from this research being primarily conducted by computer scientists, as\ccitetext{rose2019explanatory} stated.
Both strategies result in two problems.
First, these approaches are not applicable when our target user groups want support for learning their existing content.
Second, evaluations of conceptualized ALMs are conducted using specialized learning content.
The evaluation results may not always apply to common learning scenarios involving content that was not specifically designed for these ALMs.
To overcome this problem, a specification is needed that can describe existing learning content and its corresponding metadata in a system-independent way without losing system-specific information.
Another problem with ALMs that focus on isolated learning environments with separate content is that they can only provide limited support for didactic scenarios focusing on the entire learning process, such as self-regulated learning.
Instead, these approaches focus only on specific learning actions, omitting the potential to support all phases of the learning process.

\paragraph{Potential of Processing Native LMS Interaction Data Is Underestimated}
Another unused potential is working with existing interaction data.
This data, especially with regard to system-independent approaches, could be used for learning analytics across different LMSs\ccitecf{mangaroska_learning_2018}, including newly integrated LMSs.
Subsequently, the analysis results could be used to adapt the learning process.
Considering interaction data that may already be present in LMSs, ALMs are newly integrated into, can also prevent the cold start problem that some ALMs suffer from.
Another advantage of considering interaction data generated by associated LMSs, as opposed to interaction data generated for specific learning activities, is that the former typically encompasses all of the learner's interactions with the LMS.
This enables support of entire learning processes rather than specific learning actions, as described above.
We identified two possible reasons why the reviewed approaches consider native interaction data less frequently.
First, a consistent specification of this data is required for reference during extraction and processing.
However, the corresponding LMS does not support such a specification, nor do the other LMSs.
Second, the implemented ALMs do not require this data because they are based on information that cannot be derived from interaction data, but rather must be requested specifically from the learners.
For example, this is the case if the learning content is selected based solely on learning styles determined by questionnaires.

Regarding a consistent data format, specifications are already in place that can be used to capture interaction data, such as xAPI.
However, although xAPI is registered as a standard, it has not yet been adopted by all of the LMSs referenced by the reviewed approaches.
To support the widespread adoption of system-independent ALMs, LMS developers should prioritize implementing commonly agreed-upon standards.
This will allow ALM developers to rely on these standards.
Furthermore, when implementing standards such as xAPI, LMS developers should ensure that all relevant learning interactions are captured.
Generating xAPI statements for all relevant learning actions enables comprehensive ALMs and meaningful learning analytics.

\paragraph{Supporting Internal and External Evaluations for Self-Regulated Learning}
According to the COPES model of self-regulated learning, there are two types of evaluations to foster the development of self-regulated learning skills.
First, the learners themselves conduct internal evaluations.
Second, external evaluations are conducted by instructors, learning peers, or ALMs.
External evaluations can be performed by processing learner data to determine necessary adaptations.
Internal evaluations can be supported by asking learners for information to request the current state of evaluation and to foster the evaluation process.
To effectively support self-regulated learning, both must be supported.
Our results show that the majority of the reviewed approaches have already implemented this.

\subsection{Educational Surrounding Conditions and Outcomes}
\label{sec:adaptiveLearningMechanismEducationalImpact}

We extracted \ref{v:adaptationGoal}, \ref{v:adaptationBasis} and \ref{v:reportedEffects} to answer \ref{rq:educationalContext}.
\Cref{tab:scopingReviewResultsEducationalContext} in \Cref{sec:resultsTableEducationalContext} shows the results.

\subsubsection{Assessment}

\begin{figure*}
  \centering
  \includegraphics[width=.6\textwidth]{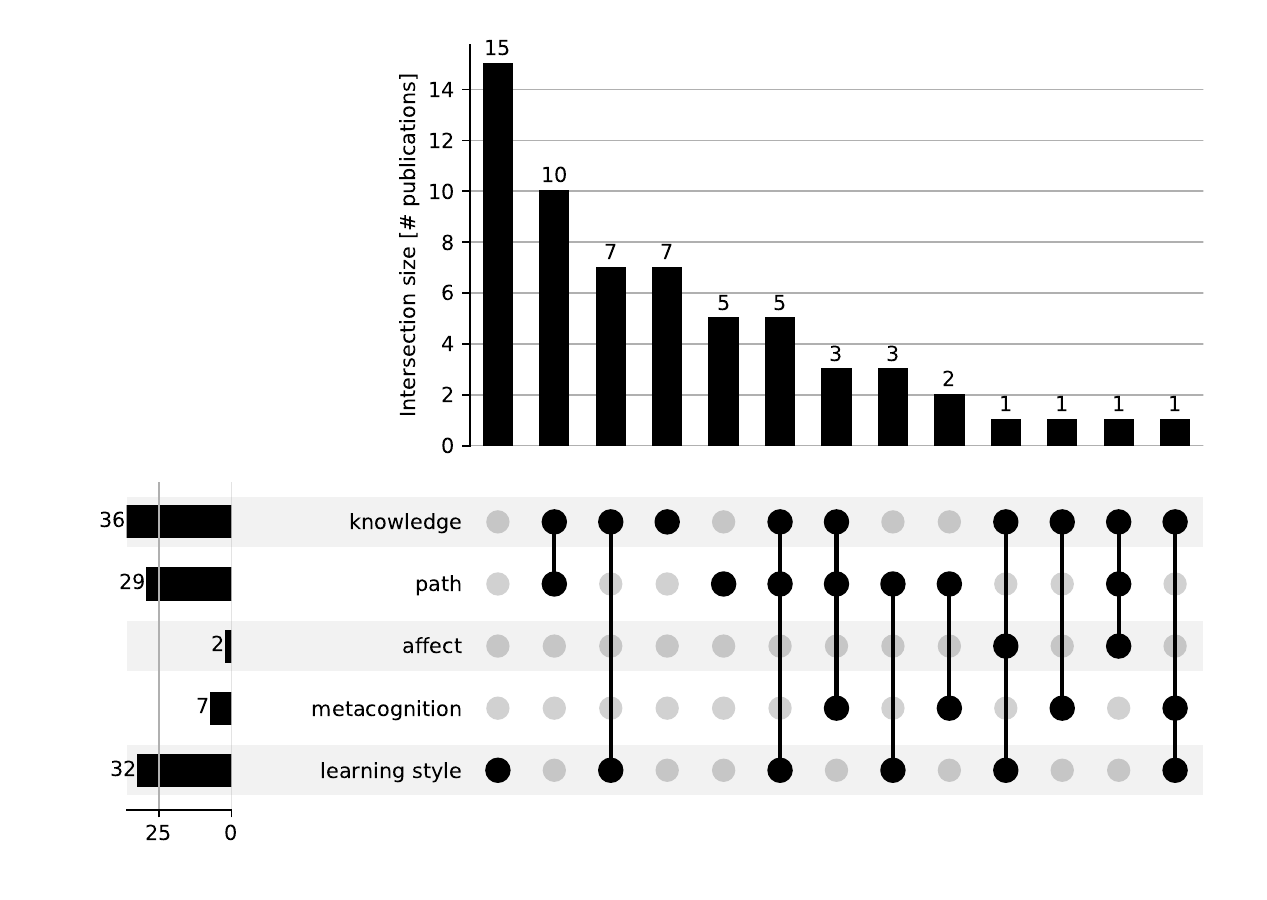}
  \caption{Considered combinations of adaptation bases of the reviewed approaches based on the categorization of\ccitetext{aleven2016instruction}.}
  \label{fig:upsetPlotBasis}
\end{figure*}

\paragraph{Adaptation Goal and Basis}
There are two groups of adaptation goals.
19.67\% of the approaches focus on authoring.
This includes developing authoring capabilities for ALMs for different contexts.
For example,\ccitetext{roselli2008integration} focus on enabling instructors to create ITSs to support learners with their homework.
Another goal of this group is to develop architectures that address certain adaptation context requirements~(\eg system independence).
The other approaches~(80.33\%) focus on developing ALMs to improve specific aspects of learning.
For example,\ccitetext{pedrazzoli2009artificial} intends to increase learners' motivation.
\ccitetext{moutafi2013mining} focus on reducing the required learning time.
However, despite the existence of concrete adaptation goals, a large number of approaches (40.82\%) focus on personalization in general.
These approaches are based on positive reports of adapting the learning process to learners' needs instead of using one-size-fits-all solutions.
Thus, they describe different desirable positive impacts of adaptive learning, but do not specify a single measurable didactic goal variable.
Implicitly, learners' grades are often assumed to be an indicator of learning efficiency and are tried to be optimized, as can be seen from the corresponding reported effects.
Optimizing learners' grades, often referred to as learning outcomes, is repeatedly (by 16.33\% of these approaches) specified as a concrete adaptation goal too, either directly or indirectly (\eg through lowering dropout rates).
Other goals, such as optimizing the match between the provided learning content and the learners' needs, or supporting specific types of learners, such as slow learners or those with dyslexia, are addressed sporadically.

\begin{figure*}
  \centering
  \includegraphics[width=.9\textwidth]{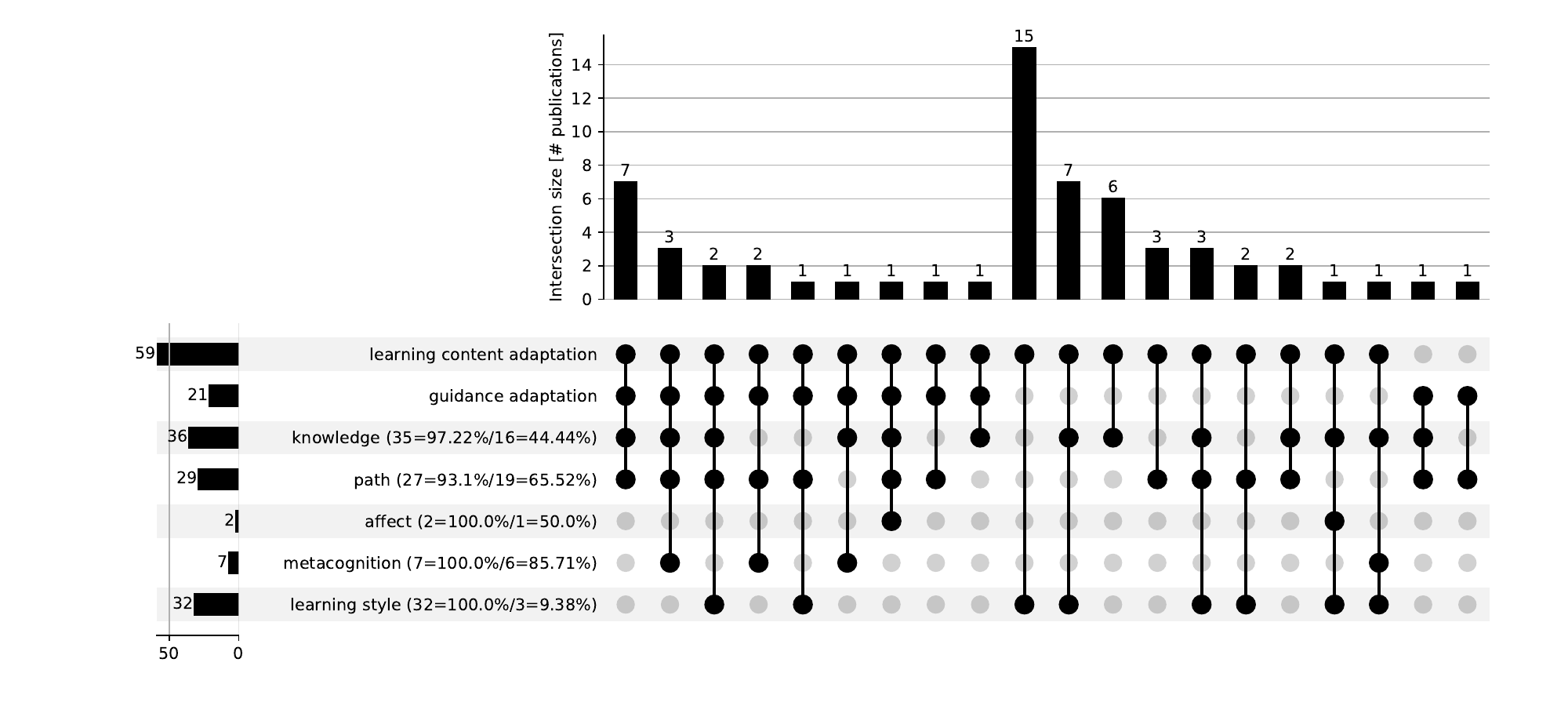}
  \caption{Considered combinations of adaptation bases of the reviewed approaches in relation to the implemented ALMs. For each adaptation basis, the number of approaches that consider this specific basis and adapt the learning content provided to learners is indicated on the left as a total number and percentage. On the right it is indicated how many approaches consider this specific adaptation basis and adapt the guidance provided to learners, showing the total number and percentage.}
  \label{fig:upsetPlotAlmBasis}
\end{figure*}

\begin{figure}[!ht]
    \centering

    \includegraphics[width=.9\textwidth]{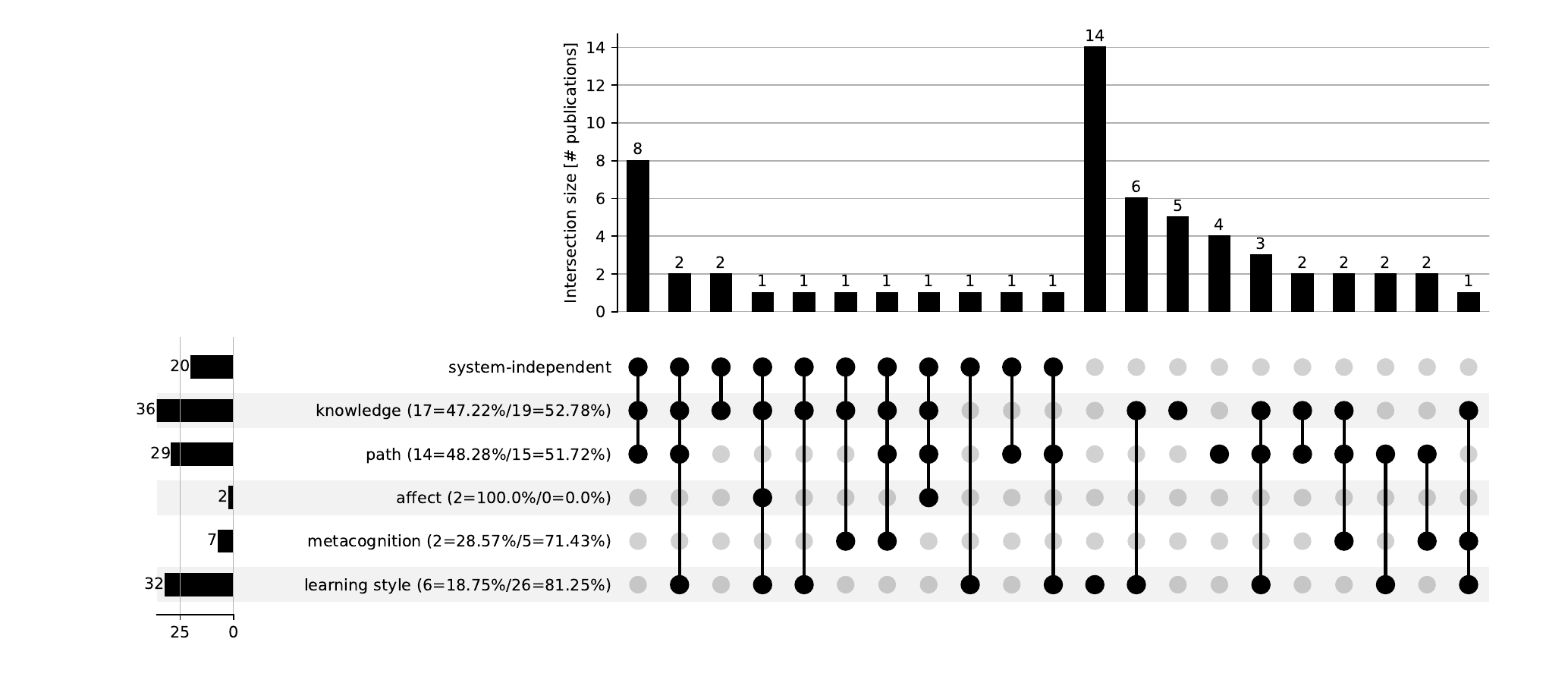}
    \caption{Considered combinations of adaptation bases of the reviewed approaches in relation to whether these approaches are system-independent or are focused on a specific LMS.}
    \label{fig:upsetPlotBasisSystemDependence}

    \nextfloat

    \begin{subfigure}{0.32\textwidth}
        \centering
        \includegraphics[height=4cm]{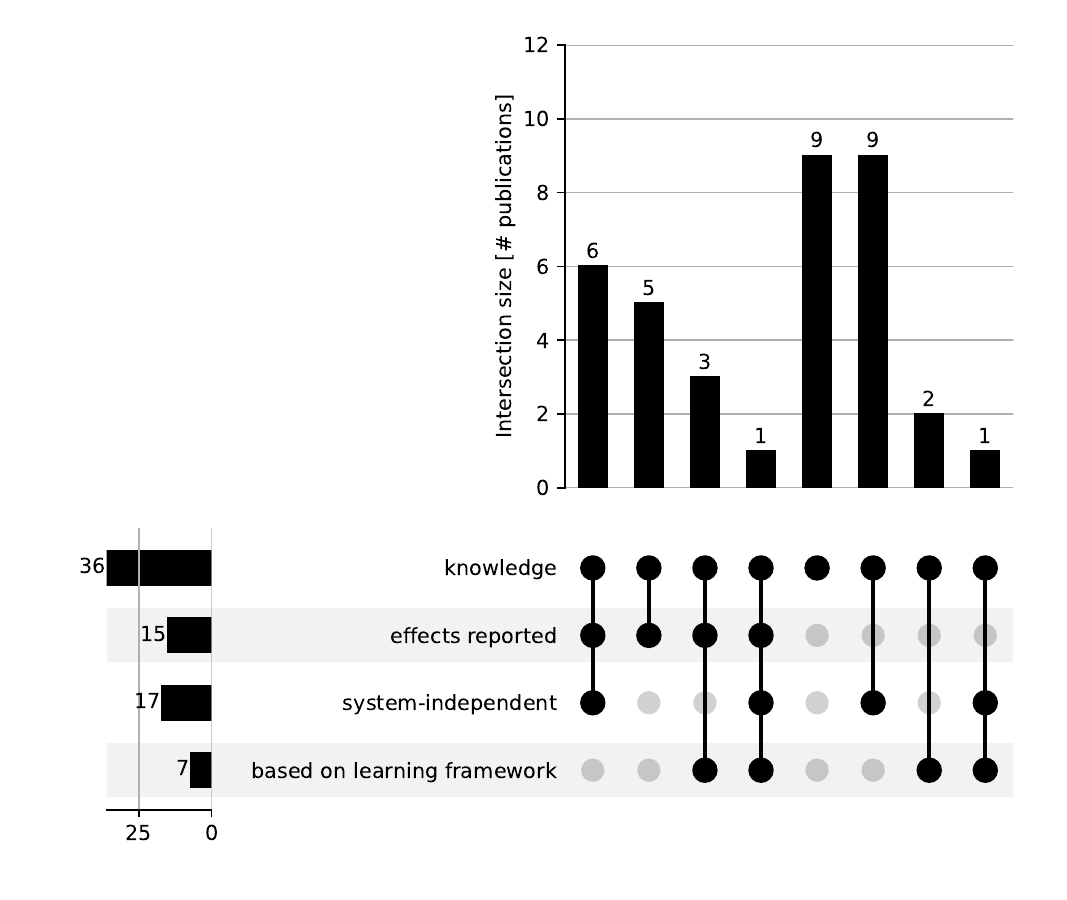}
        \caption{Adaptation basis \textit{knowledge}.}
    \end{subfigure}
    \hfill
    \hspace*{.5cm}
    \begin{subfigure}{0.3\textwidth}
        \centering
        \includegraphics[height=4cm]{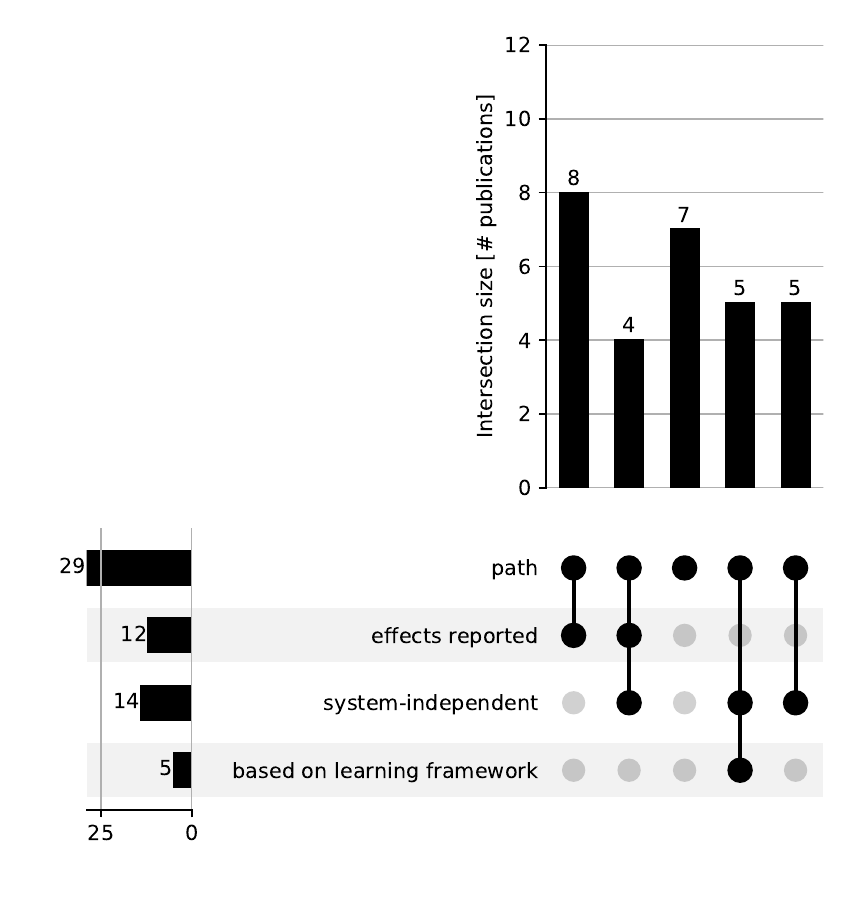}
        \caption{Adaptation basis \textit{path}.}
    \end{subfigure}
    \hfill
    \begin{subfigure}{0.28\textwidth}
        \centering
        \includegraphics[height=4cm]{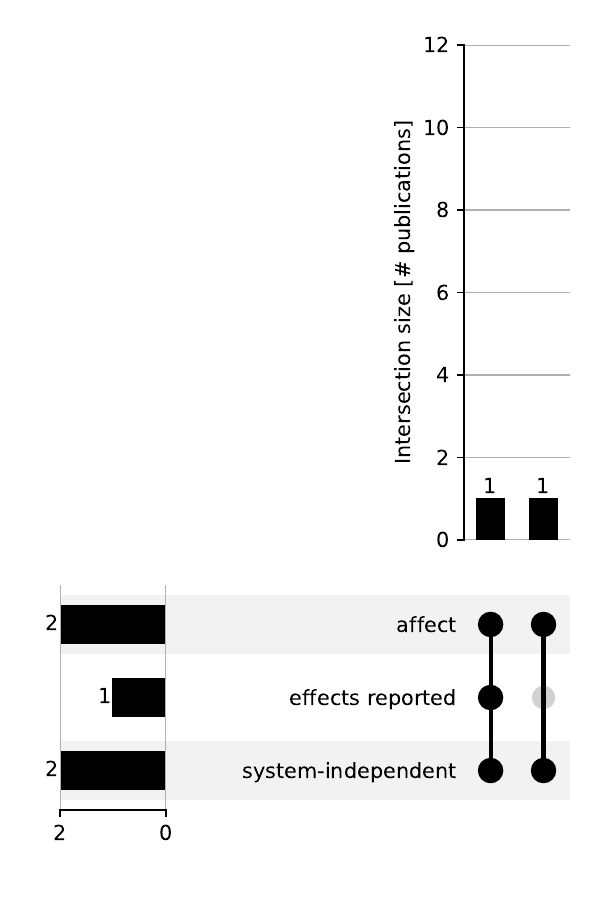}
        \caption{Adaptation basis \textit{affect}.}
    \end{subfigure}

    \medskip

    \begin{subfigure}{0.4\textwidth}
        \centering
        \includegraphics[height=4cm]{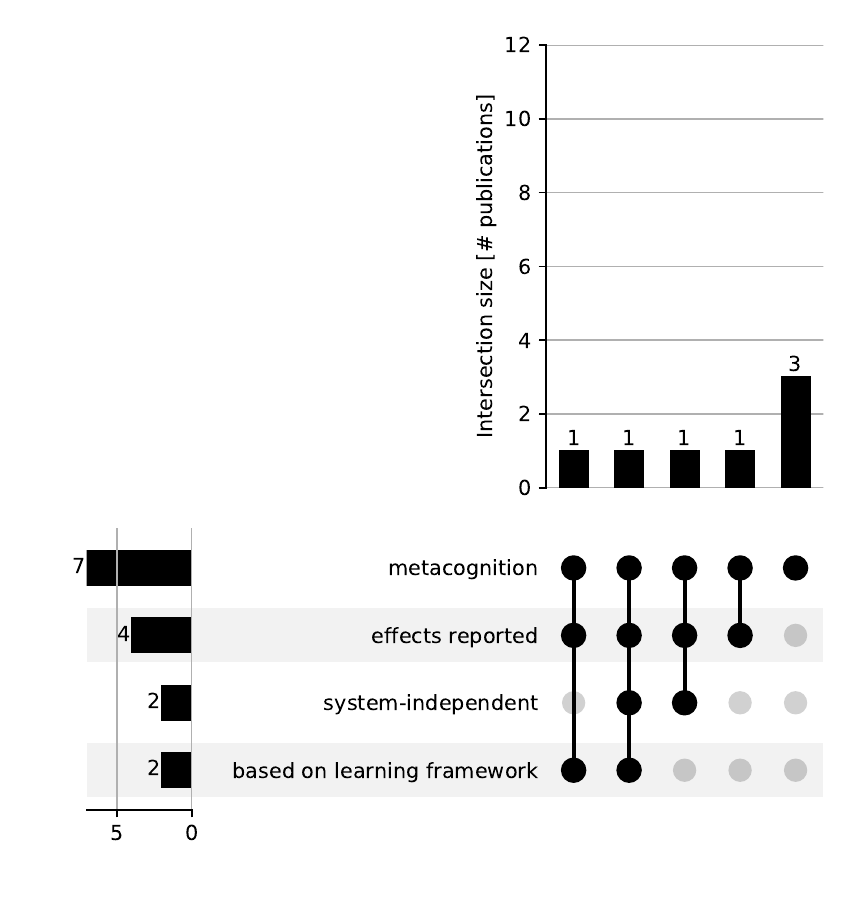}
        \caption{Adaptation basis \textit{metacognition}.}
    \end{subfigure}
    \begin{subfigure}{0.4\textwidth}
        \centering
        \includegraphics[height=4cm]{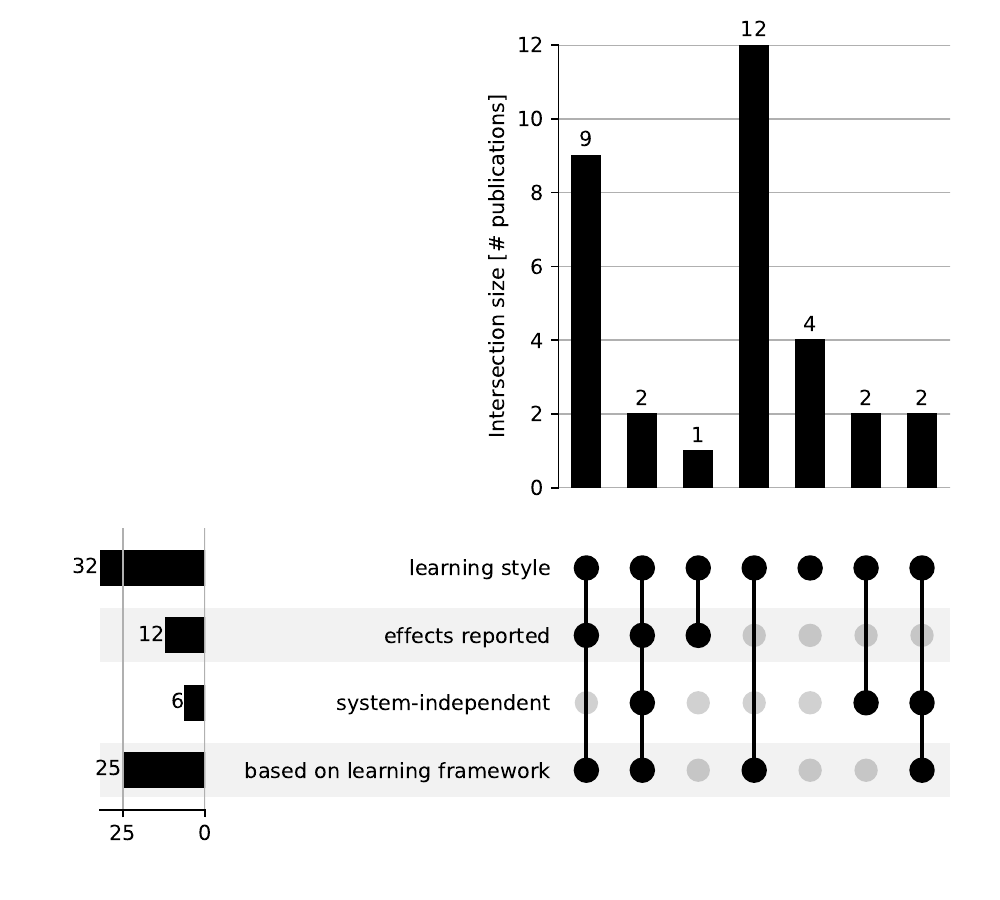}
        \caption{Adaptation basis \textit{learning style}.}
    \end{subfigure}

    \caption{The relation of the basis of adaptation, whether the corresponding approaches are system-independent or LMS-specific, whether they have reported effects, and whether they are based on an established learning framework. This information is visualized per adaptation basis. For the approaches that are based on affect, it is not visualized whether these approaches adopt an established learning framework, since we found no approach that does so.}
    \label{fig:relationAdaptationBasisSystemDependenceEffectsAndModel}
\end{figure}

\Cref{fig:upsetPlotBasis} shows the considered combinations of adaptation bases.
The most frequently considered basis (by 59.02\% of the approaches) is knowledge.
This aligns with the predominant adaptation goal of enhancing grades, which is usually associated with the acquisition of subject-specific knowledge.
In contrast, a learner's learning style can facilitate understanding of a subject, but it cannot be used to measure understanding for the purpose of systematically improving learning gains, as is done by the mastery learning framework\cciteparen{bloom1968learning}, for example.
However, the mastery learning framework assumes that all learners are capable of learning the subject in question, making learning to a matter of time and training.
If learners are unable to learn the content due to a specific type of dyslexia or accessibility issues, providing learning content according to a learner's learning style can be considered a first step in meeting their learning needs.
Furthermore, some approaches consider learning styles to improve learning efficiency.
In total, 52.46\% of the approaches consider the learners' learning style.
It is considered on its own most often.
Knowledge, in contrast, is most often considered in combination with the learners' learning path.

The considered adaptation bases are also highly connected to the applied ALMs.
This association is illustrated in \Cref{fig:upsetPlotAlmBasis}.
This plot categorizes the reviewed approaches into two groups of implemented ALMs.
Approaches that implement the adaptation (\ie\ref{v:adaptiveLearningMechanism:e:contentAdaptation}), recommendation (\ie\ref{v:adaptiveLearningMechanism:e:contentRecommendation}), or selection (\ie\ref{v:adaptiveLearningMechanism:e:contentSelection}) of learning content are categorized as learning content adaptation approaches.
Approaches that implement the adaptive provision of feedback (\ie\ref{v:adaptiveLearningMechanism:e:feedbackProvision}), prompts (\ie\ref{v:adaptiveLearningMechanism:e:promptProvision}), or support scaffolding (\ie\ref{v:adaptiveLearningMechanism:e:scaffoldingSupport}) are categorized as guidance adaptation approaches.
The adaptation of learning content appears to be significantly associated with the consideration of the learners' learning style. 
A mere 9.38\% of ALMs that consider the learners' learning style adapt the guidance provided to the learner, while 100\% adapt the learning content.
This may indicate that the guidance provided to learners based on their learning style is not suitable.
However, we also found no reported negative effects, as described below.
The same is true for the other four bases of adaptation, as most ALMs based on these adapt the learning content.
However, at least half of these approaches also adapt the guidance provided to learners.
Among those, the percentage of ALMs that adapt guidance was the lowest (44.44\%) for mechanisms based on the learners' knowledge.
The most common adaptation of the guidance provided to the learners is implemented by ALMs based on metacognition (85.71\%) and the learner's learning path (65.52\%).
One reason for this is that ALMs based on learners' paths are usually connected to tutoring processes. 
These processes are typically supported by ITSs and focus specifically on guiding learners through the learning process.
The same applies to the learning models based on metacognition, such as self-regulated learning models.

A further association exists between the considered adaptation bases and whether the reviewed approaches are system-independent or focused on a specific LMS.
This is shown in \Cref{fig:upsetPlotBasisSystemDependence}.
Most system-independent approaches are based on the learners' knowledge and path, whereas most LMS-specific approaches are based on the learners' learning style.

In addition to the considered adaptation bases, we analyzed the adopted established learning frameworks.
\Cref{fig:relationAdaptationBasisSystemDependenceEffectsAndModel} shows how many approaches adopt an established learning framework for each adaptation basis.
In summary, most ALMs are not based on any established learning framework.
When treating ALMs separately according to the considered adaptation basis, only 36.79\% adopt an established learning framework.
When analyzed per adaptation basis, 78.13\% of the mechanisms considering learners' learning style, 19.44\% of the mechanisms considering learners' knowledge, 17.24\% of the mechanisms considering learners' learning path, 28.57\% of the mechanisms considering metacognition, and 0\% of the mechanisms considering affect adopt an established learning framework.

\paragraph{Reported Adaptation Effects}

\begin{figure}[H]
    \centering
    \begin{tikzpicture}
        \begin{axis}[
          legend cell align=left,
          legend pos=north west,
          ybar,
          height=4cm,
          width=\textwidth,
          ymin=0,ymax=105,
          ytick={0,20,40,60,80,100},
          y tick label style={black},
          y label style={black},
          ylabel={Reported Effects [\%]},
          xmin=2003,xmax=2023,
          xtick={2003,2004,2005,2006,2007,2008,2009,2010,2011,2012,2013,2014,2015,2016,2017,2018,2019,2020,2021,2022,2023},
          restrict x to domain=2003:2023,
          enlarge x limits=false,
          xlabel={Publication Time [year]},
          xticklabel={'\pgfmathparse{int(mod(\tick,2000))}\ifnum\pgfmathresult<10 0\fi\pgfmathresult},
		      x tick label style={
			      /pgf/number format/1000 sep=,
            /pgf/number format/fixed,
            black,
          },
          xlabel style={black},
          xtick distance=1,
          enlarge x limits=0.05,
          grid=both
        ]
        \addplot [
          mark=*,
          smooth,
          mark size=2pt
        ] coordinates {
            (2003,0)
            (2004,0)
            (2005,0)
            (2006,0)
            (2007,50)
            (2008,0)
            (2009,0)
            (2010,20)
            (2011,0)
            (2012,100)
            (2013,33.33)
            (2014,16.67)
            (2015,60)
            (2016,0)
            (2017,50)
            (2018,100)
            (2019,80)
            (2020,50)
            (2021,66.67)
            (2022,100)
            (2023,0)
          };
        \end{axis}
    \end{tikzpicture}
    \caption{Relative frequency of publications per year reporting adaptation effects.}
    \label{fig:percentageOfPapersWithReportedEffectsByYear}
\end{figure}

\Cref{fig:percentageOfPapersWithReportedEffectsByYear} illustrates the percentage of approaches that reported effects relative to the total number of approaches reviewed per year of publication.
We can point out two peculiarities.
First, the number of approaches that reported effects is low, especially prior to 2012.
Of the reviewed approaches, only 40.98\% reported effects regarding the educational impact of the elaborated ALMs.
Several approaches mention evaluating their ALMs as future work. 
However, despite searching for related papers, we could not find corresponding evaluation results.
Second, all reported effects are positive.
There is no approach that has reported mixed or negative effects of the elaborated ALMs, nor has there been an evaluation of an elaborated ALM that has reported that no effects could be found.

\begin{figure*}
  \centering
  \includegraphics[width=.9\textwidth]{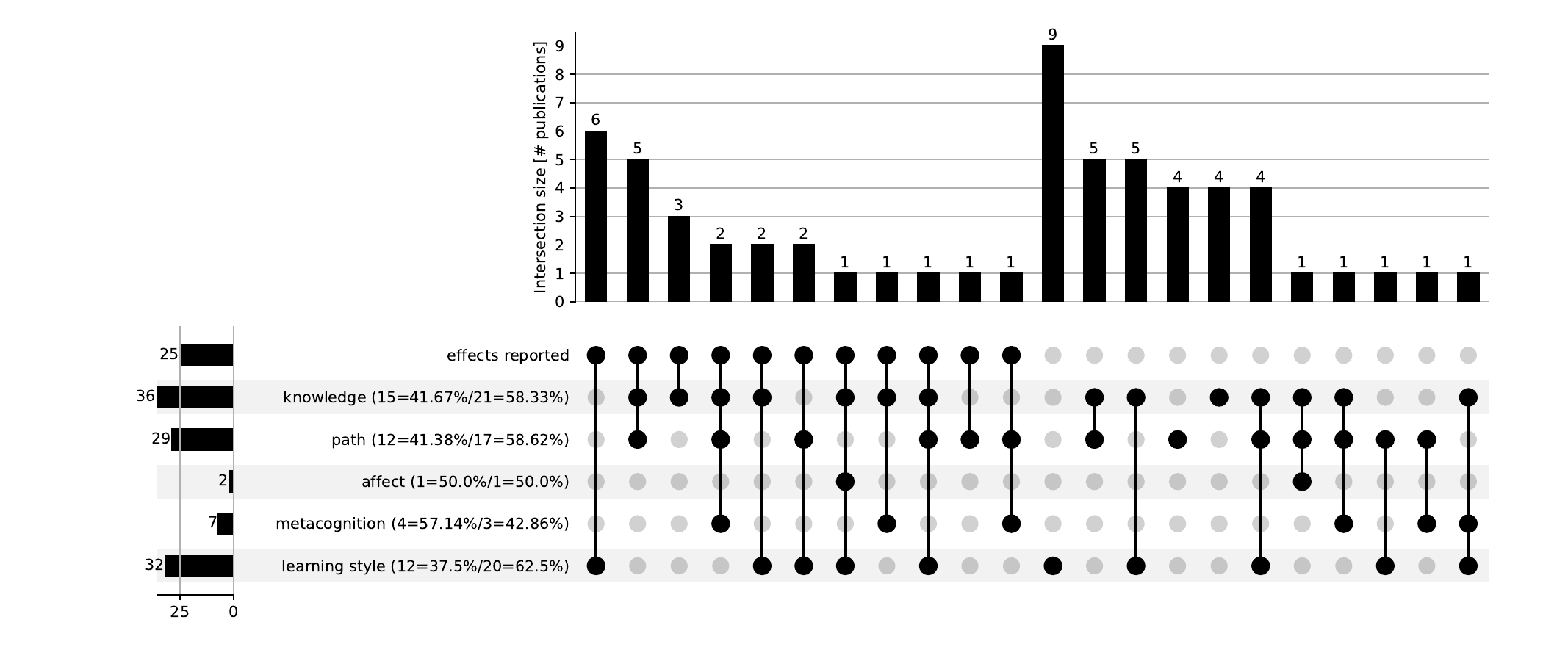}
  \caption{Considered combinations of adaptation bases of the reviewed approaches in relation to whether effects were reported for these approaches or not. For each adaptation basis, the number of approaches that consider this specific basis and reported effects is indicated on the left as a total number and percentage. On the right it is indicated how many approaches consider this specific adaptation basis and did not reported effects, showing the total number and percentage.}
  \label{fig:upsetPlotBasisEffect}
\end{figure*}

\begin{figure*}
  \centering
  \includegraphics[width=.9\textwidth]{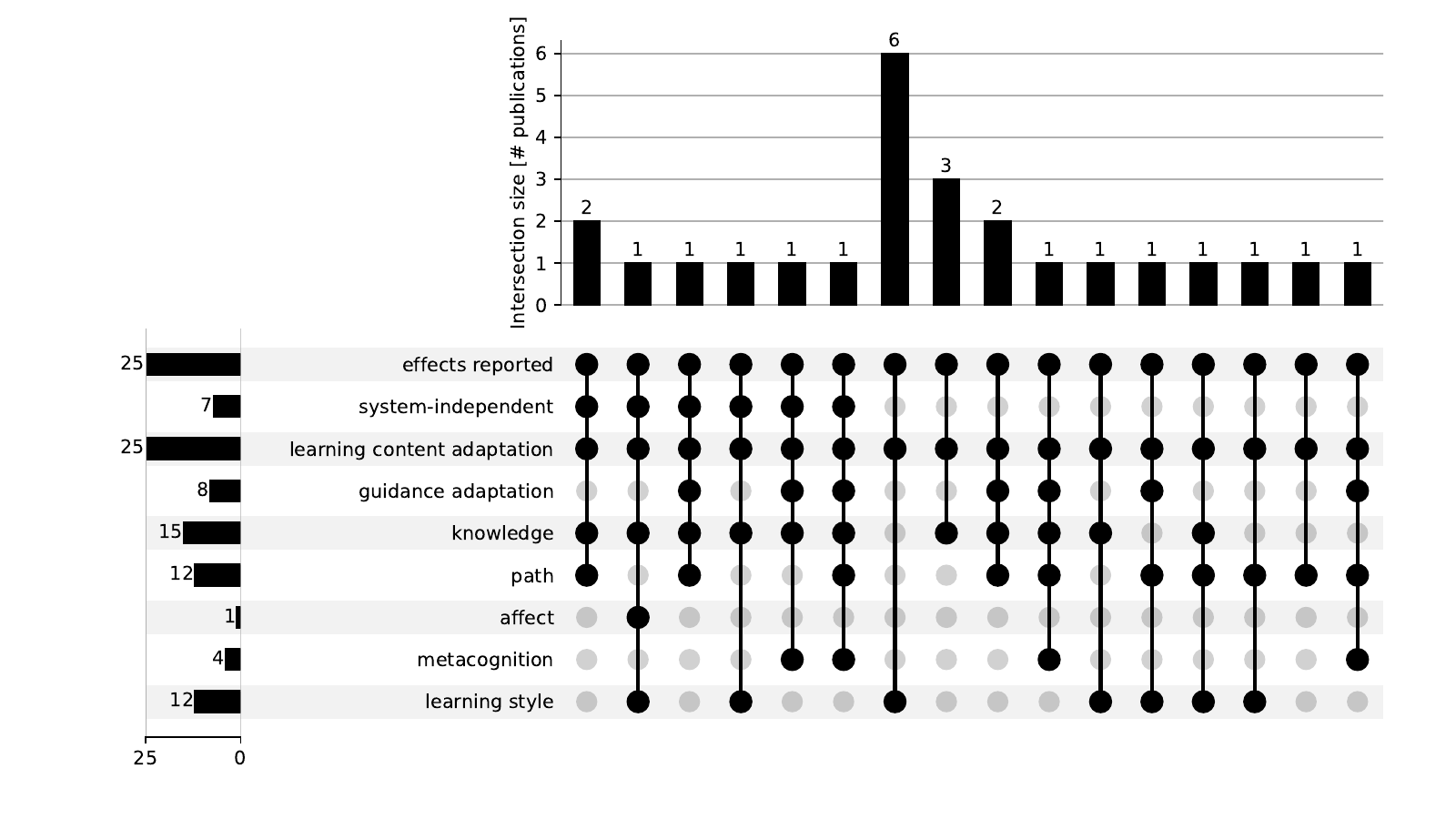}
  \caption{Implemented ALMs for which effects were reported in relation to the considered combinations of adaptation bases and whether they were implemented system-independently or for a specific LMS.}
  \label{fig:upsetPlotBasisAlmreducedSystemdependenceEffect}
\end{figure*}

The combinations of considered adaptation bases, as well as whether or not effects were reported for them, are shown in \Cref{fig:upsetPlotBasisEffect}.
When analyzing the adaptation bases separately, approximately half of the corresponding approaches reported effects, and approximately half did not.
Regarding the adaptation bases knowledge, path, and learning style, the proportion of approaches that did report effects was around 40\%, while for metacognition, the proportion was vice versa.
The results are equal when analyzing the considered combinations of adaptation bases.
40\% of the approaches that consider the combination containing only the learners' learning style, which was implemented most frequently, reported effects.
For the second most common combination of knowledge and path, 50\% of the corresponding approaches reported effects.
A peculiarity becomes apparent when analyzing the relationship between the implemented ALMs, for which effects were reported, the considered adaptation bases, and whether the approaches are system-independent or LMS-specific.
This is shown in \Cref{fig:upsetPlotBasisAlmreducedSystemdependenceEffect}.
In this plot, ALMs are categorized again by learning content and guidance adaptation approaches.
Due to the uncertainty regarding the reasons some reviewed approaches have not reported effects, we filtered out those without reported effects from the analysis to determine which ALMs succeed under which circumstances.
A total of 35\% of the system-independent approaches and 43.90\% of the LMS-specific approaches reported effects.
The most system-independent approaches that reported effects are those that adapt the learning content based on the learners' knowledge and learning path.
The most LMS-specific approaches that reported effects are those that adapt the learning content based on the learners' learning style.

\subsubsection{Discussion}

\paragraph{Restrictive Consideration of Multiple Adaptation Bases Hinders the Extensive Support of Self-Regulated Learning}
Providing effective support for self-regulated learning requires considering all conditions that influence learning processes\cciteparen{winnie1998studying}.
These range from task conditions, such as available learning resources, to cognitive conditions of learners, including learning style, knowledge, and learning tactics.
The cognitive conditions correlate with the adaptation bases proposed by\ccitetext{aleven2016instruction} that we adopted in our educational framework.
However, no approach takes all adaptation bases into account, yet it is required to extensively support self-regulated learning.
The only bases that are frequently considered in combination are learners' knowledge and the learning path.
We identified two likely reasons for this.
First, ALMs that focus on adapting the provided guidance or learning content based on the learner's knowledge usually also foster the acquisition of that knowledge, as is common with ITS.
This typically involves analyzing the learner's learning path, including errors made and requests for help.
Second, learners' knowledge of a specific subject can be estimated by tracking and evaluating their interactions.
Since this information is already related to the learners' learning path, it stands to reason that the reviewed approaches leverage this information for further adaptation.

The following are likely reasons, why these bases are not considered in combination with others.
Many approaches highlight the benefits of personalized learning, which can be achieved through adaptations based on a single basis, such as learners' learning styles.
Furthermore, these approaches argue that especially this restriction is advantageous for two reasons.
First, the fewer bases of adaptation considered, the less learning data must be processed.
Thus, these approaches can be more easily adopted and integrated into different LMSs.
Second, if multiple bases are considered, more conditions must be considered during the evaluation of ALMs.
This makes evaluations more difficult to conduct and results in less representative outcomes.

\paragraph{Consideration of Metacognition and Self-Regulated Learning Is Underestimated}
Developing self-regulated learning skills closely related to metacognition is crucial for learning success\cciteparen{zimmerman2002becoming}.
Nevertheless, only a fraction of the approaches consider metacognition as an adaptation basis.
The self-regulated learning model proposed by\ccitetext{zimmerman2000attaining} is only considered by one approach.
The COPES model\cciteparen{winnie1998studying} is not considered at all.
There are two likely reasons for this.
First, because self-regulated learning focuses on the entire learning process, the benefits of ALMs may not be apparent immediately.
Instead, learners are supported during multiple learning sessions to develop self-regulated learning skills that will benefit them later on when planning their learning process or when encountering problems.
However, most of the reviewed approaches focus solely on short-term evaluations.
Sometimes, they consider only a single learning session.
Consequently, they deem ALMs that focus on the same time range as more beneficial.
When considering learning setups like MOOCs, in which learners assess the suitability of the setup and engage in short learning sessions, such ALMs can be highly valuable.
However, during long-term learning, ALMs should be considered that focus on the entire learning process.
Second, the described self-regulated learning models focus on a didactic point of view.
From a technical point of view, these models allow for interpretation.
Other models, however, are more accurate in terms of implementation.
For example, models that focus on selecting learning content based on learners' learning styles.
When technicians implement ALMs\ccitecf{rose2019explanatory}, they may prefer these models.

Taking these two aspects into consideration, two approaches can be used to encourage the use of complex didactic scenarios that focus on the entire learning process.
First, interdisciplinary teams of technicians and didactic experts can develop ALMs in cooperation\ccitecf{proske2023towar}.
This allows for a case-specific analysis of open points regarding the implementation of didactic models, leading to educationally feasible solutions.
Second, the authoring of ALMs can be supported in a way that allows didactic experts to execute them without needing technical support.

\paragraph{System-Independent Approaches Lack Capabilities to Consider Learning Styles}
LMS-specific approaches frequently consider the learner's learning style as the basis for adaptation.
They have shown this to be beneficial.
However, system-independent approaches rarely take learning style into account.
There might be two reasons for this.
First, research on system-independent ALMs is often grounded in combining the benefits of LMSs and ITSs.
Research on LMS-specific ALMs often focuses on improving learning within existing LMSs by preventing information overload.
ITSs that focus on providing adaptive guidance through tutoring must be based on learners' knowledge and learning path.
Consequently, most system-independent approaches are based on learners' knowledge or learning path, as can be observed in \Cref{fig:upsetPlotBasisSystemDependence}.
There is only one system-independent approach that neither considers knowledge nor the learning path.
Second, most LMS-specific approaches select learning content to prevent information overload.
To do so based on learners' knowledge or learning path, functionalities are required to track and make interaction data accessible.
While not every LMS provides these functionalities, detection of learning styles, for example according to the Felder-Silverman learning style model\cciteparen{felder1988learning}, can be done easily using questionnaires\cciteparen{soloman2005index}.
Furthermore, these learning styles are usually based on evaluated and established learning frameworks.

In summary, adapting learning content based on learning styles is a worthwhile research target due to its easy applicability and didactic foundation.
The reason corresponding ALMs are not adopted as frequently by system-independent approaches as by LMS-specific approaches is that they require exact knowledge about the structure and properties of existing learning content.
When they are implemented for a specific LMS, the available learning content formats are known.
In contrast, system-independent approaches lack this knowledge because there is no established specification representing the necessary information to adapt the learning process independently of the LMS.

\paragraph{Established Learning Frameworks Lack Adoption Beyond Learning Style-Based ALMs}
Many approaches do not adopt an established learning framework. 
These results lead to two conclusions.
First, there is no single established learning framework adopted across all approaches to implement adaptive learning in digital learning environments\cciteparen{du2024personalized}.
Thus, the explicated educational framework appears suitable for aligning the reviewed approaches with established learning frameworks.
Second, although adopting established learning frameworks is appropriate for ALMs based on learning styles, it does not seem to be appropriate for the other adaptation bases.
There might be different reasons for this.
One reason may be that although there are established learning frameworks for different adaptation bases, they are not suitable for digital learning setups using LMSs.
They may not be suitable because the pedagogical concept is not suitable for this learning setup.
Alternatively, the learning setup may not provide sufficient data (\eg with regard to learning frameworks focused on affect) or intervention capabilities to adopt the learning framework.
Another reason could be that there are suitable learning frameworks but they cannot be put to practice.
This may be because these are high-level concepts that lack the detailed information necessary for computer scientists to apply them technically.
At the same time, they may lack authoring capabilities that are appropriate for instructional designers to provide this missing information or implement ALMs independently.
A final possible reason is that the related research is conducted by interdisciplinary teams of instructional designers and computer scientists, with the goal of investigating new learning frameworks.

\paragraph{Lack of Reporting on Negative Effects on Learning Prevents Derivation of Anti-Patterns}
Many approaches have not evaluated the effects of the developed ALMs on the learning process.
Furthermore, no approach has reported conducting an evaluation that found no effects or negative effects.
This phenomenon, referred to as the \textit{file drawer problem}\cciteparen{rosenthal1979file}, may have two causes.
First, an evaluation was conducted and either no effects were found, or negative effects were found, but these effects were never published.
Second, no evaluation was conducted, indeed.
The absence of evaluations may be attributed to one of three possible reasons.
First, an evaluation may be deemed non-necessary given the derivation of the developed ALMs.
As previously mentioned, many approaches argue that adaptive learning is generally more advantageous than one-size-fits-all solutions.
Subsequently, they focus on implementing well-described ALMs or adopting established learning frameworks that have been evaluated and found helpful in certain learning scenarios.
Therefore, an evaluation of the ALMS in a learning scenario that is similar to the one previously evaluated may be considered unnecessary.
Furthermore, many approaches do not specify a concrete didactic goal variable while they focus on the generic goal of personalizing learning.
Examples of such variables include the median of learners' grades, the passing rate for a specific course, and motivation measured according to a fixed scheme.
Such a variable can be measured to evaluate the appropriateness of ALMs.
By contrast, such a generic goal is either considered to have been reached by default due to the implementation of any ALMs, or it cannot be properly evaluated.
One last reason may be that many ALMs are elaborated by technicians or AI experts\cciteparen{rose2019explanatory}.
They are primarily interested in evaluating the possibility of implementing specific ALMs and the applicability in a learning scenario based on present learning process data.
Thus, they often focus on implementing established learning frameworks or work in interdisciplinary teams, leaving the evaluation of the educational suitability to didactic experts or instructional designers.
In summary, this prevented us from analyzing which characteristics of ALMs failed to positively influence the learning process, and which framing conditions should be avoided to prevent failure of specific ALMs.
\section{Discussion}
\label{sec:discussion}

The results allow us to derive several implications for research on ALMs for LMS and their practical application. 
These implications are discussed in \Cref{sec:implications}.
Afterwards, in \Cref{sec:limitations}, we discuss the limitations of our review and how we intended to address them.

\subsection{Implications}
\label{sec:implications}

\paragraph{Joint Use of ALMs for LMSs by Instructors and Researchers Remains a Challenge}
There are two challenges when conceptualizing an ALS for our target user groups.
First, there are different LMSs in use that work with data in different formats.
To make use of existing learning content and avoid re-authoring, ALMs should be usable with different existing LMSs.
This review focuses on ALMs that are integrated system-independently into LMSs.
However, this does not necessarily mean that these mechanisms consider existing learning content.
In contrast, most ALMs focused on a specific LMS do work with these LMSs' native learning content.
If the LMS on which the user is working is not the LMS being focused on, the user will either have to recreate the learning content in the other LMS or will not be able to use the approach, \eg due to the limitations of the usable LMSs of the user's institution.
This is why we focus on system-independent approaches.
The second challenge is that different ALMs are needed for different scenarios.
This may be due to differences in learning content, learning scenarios, or the fact that adjusting the ALMs is a primary goal, as is the case when the target user group of researchers applies them.
However, none of our target user groups can be expected to have the technical expertise necessary to adjust ALMs programmatically.

We proposed three hypotheses corresponding to different characteristics of state-of-the-art approaches to system-independently integrating ALMs into LMSs to be validated by this review.
These hypotheses contradict the aforementioned challenges and, if supported, can therefore be a starting point for future work in this research direction.
The first hypothesis is that ALMs that are system-independently integrated into LMSs rarely support the consideration of existing data~(\ie\ref{h:insufficientDataBasisSupport}).
Thus, they do not address the first of the described challenges.
The second hypothesis is that these approaches typically apply only a limited number of data processing mechanisms (\ie\ref{h:insufficientProcessingMechanismsSupport}).
This limits the implementable ALMs, contradicting challenge number two.
The third hypothesis is that these approaches rarely support our target user groups in authoring or adjusting the provided ALMs~(\ie\ref{h:insufficientAuthoringCapabilitiesSupport}).
This also contradicts challenge number two.
This review's results support all three hypotheses.
Thus, it can be concluded that current approaches to integrating ALMs into LMSs do not meet all the requirements of our target user groups.

\paragraph{Independence of LMS Remains a Challenge}
The results imply that system-independently integrating ALMs into LMSs remains a challenge.
There are ALMs that can be provided system-independently and that have been reported as beneficial by at least one evaluation.
However, these mechanisms often remain system-dependent in terms of data access, do not support learning with existing content, or simplify applied didactic scenarios for portability.
Furthermore, there are ALMs that have been reported as beneficial by at least one evaluation in LMS-specific setups but that could not be provided system-independently yet.
One example of this is the selection of learning content based on learning styles.
This is due to these ALMs processing system-dependent information, which cannot yet be described system-independently.

To address this issue, a system-independent standard should be introduced that describes the structure and properties of learning content and attached ALMs in conjunction.
Subsequently, LMSs implementing this standard could apply an ALM based on the specifics of existing learning content by mapping it according to standardized descriptions.
Although existing specifications, such as SCORM, aim to achieve this goal, they typically introduce new learning content formats instead of describing existing ones.
This prevents the use of existing content and impedes a comprehensive introduction.
Furthermore, implementing ALMs within LMSs according to such a standard would allow for the application of invasive ones. 
Nevertheless, this may require significant development effort for established LMSs, impeding comprehensive introduction as well.
To overcome this issue, the specified ALMs could also be implemented system-independently and applied to corresponding LMSs through adapters, as targeted by\ccitetext{verdu_integration_2017}. 
This, in turn, would require an additional specification of which functionalities the adapters must implement.

Furthermore, whether all required information for ALMs can be represented system-independently at all, may be investigated in future work.
Corresponding questions to focus on include whether information for highly learning content structure-dependent ALMs can be represented system-independently, whether there is a minimal data set that can be assumed to be fillable by each LMS, and how to handle missing data.
Additionally, the feasibility of transferring information about existing learning content to a newly specified learning content description format should be investigated.
Focusing future research on AI technologies for enriching data may be worthwhile, as it could save human resources and handle the provision of properties that cannot be derived algorithmically.

\paragraph{Comprehensive Introduction of System-Independent ALMs Depends on the Target LMSs}
Many system-independent approaches focus on their own specification for ALMs.
This is due to the lack of established standards for specifying ALMs and describing the learning content to which they can be applied.
These mechanisms are typically controlled and implemented via separate applications and applied to different LMSs through LMS-specific adaptation layers.
Thus, although these mechanisms are implemented system-independently, they are highly dependent on the LMSs to which they are applied.
To address this issue, make ALMs more system-independent, and foster their interaction with new LMSs, a few approaches base their implementations on two types of established specifications.
First, they focus on specifications for representing learning process data. 
The xAPI specification, for example, can describe interaction data.
The ALMs proposed by\ccitetext{zagorskis_eci_2020} are based on a well-defined API that provides learning process data using this specification.
Second, they focus on established mechanisms for integrating extended learning tools into existing learning environments.
The Learning Tools Interoperability (LTI) standard, which allows for the integration of learning tools in a system-independent manner, is an example of this.
To foster the introduction of system-independent ALMs, LMS developers should focus on such established specifications.
On the other hand, ALM developers should focus on adopting established specifications rather than introducing new data models or specifications.

\paragraph{Appropriateness of the Applied ALMs Lacks Comprehensive Empirical Evidence}
Many of the approaches did not report effects of the elaborated ALMs.
This finding aligns with the conclusions reported by\ccitetext{ochukut_research_2023}.
The reason why no effects were reported often remains unclear.
This makes it impossible to identify anti-patterns to prevent when applying ALMs.
One reason for this may be that the considered mechanisms do not clearly define educational goals. 
Another reason may be that these mechanisms cannot be tried out and validated by others such as didactic experts.
As described above, a unified specification format could be used to describe ALMs, making them available for others to try out and validate.
This would address the latter issue.
ALM developers should clearly define evaluable didactic goals.
Many approaches claim to aim for personalization without specifying any measurable instructional variables.
This complicates the planning of appropriate evaluations and hinders assessment of these approaches' educational value.
Reviewing the educational goals addressed by ALMs, their connections, and the application context of the corresponding mechanisms can help to identify such appropriate goals.
This could be a worthwhile analysis objective for a subsequent review.

Regarding future work on elaborating ALMs, evaluations should assess both their technical feasibility and their educational impact.
Furthermore, more attention should be paid to these evaluations, especially over time.
Although some approaches were evaluated, we frequently identified the following limitations.
Some approaches were only evaluated in the short term, sometimes limited to a single learning session.
This makes it impossible to make statements about didactic scenarios focusing on long-term learning processes, such as self-regulated learning.
Some approaches were only evaluated with a limited number of participants, such as a course.
This limits the generalizability of the results.
In order to gather evidence that either supports or refutes the didactic feasibility of different ALMs for LMSs, these patterns should be avoided.
Furthermore, evaluation results indicating no impact or a negative impact on learning outcomes should also be published.
For educators, knowing which ALMs to apply in specific digital learning scenarios is as important as knowing which to avoid.

\paragraph{Educational Theory and Practical Application of ALMs for LMSs Diverge}
While there are established learning frameworks that are adopted by ALMs based on learning styles, approaches that consider other adaptation bases rarely adopt such.
This may indicate a growing gap between educational theory and the practice of applying ALMs to LMSs.
Possible reasons for this may be a lack of appropriate learning frameworks for LMS-based learning, or missing technical information required to apply existing frameworks.
Without an evaluation of elaborated ALMs or an adoption of established learning frameworks, it is not feasible to contextualize their educational benefit.
Furthermore, even if approaches are evaluated but do not adopt an established learning framework, the generalization of the evaluation results remains difficult.
This is because evaluations often demonstrate applicability only for specific learning scenarios or subjects.
In contrast, established learning frameworks have typically been proven successful in a broader range of contexts.
Thus, developing, evaluating, and describing learning frameworks appropriate for LMS-based learning could be a worthwhile target for future research.

To enable didactic experts and educational researchers to investigate, describe, and evaluate learning frameworks, the required authoring capabilities are necessary.
These capabilities must be accompanied by functionalities that help users understand learning process data and evaluation functionalities.
Future research could investigate graphical authoring capabilities and authoring support functionalities using AI to provide these functionalities for our target user groups.
In this context, it can also be investigated whether AI could support instructors in selecting the most appropriate ALMs for a given learning subject by analyzing the available learning process data.
However, replacing authoring and the application of ALMs with such mechanisms may not be appropriate.
This is due to removing the instructor from the decision-making process.
Nevertheless, selecting ALMs to apply to a specific situation in some learning scenarios may require contextual information that cannot be derived from learning process data and that requires human expertise\cciteparen{holstein2020conceptual}.
Another reason may be that AI used in an educational context requires instruction on what adaptation is useful from a didactic perspective in different scenarios.

Selecting or suggesting the most appropriate ALMs for a given scenario using AI would also require the aforementioned authoring functionalities.
The reason for this is that instructors may still want to understand how these mechanisms work and what data they process to build trust in them.
For this purpose, instructors would need authoring capabilities to investigate the functionality of ALMs, and the AI functionalities would require these capabilities to visualize them.

\paragraph{Supporting Self-Regulated Learning Lacks Sufficient Consideration}
Our analysis identified several factors hindering the effective support of self-regulated learning.
First, existing approaches only allow for considering a limited number of learning conditions at a time.
This prevents the consideration of interrelated learning aspects that influence learning success, which is crucial for effectively supporting the development of self-regulated learning skills.
Second, especially the system-independent approaches often create and support learning within learning environment islands with separate learning content and interaction tracking functionalities.
Thus, instead of supporting the entire learning process within an LMS, which is essential for developing self-regulated learning skills, these approaches only support individual learning activities.
Third, the reviewed approaches primarily focus on supporting performance according to the COPES and Zimmerman models, but they mostly neglect forethought and self-reflection.
This is evident through a strong emphasis on program-controlled ALMs and a neglect of learner autonomy as a critical component in developing self-regulated learning skills.

In addition to the technical restrictions due to missing specifications and standards, our analysis identified two likely reasons that contribute to these factors.
The first reason is the lack of awareness of the importance of metacognition and self-regulation for successful learning, especially in digital learning environments\ccitecf{zimmerman2002becoming}.
Only a few of the reviewed approaches consider metacognition as a basis for adaptation.
The self-regulated learning model proposed by\ccitetext{zimmerman2000attaining} is considered only in a single approach.
No approach considers the COPES model\cciteparen{winnie1998studying}.
This may also be due to these models being accurate from a didactic standpoint, yet insufficient for the technical implementation of ALMs.
Thus, if technicians develop ALMs\ccitecf{rose2019explanatory}, they may abandon adopting these models.
Nevertheless, simplifying these models to the extent that their didactic value is diminished, in order to facilitate technical implementability, is not a viable option.
Therefore, similar to\ccitetext{rose2019explanatory}, we recommend developing ALMs within interdisciplinary teams of computer scientists and instructional designers.
This will combine didactic expertise with the technical knowledge and skills necessary to implement the corresponding support mechanisms.

The second reason may be that supporting self-regulated learning may become valuable only if the corresponding learning process is sufficiently complex and takes a certain amount of time.
This goes beyond learning a single subject by heart and may be insufficient for short learning sessions.
Subsequently, it is difficult to handle the corresponding ALMs in small, short-term evaluations.
However, supporting self-regulated learning is of increasing value for the learning process, especially with regard to system-independent approaches that focus on supporting learning across learning sessions and LMSs.
Thus, we conclude that future work should focus more intensively on conceptualizing support for self-regulated learning.
Our review results indicate the presence of a research gap in this regard.

\subsection{Limitations}
\label{sec:limitations}
Even though we carefully conducted our scoping review, some limitations remain.
\one Only papers included in the selected primary databases (Scopus and Web of Science) could be considered.
This is especially a limitation if records of relevant journals or conferences in the field are not indexed in the databases considered.
We have addressed this issue in two ways.
First, during the selection of these databases, we compiled a list of journals and conferences that we considered relevant based on the results of a general literature search, and checked whether records of these journals and conferences are contained in these databases.
Second, we searched Google Scholar and conducted snowball searches, which indicated that no additional papers on this topic are available.

\two The selection of papers highly depends on the terminology used, as the field of research corresponding to ALMs, or technology-enhanced learning in general, is on the borderline between different research disciplines.
To address this problem, we examined the terms used in research related to our review topic by analyzing related work and reviews on the terminology of related work to ensure that we include the most common terms of the relevant disciplines in our search string.

\three The selection and data charting process was performed by the first author of this paper only.
To minimize the risk of bias, we specified the inclusion criteria before conducting the review, refined them during team discussions, and attempted to resolve ambiguities that arose during the data selection or data charting process during team discussions.
For this purpose, we proceeded in case of ambiguities as follows.
First, the first and either the second or third authors of this paper captured all the variables for the approach in question.
Afterwards, we checked which variables had different values.
For these variables, either the second or third author, who was not involved in the first step, captured corresponding information as well.
Finally, we used the most frequently captured values for each variable.

\four The review only considers implemented approaches.
Therefore, we excluded papers that contained only conceptual ideas.
To minimize the risk of excluding approaches that implemented the proposed concepts without describing the implementation in the extracted paper, we also considered related papers.
The purpose of this methodology was to exclude purely conceptual approaches because their applicability had not yet been evaluated.
These approaches are outside the scope of this review.

\five The information extraction was highly dependent on the quality of the description provided by the selected papers.
This is due to the wide range of information covered by the research questions, as well as the lack of a standardized methodology for describing that information.
We attempted to resolve potential ambiguities by asking an author of the considered paper or during team discussions.

\six We had to rely on the information in the selected papers regarding the analysis of effects because we could not verify their accuracy.
To ensure we only include meaningful results based on conscientious evaluations, we only considered peer-reviewed papers.
\section{Future Prospects}
\label{sec:futureProspects}

Based on this review, future work should focus on the interrelated psychological and technological aspects of adaptive learning for LMSs, addressing the discussed implications.
We focus on the challenges that arise from the proposed hypotheses for the target user groups regarding authoring and provisioning.
To do so, we investigate the possibilities for a user-friendly modeling environment for ALMs for LMSs that fulfill the requirements of both user groups.
To work toward such a modeling approach, two questions need to be addressed.

\paragraph{Conceptual Model and Architecture}
First, how can ALMs for LMSs be specified system-independently?
As this review shows, existing processable definitions of ALMs are either not system-independent or do not consider all defined processing mechanisms, thus limiting the implementable ALMs.
Consequently, our research focuses on two areas.
First, we investigate how ALMs can be described in a machine-processable format that can be used in a user-friendly modeling process and that takes the specified processing mechanisms into account.
Second, we investigate how to conceptualize an architecture that enables system-independent modeling and provisioning of ALMs, considering system-specific content and interaction data.

While there is already research activity in the direction of feasible architectures\cciteeg{fuchs_intui_2022,kapenieks_eci_2020,kucharskiArchitecturalDesignAdaptive2023}, the conceptual model requires a specification of the pedagogical model of the three original models of ALSs\cciteparen{meier2019ki} in principle, considering the requirements our target user groups.

\paragraph{Support for Modeling and Provision for Both Target User Groups}
Second, how can our target user groups model ALMs that consider existing data?
In order to make ALMs fully accessible to our target user groups, who may be technically inexperienced, the entire process from conceptualization to provision must be supported in a user-friendly way.
This process to be supported includes three steps, analogous to the adaptation process itself\cciteparen{argotte-ramosDynamicCourseEditor2009}.
The first step is understanding the data that will be considered during the adaptation process, as well as the mechanisms for accessing and referencing that data.
In this regard, we investigate how to make the generated interaction data, its structure, the extracted learning content, and the corresponding metadata understandable.
Furthermore, we investigate mechanisms for requesting and referencing this data during the adaptation process that do not require programming skills.
The second step focuses on user-friendly modeling of ALMs.
In this regard, we investigate how to model how the previously described data is used by the specified processing mechanisms to apply the intended adaptation.
The third step focuses on the adaptation during the learning process itself.
In this regard, we investigate effective observation methods for adaptation processes.
These allow modelers to verify that the modeled ALMs result in the desired adaptation of the learning process and, consequently, the desired learning outcomes.
Furthermore, we investigate how to enable our target user groups to simulate ALMs before applying them to real learning processes, in order to verify or evaluate their functionality.
\section{Conclusion}
\label{sec:conclusion}

In this scoping review, we analyzed the methods proposed by the literature for integrating ALMs into LMSs.
Although it focuses on system-independent approaches, we also included LMS-specific ones in our analysis.
We selected 61 approaches for further analysis out of 4973 papers extracted from 2003 to 2023.
We analyzed the approaches in terms of how they integrate ALMs into LMSs, how the integrated mechanisms are provided in the LMSs, how the ALMs are specified, on which data base they operate, the educational conditions surrounding the application of learning process adaptations, and the educational outcomes.
Afterwards, we evaluated the analysis results based on three hypotheses.
The first hypothesis was that ALMs that are system-independently integrated into LMSs can rarely consider existing data.
The second hypothesis was that these ALMs typically support only a limited number of data processing mechanisms.
The third hypothesis was that our target user groups (\ie instructors and instructional designers) rarely have the ability to model or adapt these ALMs.
Instead, predefined mechanisms are proposed that can be parameterized by these users at most.
The presented results support all hypotheses.
Furthermore, regarding our target user groups interested in modeling and providing ALMs for LMSs system-independently, we discussed these groups' main requirements regarding the modeling and provision process.
We also discussed the challenges arising from these requirements and how they contradict with the proposed hypotheses.
Finally, we discussed the implications of these results and outlined the next steps for further research.
We focus on the following two main steps.
First, we conduct research towards a conceptual model for the system-independent specification of ALMs and an architecture that enables the system-independent provision of these mechanisms in LMSs.
Second, we investigate how to enable our target user groups to carry out the process from the system-independent conceptualization of ALMs to the provision in the intended LMS on their own. 
\subsection*{Declarations}

\textbf{Conflict of interests} On behalf of all authors, the corresponding author states that there is no conflict of interest.

\appendix

\section{Educational Surrounding Conditions and Outcomes}
\label{sec:resultsTableEducationalContext}

{
\small
\begin{longtable}{|p{.205\textwidth}|p{.795\textwidth}|}
      \caption{Summary of the results corresponding to \ref{v:adaptationGoal}, \ref{v:adaptationBasis}, and \ref{v:reportedEffects}. The approaches are listed in order of publication year, with the most recent one appearing last.}
      \label{tab:scopingReviewResultsEducationalContext}\\
      \hline
      \rowcolor{gray!50} %
      \multicolumn{2}{|l|}{\ccitetext{santos2004overview}} \\ \hline %
      \goal & improve online learning effectivity \\ \hline %
      \basis & \abpath \\ \hline %
      \effect & no effects reported \\ \hline %
      \rowcolor{gray!50} %
      \multicolumn{2}{|l|}{\ccitetext{watson2005steps}} \\ \hline %
      \goal & personalize learning \\ \hline %
      \basis & \abknowledge, \abpath, \abmetacognition, \abstyle \\ \hline %
      \effect & no effects reported \\ \hline %
      \rowcolor{gray!50} %
      \multicolumn{2}{|l|}{\ccitetext{arapi2007supporting}} \\ \hline %
      \goal & enable the authoring of pedagogically sound personalization mechanisms that can be reused by other instructors \\ \hline %
      \basis & \abknowledge, \abpath, \abstyle \\ \hline %
      \effect & no effects reported \\ \hline %
      \rowcolor{gray!50} %
      \multicolumn{2}{|l|}{\ccitetext{cobos2007learning}} \\ \hline %
      \goal & personalize learning according to teacher-chosen learning style models \\ \hline %
      \basis & \abstyle{}~(\fslsm, \halsm) \\ \hline %
      \effect & learners stated that the personalized material helped them learn and pass the exam \\ \hline %
      \rowcolor{gray!50} %
      \multicolumn{2}{|l|}{\ccitetext{roselli2008integration}} \\ \hline %
      \goal & enable the authoring of ITS that can support learners with their homework and report their performance to instructors, who can then work specifically with learners who need support \\ \hline %
      \basis & \abpath, \abmetacognition \\ \hline %
      \effect & authoring efficiency was evaluated, no effects for learners were reported \\ \hline %
      \rowcolor{gray!50} %
      \multicolumn{2}{|l|}{\ccitetext{serce_moda_2008}} \\ \hline %
      \goal & personalize learning \\ \hline %
      \basis & \abstyle{}~(\fslsm) \\ \hline %
      \effect & no effects reported \\ \hline %
      \rowcolor{gray!50} %
      \multicolumn{2}{|l|}{\ccitetext{pedrazzoli2009artificial}} \\ \hline %
      \goal & increase learner performance and motivation \\ \hline %
      \basis & \abknowledge, \abpath, \abmetacognition \\ \hline %
      \effect & no effects reported \\ \hline %
      \rowcolor{gray!50} %
      \multicolumn{2}{|l|}{\ccitetext{adorni2010caddie}} \\ \hline %
      \goal & enable instructional design of personalized learning processes \\ \hline %
      \basis & \abknowledge \\ \hline %
      \effect & increase in the number of amassed competencies during the learning process \\ \hline %
      \rowcolor{gray!50} %
      \multicolumn{2}{|l|}{\ccitetext{graf2010flexible}} \\ \hline %
      \goal & personalize learning \\ \hline %
      \basis & \abstyle{}~(\fslsm) \\ \hline %
      \effect & no effects reported \\ \hline %
      \rowcolor{gray!50} %
      \multicolumn{2}{|l|}{\ccitetext{limongelli2010module}} \\ \hline %
      \goal & personalize learning \\ \hline %
      \basis & \abknowledge, \abstyle{}~(\fslsm) \\ \hline %
      \effect & no effects reported \\ \hline %
      \rowcolor{gray!50} %
      \multicolumn{2}{|l|}{\ccitetext{peter2010adaptable}} \\ \hline %
      \goal & personalize learning \\ \hline %
      \basis & \abstyle{}~(\varklsm) \\ \hline %
      \effect & no effects reported \\ \hline %
      \rowcolor{gray!50} %
      \multicolumn{2}{|l|}{\ccitetext{santos_web-based_2010}} \\ \hline %
      \goal & enable the authoring of personalized learning environments using SCORM \\ \hline %
      \basis & \abknowledge, \abpath{}~(\all) \\ \hline %
      \effect & focus on authoring, no effects for learners were reported \\ \hline %
      \rowcolor{gray!50} %
      \multicolumn{2}{|l|}{\ccitetext{hammadadaptive}} \\ \hline %
      \goal & personalize learning \\ \hline %
      \basis & \abknowledge, \abstyle \\ \hline %
      \effect & no effects reported \\ \hline %
      \rowcolor{gray!50} %
      \multicolumn{2}{|l|}{\ccitetext{rossi2011mapit}} \\ \hline %
      \goal & answer learners' common learning-process-related questions automatically and subsequently free tutors from repetitive question answering. \\ \hline %
      \basis & \abpath \\ \hline %
      \effect & no effects reported \\ \hline %
      \rowcolor{gray!50} %
      \multicolumn{2}{|l|}{\ccitetext{yaghmaie_context-aware_2011}} \\ \hline %
      \goal & create reusable learning process personalisation \\ \hline %
      \basis & \abstyle{}~(\kolblsm~or \hmlsm ) \\ \hline %
      \effect & no effects reported \\ \hline %
      \rowcolor{gray!50} %
      \multicolumn{2}{|l|}{\ccitetext{al2012intelligent}} \\ \hline %
      \goal & personalize collaborative learning \\ \hline %
      \basis & \abstyle \\ \hline %
      \effect & learners stated that the established connections with learning peers and the instructor were helpful \\ \hline %
      \rowcolor{gray!50} %
      \multicolumn{2}{|l|}{\ccitetext{de_bra_grapple1_2013}} \\ \hline %
      \goal & enable instructors to define adaptation mechanisms for LMSs that support adaptive information exploration and individual, self-paced learning \\ \hline %
      \basis & \abknowledge, \abpath \\ \hline %
      \effect & focus on authoring, several experiments were conducted with modeled adaptation mechanisms that were considered helpful, but the extent of the benefit largely depends on the mechanisms created \\ \hline %
      \rowcolor{gray!50} %
      \multicolumn{2}{|l|}{\ccitetext{moutafi2013mining}} \\ \hline %
      \goal & personalize learning to reduce the learning time required and interference with the instructor \\ \hline %
      \basis & \abknowledge, \abstyle{}~(\fslsm) \\ \hline %
      \effect & no effects reported \\ \hline %
      \rowcolor{gray!50} %
      \multicolumn{2}{|l|}{\ccitetext{santos_phd_2013}} \\ \hline %
      \goal & enable personalization of learning in existing learning environments \\ \hline %
      \basis & \abknowledge, \abpath{}~(\all) \\ \hline %
      \effect & focus on authoring and reusability of authored mechanisms, extent of the benefit largely depends on the mechanisms created \\ \hline %
      \rowcolor{gray!50} %
      \multicolumn{2}{|l|}{\ccitetext{batanero2014considering}} \\ \hline %
      \goal & improve the accessibility of online education \\ \hline %
      \basis & \abstyle \\ \hline %
      \effect & no effects reported \\ \hline %
      \rowcolor{gray!50} %
      \multicolumn{2}{|l|}{\ccitetext{bhaskaran2014research}} \\ \hline %
      \goal & personalize online and offline learning according to learners' performance \\ \hline %
      \basis & \abknowledge, \abpath \\ \hline %
      \effect & learners stated that they could learn online and offline, that their performance was fairly classified, and that they considered the system helpful for the learning process \\ \hline %
      \rowcolor{gray!50} %
      \multicolumn{2}{|l|}{\ccitetext{giuffra_palomino_intelligent_2014}} \\ \hline %
      \goal & personalize learning by aggregating the benefits of ITSs and LMSs \\ \hline %
      \basis & \abknowledge \\ \hline %
      \effect & no effects reported \\ \hline %
      \rowcolor{gray!50} %
      \multicolumn{2}{|l|}{\ccitetext{pratas2014adapt}} \\ \hline %
      \goal & personalize learning based on proven pedagogical concepts to ensure that learning outcomes do not depend solely on learner motivation \\ \hline %
      \basis & \abstyle{}~(\fslsm, \varklsm) \\ \hline %
      \effect & no effects reported \\ \hline %
      \rowcolor{gray!50} %
      \multicolumn{2}{|l|}{\ccitetext{radwan2014adaptive}} \\ \hline %
      \goal & personalize learning \\ \hline %
      \basis & \abstyle{}~(\mbti) \\ \hline %
      \effect & no effects reported \\ \hline %
      \rowcolor{gray!50} %
      \multicolumn{2}{|l|}{\ccitetext{surjono2014evaluation}} \\ \hline %
      \goal & personalize learning \\ \hline %
      \basis & \abstyle{}~(\fslsm, \varklsm) \\ \hline %
      \effect & no effects reported \\ \hline %
      \rowcolor{gray!50} %
      \multicolumn{2}{|l|}{\ccitetext{munoz2015ontosakai}} \\ \hline %
      \goal & increase learning benefit \\ \hline %
      \basis & \abpath \\ \hline %
      \effect & no effects reported \\ \hline %
      \rowcolor{gray!50} %
      \multicolumn{2}{|l|}{\ccitetext{nafea2015novel}} \\ \hline %
      \goal & personalize learning \\ \hline %
      \basis & \abknowledge, \abpath, \abstyle (\fslsm) \\ \hline %
      \effect & no effects reported \\ \hline %
      \rowcolor{gray!50} %
      \multicolumn{2}{|l|}{\ccitetext{ovalle2015proposal}} \\ \hline %
      \goal & increase competition between learners and foster metacognition \\ \hline %
      \basis & \abpath, \abmetacognition \\ \hline %
      \effect & the learners' awareness of and understanding of their own learning failures and performance has increased \\ \hline %
      \rowcolor{gray!50} %
      \multicolumn{2}{|l|}{\ccitetext{sein2015design}} \\ \hline %
      \goal & foster the implementation of personalized learning in real-world scenarios by enabling non-technical users to apply it \\ \hline %
      \basis & \abknowledge, \abpath \\ \hline %
      \effect & the learners' motivation and final grades have increased \\ \hline %
      \rowcolor{gray!50} %
      \multicolumn{2}{|l|}{\ccitetext{sweta2015adaptive}} \\ \hline %
      \goal & enhance the learning process, learner's efficiency and learner engagement \\ \hline %
      \basis & \abstyle{}~(\fslsm) \\ \hline %
      \effect & the learner's satisfaction and grades have increased, the time required to acquire specific knowledge has decreased \\ \hline %
      \rowcolor{gray!50} %
      \multicolumn{2}{|l|}{\ccitetext{aleven_embedding_2016}} \\ \hline %
      \goal & Enable instructors to create ITSs for implementation in LMSs to improve learning through tutoring \\ \hline %
      \basis & \abknowledge, \abpath{}~(\all) \\ \hline %
      \effect & focus on authoring, no effects for learners were reported \\ \hline %
      \rowcolor{gray!50} %
      \multicolumn{2}{|l|}{\ccitetext{de_amorim_towards_2016}} \\ \hline %
      \goal & personalize learning \\ \hline %
      \basis & \abknowledge \\ \hline %
      \effect & no effects reported \\ \hline %
      \rowcolor{gray!50} %
      \multicolumn{2}{|l|}{\ccitetext{luccioni_sti-dico_2016}} \\ \hline %
      \goal & enable the learning of a specific skill (\ie the use of French electronic dictionaries) \\ \hline %
      \basis & \abknowledge, \abpath \\ \hline %
      \effect & no effects reported \\ \hline %
      \rowcolor{gray!50} %
      \multicolumn{2}{|l|}{\ccitetext{rani2016ontological}} \\ \hline %
      \goal & personalize learning \\ \hline %
      \basis & \abknowledge, \abpath, \abstyle (\varklsm) \\ \hline %
      \effect & no effects reported \\ \hline %
      \rowcolor{gray!50} %
      \multicolumn{2}{|l|}{\ccitetext{aleven2017integrating}} \\ \hline %
      \goal & increase quality of ITSs to foster personalized learning but simultaneously decrease development costs \\ \hline %
      \basis & \abknowledge, \abpath{}~(\all), \abaffect \\ \hline %
      \effect & focus on authoring, no effects for learners were reported \\ \hline %
      \rowcolor{gray!50} %
      \multicolumn{2}{|l|}{\ccitetext{lagman2017extracting}} \\ \hline %
      \goal & prevent learners with difficulties from failing \\ \hline %
      \basis & \abknowledge{}~(\mastery) \\ \hline %
      \effect & no effects reported \\ \hline %
      \rowcolor{gray!50} %
      \multicolumn{2}{|l|}{\ccitetext{oskouei2017proposing}} \\ \hline %
      \goal & support learning of a specific subject (\ie English as a foreign language) \\ \hline %
      \basis & \abpath \\ \hline %
      \effect & learners' satisfaction and learning performance increased, especially with regard to remembering words \\ \hline %
      \rowcolor{gray!50} %
      \multicolumn{2}{|l|}{\ccitetext{verdu_integration_2017}} \\ \hline %
      \goal & lower dropout rates resulting from isolated learning experiences and the subsequent frustration \\ \hline %
      \basis & \abknowledge, \abpath, \abmetacognition \\ \hline %
      \effect & learners' satisfaction and grades have improved, learners stated that they perceived the support provided by the system as useful and caring \\ \hline %
      \rowcolor{gray!50} %
      \multicolumn{2}{|l|}{\ccitetext{cai2018adaptive}} \\ \hline %
      \goal & personalize learning \\ \hline %
      \basis & \abknowledge \\ \hline %
      \effect & learners' engagement, grades, and passing rates have improved \\ \hline %
      \rowcolor{gray!50} %
      \multicolumn{2}{|l|}{\ccitetext{karagiannis2018adaptive}} \\ \hline %
      \goal & personalize learning \\ \hline %
      \basis & \abstyle{}~(\fslsm) \\ \hline %
      \effect & mechanism positively affected learners' motivation and performance \\ \hline %
      \rowcolor{gray!50} %
      \multicolumn{2}{|l|}{\ccitetext{perivsic2018semantic}} \\ \hline %
      \goal & improve learning outcomes and learner satisfaction with the learning process \\ \hline %
      \basis & \abstyle{}~(\fslsm) \\ \hline %
      \effect & learner performance increased, and learners stated that they perceived the implemented mechanisms as useful \\ \hline %
      \rowcolor{gray!50} %
      \multicolumn{2}{|l|}{\ccitetext{alsobhi2019adaptation}} \\ \hline %
      \goal & align the learning experience with the specific needs of learners who are dyslexic \\ \hline %
      \basis & \abpath, \abstyle{}~(\fslsm) \\ \hline %
      \effect & learner performance increased \\ \hline %
      \rowcolor{gray!50} %
      \multicolumn{2}{|l|}{\ccitetext{joseph2019adaptive}} \\ \hline %
      \goal & support slow learners during the learning process \\ \hline %
      \basis & \abknowledge, \abpath, \abstyle{}~(\fslsm) \\ \hline %
      \effect & performance of slow learners increased \\ \hline %
      \rowcolor{gray!50} %
      \multicolumn{2}{|l|}{\ccitetext{joy2019ontology}} \\ \hline %
      \goal & personalize learning \\ \hline %
      \basis & \abknowledge, \abstyle{}~(\fslsm) \\ \hline %
      \effect & no effects reported \\ \hline %
      \rowcolor{gray!50} %
      \multicolumn{2}{|l|}{\ccitetext{khosravi2019ripple}} \\ \hline %
      \goal & foster higher-order learning by letting learners create learning content and reduce learning content creation costs at the same time \\ \hline %
      \basis & \abknowledge, \abmetacognition{}~(\sap) \\ \hline %
      \effect & learners grades improved \\ \hline %
      \rowcolor{gray!50} %
      \multicolumn{2}{|l|}{\ccitetext{troussas2019injecting}} \\ \hline %
      \goal & personsalize learning \\ \hline %
      \basis & \abpath, \abstyle{}~(\varklsm) \\ \hline %
      \effect & learners had to rate the functional suitability, comparability, reliability, usability and portability, all of which were generally rated highly \\ \hline %
      \rowcolor{gray!50} %
      \multicolumn{2}{|l|}{\ccitetext{arsovic_e-learning_2020}} \\ \hline %
      \goal & improve learning outcomes \\ \hline %
      \basis & \abknowledge, \abstyle{}~(\fslsm) \\ \hline %
      \effect & grades improved \\ \hline %
      \rowcolor{gray!50} %
      \multicolumn{2}{|l|}{\ccitetext{louhab2020novel}} \\ \hline %
      \goal & improve learning results \\ \hline %
      \basis & \abknowledge{}~(\flippedca) \\ \hline %
      \effect & the approach was met with a high level of acceptance, and there was an improvement in the learners' grades \\ \hline %
      \rowcolor{gray!50} %
      \multicolumn{2}{|l|}{\ccitetext{sayed2020towards}} \\ \hline %
      \goal & improve learning effectivity \\ \hline %
      \basis & \abknowledge{}~(\bloom) \abstyle{}~(\varklsm) \\ \hline %
      \effect & no effects reported \\ \hline %
      \rowcolor{gray!50} %
      \multicolumn{2}{|l|}{\ccitetext{zagorskis_eci_2020}} \\ \hline %
      \goal & provide a standardized ALS architecture and interfaces that can be used by LMSs to personalize learning to increase learning efficiency \\ \hline %
      \basis & \abknowledge, \abpath, \abstyle \\ \hline %
      \effect & focus on architectural design, no effects for learners were reported \\ \hline %
      \rowcolor{gray!50} %
      \multicolumn{2}{|l|}{\ccitetext{fakoya2021automatic}} \\ \hline %
      \goal & personalize learning \\ \hline %
      \basis & \abstyle{}~(\fslsm) \\ \hline %
      \effect & grades improved \\ \hline %
      \rowcolor{gray!50} %
      \multicolumn{2}{|l|}{\ccitetext{pagano2021training}} \\ \hline %
      \goal & decrease learning time \\ \hline %
      \basis & \abknowledge \\ \hline %
      \effect & simulation indicates time saving, no effects for real learners were reported \\ \hline %
      \rowcolor{gray!50} %
      \multicolumn{2}{|l|}{\ccitetext{sychev_compprehension_2021}} \\ \hline %
      \goal & foster comprehension level understanding according to bloom's taxonomy \\ \hline %
      \basis & \abknowledge{}~(\bloom), \abpath \\ \hline %
      \effect & grades improved \\ \hline %
      \rowcolor{gray!50} %
      \multicolumn{2}{|l|}{\ccitetext{apoki_modular_2022}} \\ \hline %
      \goal & provide learning content that is appropriate for the larner \\ \hline %
      \basis & \abknowledge, \abaffect, \abstyle{}~(\fslsm) \\ \hline %
      \effect & the provision of appropriate learning content was confirmed by learners \\ \hline %
      \rowcolor{gray!50} %
      \multicolumn{2}{|l|}{\ccitetext{bradavc2022design}} \\ \hline %
      \goal & improve learning performance \\ \hline %
      \basis & \abknowledge, \abstyle{}~(\varklsm) \\ \hline %
      \effect & grades improved \\ \hline %
      \rowcolor{gray!50} %
      \multicolumn{2}{|l|}{\ccitetext{cardenas2022personalised}} \\ \hline %
      \goal & improve grades and learner satisfaction \\ \hline %
      \basis & \abknowledge{}~(\mastery), \abpath, \abmetacognition{}~(\srl) \\ \hline %
      \effect & grades improved and learner satisfaction was higher than 80\% \\ \hline %
      \rowcolor{gray!50} %
      \multicolumn{2}{|l|}{\ccitetext{yilmaz2022smart}} \\ \hline %
      \goal & improve learning success \\ \hline %
      \basis & \abknowledge{}~(\mastery), \abpath \\ \hline %
      \effect & learners stated that the system helped them to reach their learning goals \\ \hline %
      \rowcolor{gray!50} %
      \multicolumn{2}{|l|}{\ccitetext{kaouni2023design}} \\ \hline %
      \goal & provide learning content that is appropriate for the larner \\ \hline %
      \basis & \abstyle{}~(\fslsm) \\ \hline %
      \effect & no effects reported \\ \hline %
      \rowcolor{gray!50} %
      \multicolumn{2}{|l|}{\ccitetext{kouvara2023discovering}} \\ \hline %
      \goal & provide learning content that is appropriate for the larner \\ \hline %
      \basis & \abstyle \\ \hline %
      \effect & no effects reported \\ \hline %
      \rowcolor{gray!50} %
      \multicolumn{2}{|l|}{\ccitetext{pardos2023conducting}} \\ \hline %
      \goal & enable instructors and researchers to develop suitable ITS for their learning content, thereby fostering personalised learning \\ \hline %
      \basis & \abknowledge{}~(\mastery), \abpath{}~(\all) \\ \hline %
      \effect & focus on authoring, no effects for learners were reported \\ \hline %
      \rowcolor{gray!50} %
      \multicolumn{2}{|l|}{\ccitetext{senthil2023towards}} \\ \hline %
      \goal & preventing information overload by providing learning content that is appropriate for the larner \\ \hline %
      \basis & \abpath \\ \hline %
      \effect & no effects reported \\ \hline %
\end{longtable}
}  
\printbibliography[title={References}]

\end{document}